\documentclass[useAMS,usenatbib]{mn2e}
\usepackage{subfigure}
\usepackage{graphicx, amssymb}
\usepackage[fleqn]{amsmath}
\usepackage{epsfig}
\usepackage{color}

\title[Star Formation in the First Galaxies - II]{Star Formation in the First Galaxies - II: 
Clustered Star Formation and the Influence of Metal Line Cooling}

\author[C. Safranek-Shrader et al.]
{Chalence~Safranek-Shrader\thanks{E-mail: ctss@astro.as.utexas.edu}, Milo\v s~Milosavljevi\'c, Volker~Bromm\\
Department of Astronomy, University of Texas at Austin, Austin, Texas, USA
 }

\newcommand{\kelvin}{\mathrm{K}}
\newcommand{\cc}{\mathrm{cm}^{-3}}
\newcommand{\msun}{M_{\odot}}

\newcommand{\zsun}{Z_{\odot}}

\newcommand{\htwo}{\mathrm{H}_2}
\newcommand{\hd}{\mathrm{HD}}

\newcommand{\lya}{\mathrm{Ly}\alpha}
\newcommand{\tcmb}{T_{\mathrm{CMB}}}
\newcommand{\fshhtwo}{f_{\mathrm{shield},\mathrm{H}_2}}

\newcommand{\tff}{t_{\mathrm{ff}}}
\newcommand{\kms}{\mathrm{km}\,\mathrm{s}^{-1}}
\newcommand{\kb}{k_{\mathrm{B}}}

\newcommand{\ev}{\mathrm{eV}}

\newcommand{\pc}{\mathrm{pc}}
\newcommand{\au}{\mathrm{AU}}
\newcommand{\Mpc}{\mathrm{Mpc}}
\newcommand{\yrs}{\mathrm{yrs}}
\newcommand{\yr}{\mathrm{yr}}
\newcommand{\tvir}{T_{\mathrm{vir}}}
\newcommand{\jlw}{J_{\mathrm{LW},21}}
\newcommand{\cs}{c_{\mathrm{s}}}

\newcommand{\lj}{L_{\mathrm{J}}}

\newcommand{\mj}{M_{\mathrm{J}}}

\newcommand{\mach}{\mathcal{M}}

\newcommand{\omegab}{\Omega_{\mathrm{b}}}
\newcommand{\omegal}{\Omega_{\Lambda}}
\newcommand{\omegam}{\Omega_{\mathrm{m}}}

\newcommand{\tcool}{t_{\mathrm{cool}}}

\newcommand{\racc}{r_{\mathrm{acc}}}
\newcommand{\msunperyr}{M_{\odot}\,\mathrm{yr}^{-1}}

\newcommand{\rsoft}{r_{\mathrm{soft}}}
\newcommand{\myr}{\mathrm{Myr}}

\newcommand{\rhoj}{\rho_{\mathrm{J}}}

\newcommand{\vect}[1]{\boldsymbol{#1}}

\newcommand{\abunde}{x_{\mathrm{e}}}

\begin{document}

\label{firstpage}

\maketitle
\topmargin-1cm

\begin{abstract}

We present results from three cosmological simulations, only differing in gas metallicity, that focus on the impact of metal fine-structure line cooling on stellar cluster formation in a high-redshift atomic cooling halo. Sink particles allow the process of gas hydrodynamics and accretion onto cluster stars to be followed for $\sim4$ Myr corresponding to multiple local free-fall times. At metallicities at least $10^{-3}\,\zsun$, gas is able to reach the CMB temperature floor and fragment pervasively resulting in a stellar cluster of size $\sim1\,\pc$ and total mass $\sim1000\,\msun$. The masses of individual sink particles vary, but are typically $\sim100\,\msun$, consistent with the Jeans mass at $\tcmb$, though some solar mass fragments are also produced. Below $10^{-4}\,\zsun$, fragmentation is strongly suppressed on scales greater than $0.01\,\pc$ and total stellar mass is lower by a factor of $\sim3$ than in the higher metallicity simulations. The sink particle accretion rates, and thus their masses, are determined by the mass of the gravitationally unstable gas cloud and prolonged gas accretion over many Myr, exhibiting features of both monolithic collapse and competitive accretion. Even considering possible dust induced fragmentation that may occur at higher densities, the formation of a bona fide stellar cluster seems to require metal line cooling and metallicities of at least $\sim 10^{-3}\,\zsun$.

\end{abstract}


\section{Introduction}

The process of star formation is extremely complex, with gas thermodynamics, supersonic turbulence, self gravity, magnetic fields, and stellar feedback all playing important roles on a variety of spatial and temporal scales. It is observed that the vast majority of present-day stars form as part of small associations or clusters inside supersonically turbulent molecular clouds \citep[e.g.,][]{Lada03}. The stellar initial mass function (IMF) peaks around $\sim0.1-0.5\,\msun$ \citep[e.g.,][]{Kroupa02,Chabrier03} and has a power-law tail extending to larger masses \citep{Salpeter55}. Compared with the free-fall time evaluated at the characteristic density of the parent molecular clouds, star formation is a slow and inefficient process \citep[e.g.,][]{Zuckerman74,Evans09}. Any comprehensive theory of star formation should be able to explain these characteristics in a single unified framework \citep[see][]{McKee07}.

Supersonic turbulence, along with self-gravity, is believed to be the main process controlling star formation \citep{MacLow04}. It creates a complex, self-similar network of sheet-like and filamentary shock compressed density fluctuations. When one of these fluctuations becomes Jeans unstable it collapses and forms one, or possibly many, protostellar objects \citep{Elmegreen93,Padoan99,MacLow04}. This process is known as `gravoturbulent fragmentation' and is pivotal for understanding how stars and stellar clusters form. 
The character of turbulence \citep{Klessen00a,Klessen01a,Padoan02}, cloud geometry \citep{Bonnell93}, rotation level, and magnetic field strength \citep{Heitsch01,VazquezSemadeni05,Padoan11,Federrath12} will all influence gas fragmentation. The thermodynamical behavior of collapsing turbulent gas, set by various heating and cooling processes, is also crucial in moderating fragmentation \citep[e.g.,][]{Larson85,Larson05}. In simulations utilizing a polytropic equation of state, $P\propto\rho^{\gamma}$, \citet{Li03} demonstrated that the degree of fragmentation increases with decreasing polytropic exponent $\gamma$ (corresponding to an increase in the gas cooling rate). \citet{Jappsen05} further demonstrated that the characteristic fragmentation mass scale is approximately the Jeans mass at the point where $\gamma$ first exceeds unity after dipping below unity. Variations in $\gamma$ have also been shown to have strong effects on gas morphology \citep[e.g.,][]{Peters12}, the gas density probability distribution function (PDF) of supersonically turbulent gas \citep{Scalo98,VazquezSemadeni06}, and may also affect the shape of the stellar IMF \citep{Spaans00}.

The theory of Galactic star formation is only partially applicable to the formation of the first, so-called Population III (Pop III), stars \citep[reviewed in][]{Bromm13}. These objects formed in $10^{5-6}\,\msun$ dark matter `minihaloes' at redshifts $z\sim15-40$ \citep{Couchman86,Haiman96,Tegmark97}. Given the complete lack of metals, gas cooling was provided by molecular hydrogen ($\htwo$), a relatively inefficient cooling agent. The thermalization and energy spacing of the lowest rotational levels of $\htwo$ set a characteristic density, $n\sim10^4\,\cc$, and temperature, $T\sim200\,\kelvin$, where primordial gas ends its initial free-fall collapse phase. Simulations have shown the resulting star-forming clumps to have masses $\gtrsim100\,\msun$, in accord with the Jeans mass at this point \citep{Abel00,Bromm02,Yoshida06,OShea07}. The final stellar mass is determined by subsequent accretion onto the initial protostellar hydrostatic core. Recent simulations revise this characteristic mass downwards by showing that the Pop III protostellar disk can fragment at high densities \citep{Machida08,Stacy10,Greif11,Greif12,Clark11,Clark11a}, or that prostostellar radiation can suppress gas accretion \citep{Stacy12,Hosokawa11,Susa13}, but still support the general conclusion that Pop III stars were typically more massive than later stellar generations.

What caused the transition from solitary, high-mass Pop III star formation to clustered, low-mass star formation is an important open question in cosmology. The introduction of metals by the first supernovae is generally thought to have driven the transition, given the enhanced cooling and thus fragmentation in metal-enriched gas. This idea was explored in \citet{Omukai00} who showed an increase in metallicity leads to larger cooling rates and thus lower temperatures during gaseous collapse.  \citet{Bromm01} found that when metallicity is greater than a critical metallicity of $Z\approx10^{-3.5}\,\zsun$, gaseous fragmentation occurs due to metal fine-structure line emission. Later studies that explored a metal fine-structure `line induced' Pop III-II transition \citep[e.g.,][]{Bromm03,Santoro06,Smith07,SafranekShrader10} all found similar values for the metallicity needed for the star formation to transition to an efficiently fragmenting mode.

A separate set of studies suggest that the Pop III-II mode transition is not induced by metal line emission, but instead by the thermal coupling of gas and dust grains \citep[e.g.,][]{Omukai05,Schneider06,Clark08,Dopcke11,Dopcke13}. This `dust induced' fragmentation occurs at a much higher density, $n\sim10^{10-12}\,\cc$, than line-induced fragmentation when the thermal Jeans mass is of order of $0.1-1\msun$. Studies have shown that the metallicity needed for dust-gas coupling to significantly affect the gas thermodynamical evolution and induce a star formation transition is $Z\approx10^{-6}\,\zsun$, much lower than the metallicity needed for a line-induced transition. This depends, though, on the creation, abundance, and composition of dust at high redshifts \citep[e.g.,][]{Dwek11,Schneider12a} which remains very uncertain.

More realistic three-dimensional simulations have also addressed the Pop III-II transition. \citet{Clark08}, using a piecewise equation of state, showed dust induced fragmentation does indeed induce solar-mass scale fragmentation at high densities and emphasized the importance of angular momentum in regulating fragmentation. \citet{Dopcke11,Dopcke13} confirmed this finding utilizing a non-equilibrium chemical network. Starting from cosmological initial conditions, \citet{Smith09} suggested the interplay of metal fine-structure line cooling and the cosmic microwave background (CMB) temperature floor, $\tcmb = 2.725\,\kelvin(1+z)$, strongly controls gas fragmentation and leads to three distinct modes of metal-enriched star formation at high redshifts. The smoothly decreasing CMB temperature sets a lower limit to which gas can radiatively cool, which may regulate the process of star formation \citep{Clarke03,Schneider10} and imprint observational signatures in the chemical abundances of Galactic metal-poor stellar populations \citep{Tumlinson07,Ballin10}. The CMB may also considerably increase the opacity mass limit for fragmentation given the strong temperature dependance of the mean dust opacity \citep{Low76}.

Distinguishing between a dust induced and line induced star formation transition can be accomplished by observing chemical abundance patterns in metal-poor stars. \citet{Frebel07} compiled observations of metal-poor stars and showed that none has oxygen and carbon abundances smaller than the critical value needed for a line induced transition. The recently discovered star of \citet{Caffau11}, though, has such small carbon and oxygen abundances that only dust cooling induced fragmentation could explain its existence \citep{Schneider12,Klessen12}, though the possibility has been raised that its true metal abundance is higher than observed \citep{MacDonald13}. Additionally, it should be stressed that disk fragmentation into sub-solar mass clumps can occur even in completely metal-free gas \citep{Greif11,Clark11}, obviating the need for any sort of metal-enhanced cooling for low mass stellar object production.

Though the level of metal enrichment in gas plays a key thermodynamic role, numerous other factors likely contributed to effecting the transition to Pop II star formation. For example, in a series of papers \citet{Jappsen07,Jappsen09a,Jappsen09} argued that there is no metallicity threshold for a Pop III-II transition -- instead, the cosmic hydrodynamic context setting the initial conditions for star formation will have the largest effect on how fragmentation proceeds. This may be especially relevant since the first metal-enriched stellar clusters likely formed in atomic cooling haloes, rather than relatively low-mass minihaloes, with potential wells deep enough to sustain supersonic turbulent motions \citep[e.g.,][]{Wise07,Greif08}, though this is dependent on how exactly Pop III stars ended their lives.

The formation of stellar clusters comprised of low mass stars, rather than solitary massive stars, has profound cosmological implications \citep{Tornatore07,Maio10,Aykutalp11,Latif12,Wise12}. These earliest clusters played a key role in cosmic reionization \citep{Bouwens11,Finkelstein12,Kuhlen12,Robertson13} and may still survive today in the form of metal-poor globular clusters, as part of the recently discovered ultra-faint dwarf (UDF) Milky Way satellites \citep[e.g.,][]{Belokurov07,Frebel12}, or individually as metal-poor stars residing in the Galactic halo. If sufficiently luminous they are capable of being detected by upcoming telescopes designed to probe redshifts $\gtrsim10$, such as the \emph{James Webb Space Telescope} (JWST) \citep{Pawlik11,Zackrisson11,Zackrisson12,Dunlop13}. Additionally, the long term chemical evolution of a star cluster imprints unique chemical signatures onto its members that will be useful for decoding the enrichment patterns of the first supernovae \citep[e.g.,][]{BlandHawthorn10}.

\begin{figure*}
\begin{center}$
\begin{array}{cc}
\includegraphics[scale=0.264, trim = 0 0 15 0]{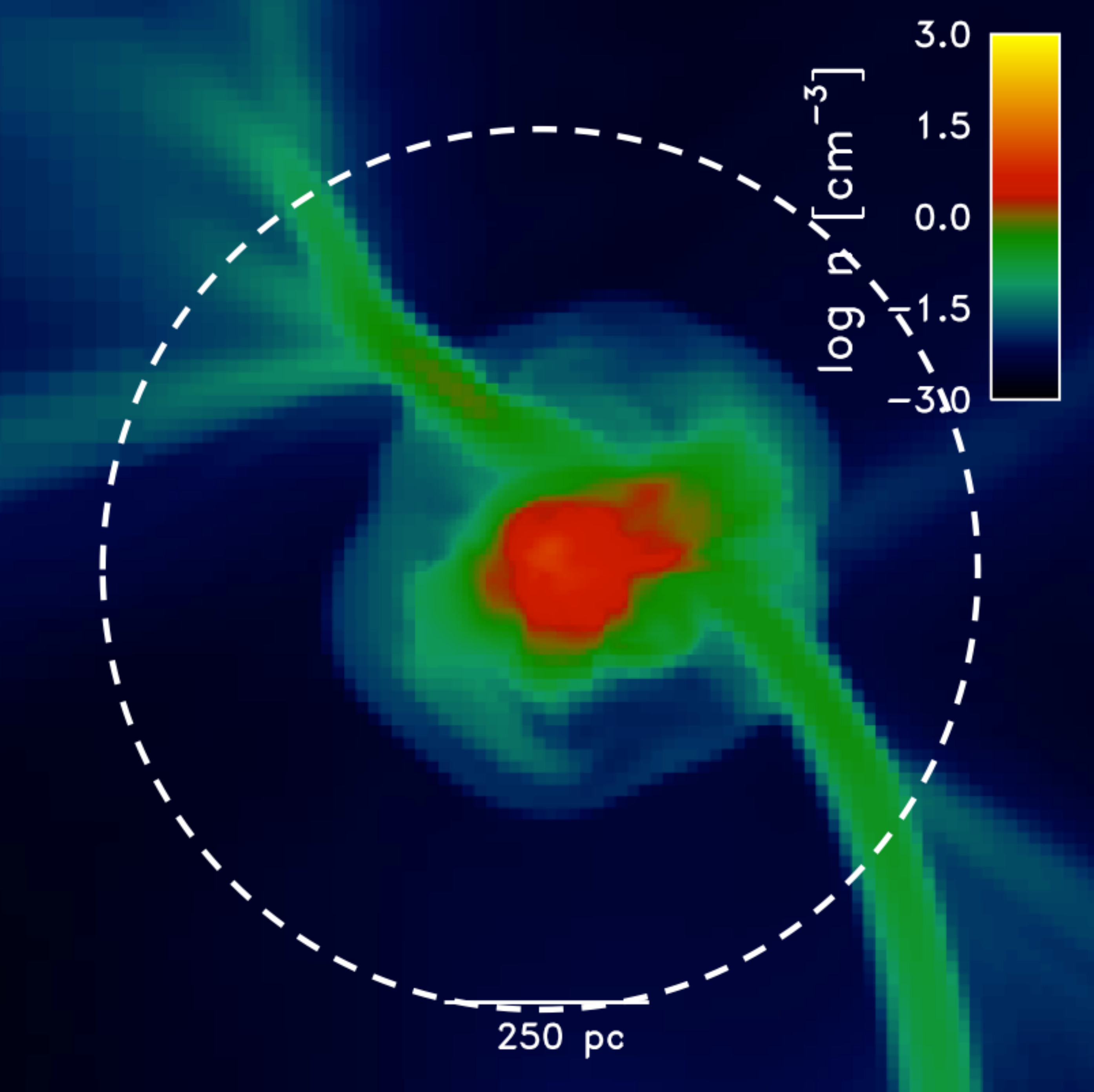} &
\includegraphics[scale=0.265, trim = 0 0 15 0]{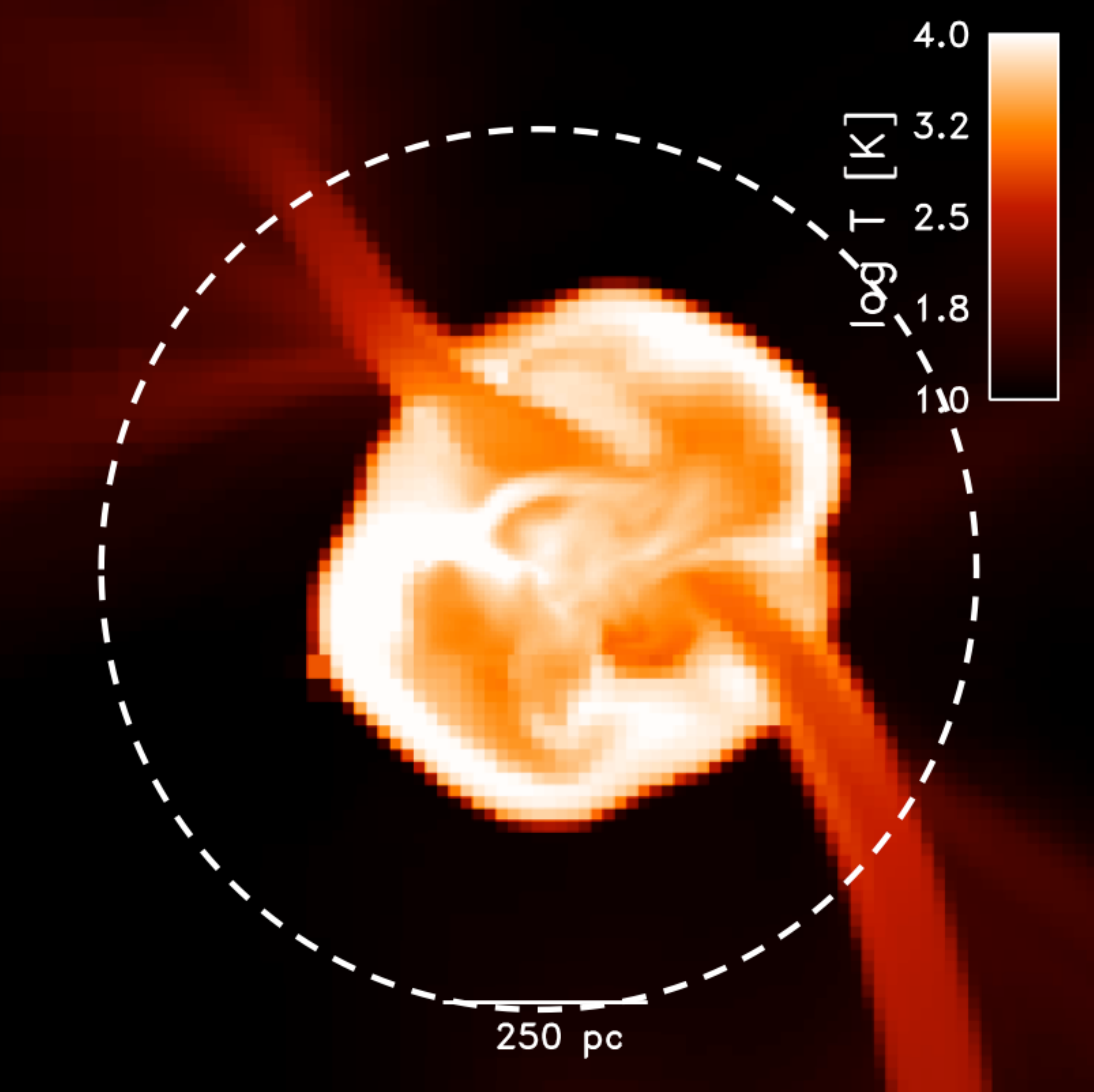} \\
\end{array}$
\end{center}
\caption{Slices of gas density (left) and temperature (right) when the metallicity of the gas was first made non-zero. The dashed circle denotes the region in which the gas metallicity was first set to a non-zero value. Cold accretion streams that penetrate the virial shock of the halo and account for most of gas accretion are readily apparent.}
\label{fig:halo_slice}
\end{figure*}

This work is focused on the process of gas fragmentation, and thus star formation, in high-redshift atomic cooling haloes that have been enriched to some non-zero, but sub-solar, metallicity. We present the results of three high-resolution cosmological simulations, achieving a spatial resolution of $\sim0.01$ physical parsecs and maximum physical density of $\sim10^{6-7}\,\cc$. We impose a strong Lyman-Werner (LW) radiation field that photodissociates molecular hydrogen across the computational box, delaying gaseous collapse until the assembly of a $M_{\mathrm{vir}}\approx2\times10^{7}\,\msun$ halo capable of cooling by $\lya$ line emission at $z\approx16$. When we identify the conditions for runaway gaseous collapse in a halo, we endow the gas in its vicinity with a non-zero metallicity and observe the subsequent metal-cooling induced collapse. At high densities, we introduce sink particles which serve as proxies for stellar associations that would have formed at gas densities not directly resolved in the simulations. We evolve this star formation region for multiple free-fall times to observe the long-term fragmentation process, measure the efficiency with which gas is converted into stars, and study mass spectrum of high-density gaseous clumps. Using this approach we aim to better understand the star formation properties of metal enriched gas and the role that metal fine-structure line cooling plays in the first instances of clustered star formation in the Universe.

We organize this paper as follows. In Section \ref{sec:numerical_setup} we describe our numerical setup and physical ingredients that enter into our study. In Section \ref{sec:results} we describe the results and analysis of our simulations. In Section \ref{sec:caveats} we discuss the potential limitations of our simulations. Finally, in Section \ref{sec:summary} we discuss the implications of our findings and provide our conclusions.

Throughout this paper we assume cosmological parameters consistent with the \emph{Wilkinson Microwave Anisotropy Probe} (WMAP) 7 year results \citep{Komatsu11}: $\omegal = 0.725$, $\omegab =  0.0458$, $\omegam = 0.275$, $h = 0.704$, $\sigma_8 = 0.810$, and $n_{\mathrm{s}} = 0.967$. Additionally, all quantities will be expressed in physical, rather than comoving, units unless explicitly stated otherwise.

\section{Numerical Setup}
\label{sec:numerical_setup}

\subsection{Parameter Choices and Initial Conditions}
\label{sec:ics}

We run our simulation using the adaptive mesh refinement (AMR) code FLASH \citep{Fryxell00}, version 4. Our cosmological initial conditions are identical to those used in \citet{SafranekShrader12}. Specifically, we use a $1\, \Mpc^3$ (comoving) box initialized at $z=145$ with a base grid resolution of $128^3$ and two nested grids for an effective resolution of $512^3$. This results in an effective dark matter particle mass of $230\,\msun$. 

Our grid refinement/derefinement scheme is also identical to \citet{SafranekShrader12}, though now we resolve the Jeans length,
\begin{equation}
\lj = \left(\frac{\pi \cs^2}{G\rho}\right)^{1/2} \mbox{,}
\label{eq:lj}
\end{equation}
by at least 24 grid cells. This comfortably satisfies the Truelove criterion \citep{Truelove97} for avoiding artificial fragmentation, though it may not be sufficient to resolve the true character of turbulence, particularly small-scale vortical motions ultimately responsible for magnetic field amplification \citep{Sur10,Federrath11,Turk12}. We derefine the grid if the Jeans length is resolved by more than 48 grid cells.

When the grid is highly refined, dark matter particles begin to become very coarsely sampled when their mass is mapped onto the grid for the purpose of the gravitational potential calculation. To alleviate this we spatially smooth the gravitational influence of dark matter \citep[in a manner similar to][]{Richardson13} above a refinement level of $12$, resulting in a dark matter smoothing length of $\approx 60$ comoving pc. At the redshift at which we study metal-enchanced fragmentation, this corresponds to $3.7$ physical pc, roughly the radial length scale on which baryons begin to dominate the gravitational potential.

\begin{figure}
\centering
\subfigure{
\includegraphics[scale=0.45, clip, trim = 35 10 0 20 ]{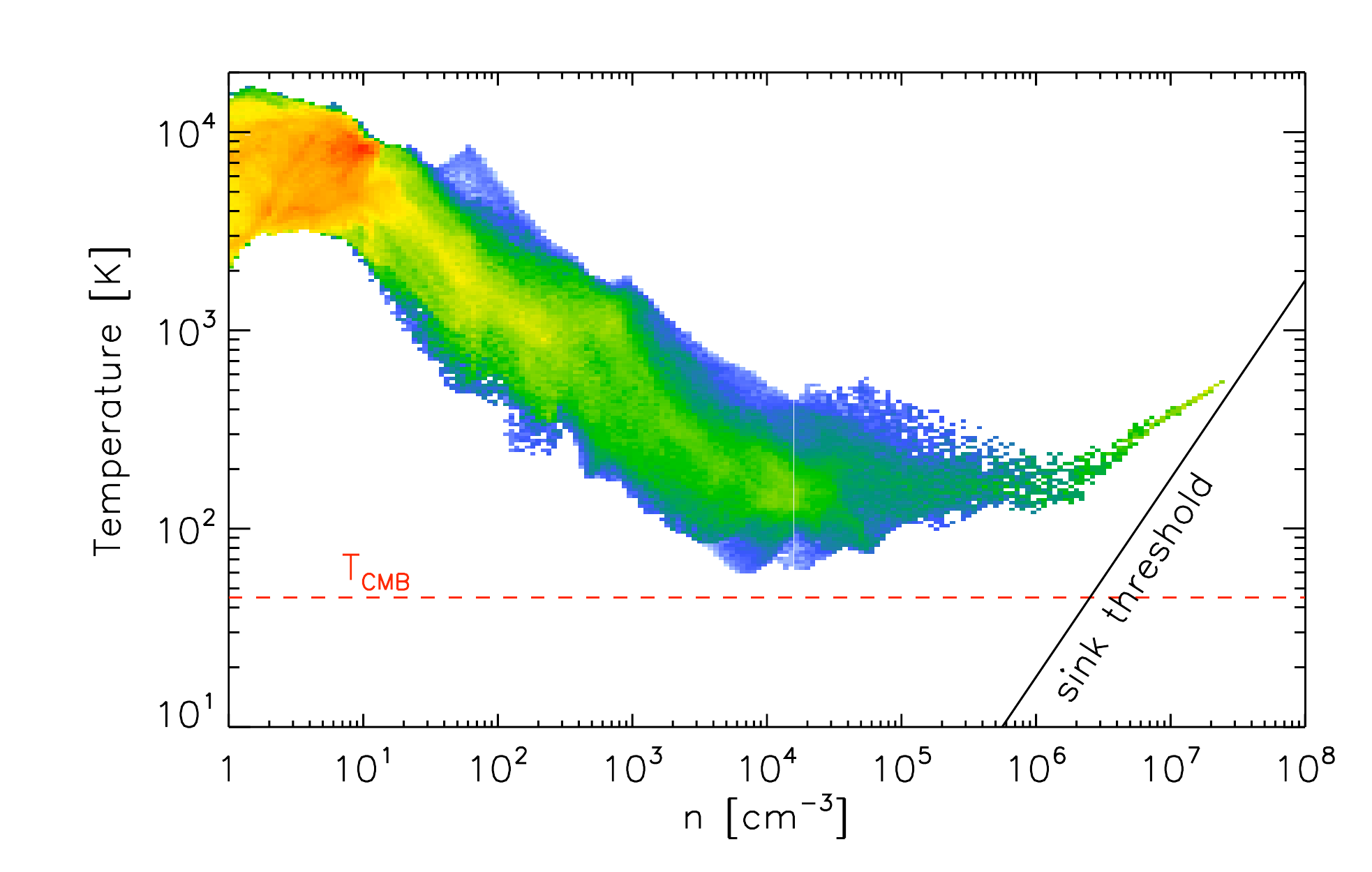}
\label{fig:subfig_mn4}
}
\subfigure{
\includegraphics[scale=0.45, clip, trim = 35 10 0 20]{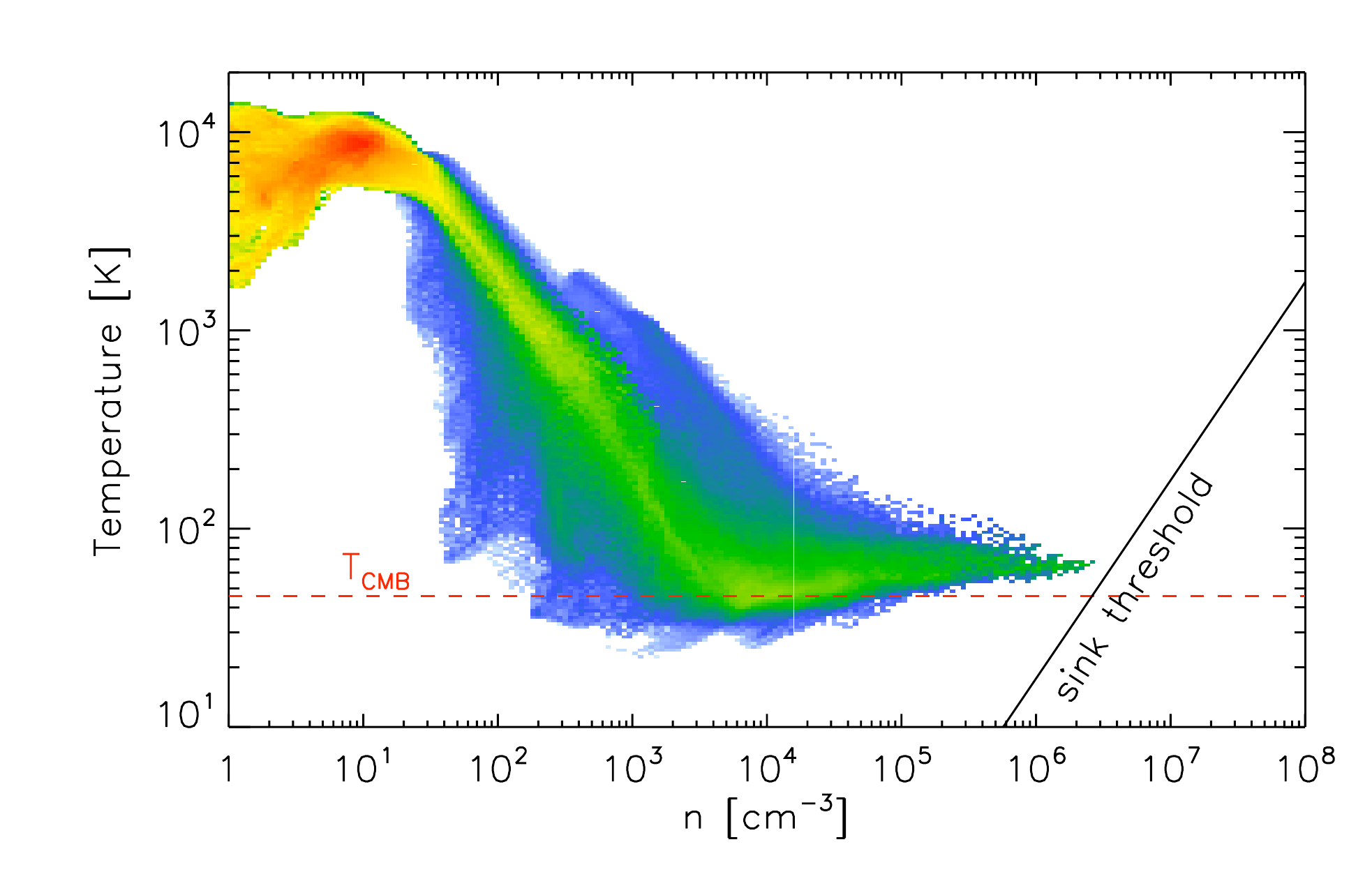}
\label{fig:subfig_mn3}
}
\subfigure{
\includegraphics[scale=0.45, clip, trim = 35 10 0 20]{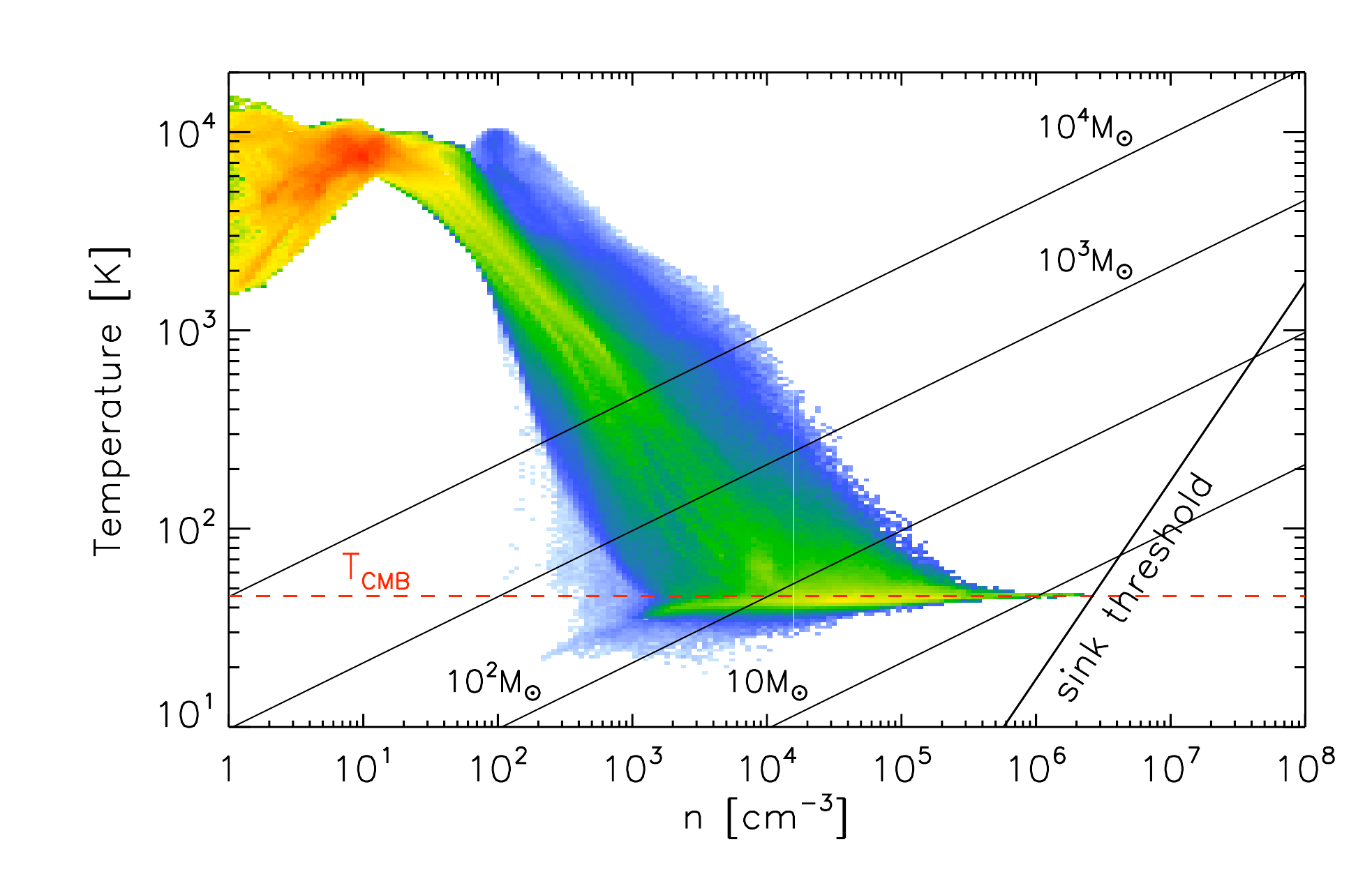}
\label{fig:subfig_mn2}
}
\caption[]{ Density-temperature phase diagrams for gas within $50\,\pc$ of the densest point as the first sink particle forms for the $10^{-4}$ (top), $10^{-3}$ (middle), and $10^{-2}\,Z_\odot$ (bottom) runs. Color corresponds to the gas mass at a given density and temperature. We also show the CMB temperature (red dashed horizontal line) and the temperature-dependent density threshold for sink particle creation (steep solid line; see Equation \ref{eq:rho_j}). Additionally, in the bottom panel, we show lines of constant Jeans mass. In the higher metallicity simulations gas first reaches $\tcmb$ when $\mj\sim100\,\msun$, consistent with the typical sink particle mass forming in these simulations.}
\label{fig:dens_temp_multiplot}
\end{figure}

Because we focus on haloes capable of $\lya$ cooling, we suppress gravitational collapse in minihaloes that relies on molecular hydrogen cooling. We do this by including a background Lyman-Werner (LW) radiation field that is capable of destroying $\htwo$ by the two-step Solomon process \citep{Stecher67}. The $\htwo$ photodissociation rate from LW radiation is \citep[e.g.,][]{Abel97,Glover07} 
\begin{equation}
k_{\htwo} = 1.38\times10^{-12}\,\jlw\fshhtwo\,\mathrm{s}^{-1} \mbox{,}
\label{eq:kh2}
\end{equation}
where $\jlw$ is the radiation intensity at the centre of the Lyman-Werner bands, $12.4\,\ev$, in units of $10^{-21}\,\mathrm{erg}\, \mathrm{s}^{-1}\,\mathrm{cm}^{-2}\,\mathrm{Hz}^{-1}\,\mathrm{sr}^{-1}$ and $\fshhtwo\leq1$ is a dimensionless factor which accounts for $\htwo$ self-shielding. By setting $\jlw=100$, we effectively guarantee the suppression of cooling in minihaloes at all redshifts of interest \citep[see][]{SafranekShrader12}. We model the effect of $\htwo$ self-shielding in an approximate way by writing the $\htwo$ self-shielding factor as \citep{Draine96}
\begin{equation}
\fshhtwo = \mathrm{min}\left[1.0,\left(\frac{N_{\htwo}}{10^{14}\,\mathrm{cm}^{-2}}\right)^{-0.75}\right] \mbox{,}
\label{eq:fsh} 
\end{equation}
and approximating $N_{\htwo}$, the $\htwo$ column density, as 
\begin{equation}
N_{\htwo} = 0.1\,n_{\htwo}\,\lj \mbox{,}
\label{eq:Nh2}
\end{equation}
where $n_{\htwo}$ and $\lj$ are, respectively, the local $\htwo$ number density and Jeans length. This allows the $\htwo$ abundance to increase once metal cooling has allowed the gas to collapse to high densities \citep[see][for a more sophisticated approach]{WolcottGreen11}.


\subsection{Metal Enrichment Strategy}
\label{sec:metal_enrichment}

The dispersal of metals synthesized in the first stars is a complicated process moderated by the interplay of supernova ejection, gravitational reassembly, and turbulent mixing with primordial gas \citep[e.g.,][]{Karlsson13}. The outcome of primordial metal enrichment is strongly dependent on the mass of the first stars. A non-rotating, metal-free star with a mass between $140$ and $260\,\msun$ is believed to end life as a highly energetic ($E_{\mathrm{kin}}\sim10^{52}\,\mathrm{ergs}$) pair-instability supernova (PISN) \citep[e.g.,][]{Heger03}, though rotationally induced mixing may bring the lower limit down by nearly a factor of two \citep{Chatzopoulos12}. Since this explosion completely disrupts the host minihalo, the onset of metal-enriched star formation is delayed until a $\sim10^8\,\msun$ halo forms around a redshift of $\sim10$ \citep{Wise08,Greif10,Wise12} where the metal ejecta would likely reassemble.

More recent simulations of Pop III star formation, though, have suggested stellar masses exceeding $\sim50\,\msun$ are unlikely given radiative feedback that acts to shut off gas accretion \citep[e.g.,][]{Hosokawa11,Stacy12}. These more moderate mass Pop III stars end their lives as less energetic core-collapse supernovae that may not completely disrupt the host minhalo. \citet{Ritter12} simulated the HII region, supernova, and metal transport from a $40\,\msun$ Pop III star ending its life as an $E_{\mathrm{kin}}=10^{51}\,\mathrm{erg}$ core collapse supernova. While only about a half metal ejecta expanded beyond the minihalo's virial radius, within $10\,\myr$ the ejecta began to fall back into the center of the host minihalo, suggesting the first burst of metal enriched star formation may have followed closely in the wake of the Pop III star formation already in cosmic minihaloes \citep[see also][]{Whalen13}.

With this in mind, to emulate the complex process of metal ejection and transport in the early Universe, we first run our metal-free cosmological simulation until a halo with a virial temperature of $\tvir\sim10^{4}\,\kelvin$ capable of cooling by $\lya$ emission has formed in the simulation. At the point at which $\lya$ cooling is effective enough for gas in the centre of the halo to begin to isothermally collapse, we increase the metallicity within $1.5\,R_{\mathrm{vir}}$ of the halo to a non-zero value, either $10^{-4}$, $10^{-3}$, or $10^{-5}\,Z_\odot$. Figure \ref{fig:halo_slice} shows a slice of gas density and temperature through the selected dark matter halo at this point of virialization and denotes the spatial region where we increased the metallicity of the gas with a dashed circle.

This idealized approach is not meant to realistically reproduce the results of simulations that track the injection and transport of metals from the first supernovae, such as \citet{Wise08}, \citet{Greif10}, and \citet{Ritter12}. These studies have shown there to be significant metal inhomogenities on kiloparsec scales around atomic cooling haloes. Mixing between primordial and metal-enriched gas, though, is much more effective at higher densities near the centers of atomic cooling halos. One should be aware, though, that the transport of passive scalars in grid-based codes is well known to overpredict the diffusion speed and degree of mixing \citep[e.g.,][]{Plewa99, Ritter12}. The goal of the present paper is to isolate the thermodynamic impact of metals on protostellar collapse and star cluster formation. Therefore we opted to perform controlled numerical experiments separating out potential effects of inhomogeneous metal dispersal by introducing metals artificially, assuming perfect mixing. Furthermore, all gaseous fragmentation, the focus of this paper, occurs within a $\sim5$ parsec region in the centre of the target halo well after the metals are introduced ($t\sim4\,\myr$) where turbulent motions are expected to homogenize metals on the order of a dynamical time. We discuss the possible implications of this idealized approach in Section \ref{sec:caveats}. 



\subsection{Chemical and Radiative Processes}
\label{sec:radiative_processes}

We employ a non-equilibrium chemistry solver that tracks the abundances of the following primordial chemical species: H, $\mathrm{H}^-$, $\mathrm{H}^+$, $\mathrm{e}^-$, $\htwo$, $\htwo^+$, He, $\mathrm{He}^+$, $\mathrm{He}^{++}$, D, $\mathrm{D}^+$, and HD. We additionally include the following gas-phase metal species: C, $\mathrm{C}^+$, Si, $\mathrm{Si}^+$, O, and $\mathrm{O}^+$ with the solar abundance pattern. Even a modest intergalactic ultraviolet (UV) radiation field is effective in keeping species with a first ionization potential less than $13.6\,\ev$, such as carbon and silicon, singly ionized. Above $13.6\,\ev$, it is a reasonable assumption that neutral hydrogen significantly attenuates the radiation field, especially before the epoch of reionization. We include this effect in our chemical network by setting the photoionization rate of neutral carbon and silicon to values corresponding to our choice of the LW background, $\jlw=100$, assuming a $T=10^4\,\kelvin$ blackbody spectral shape. The ionization state of oxygen is determined by collisional processes since its ionization potential is above that of neutral hydrogen's and thus does not experience a significant photoionizing flux.

In low density ($n<10^{8}\,\cc$) molecule-free gas, the most significant thermodynamic effect from heavy elements is fine-structure line emission from forbidden transitions of carbon, oxygen, and silicon. To model fine structure cooling, we follow the method of \citet{Glover07}, which we briefly review here.

For a given metal fine-structure coolant, the net volumetric energy rate of change due to photon emission and absorption can be written as
\begin{equation}
\dot{e} = \Gamma - \Lambda \mbox{,}
\label{eq:fs1}
\end{equation}
where $\Lambda$ and $\Gamma$ are, respectively, the volumetric cooling and heating rates due to all available electronic transitions, stimulated emission, and absorption. The cooling rate for a single fine-structure coolant can be written as
\begin{equation}
\Lambda = \sum_{i\rightarrow j}(A_{ij} + B_{ij}I_{ij})h\nu_{ij}n_i \mbox{,}
\label{eq:fs2}
\end{equation}
and the heating rate as
\begin{equation}
\Gamma = \sum_{j\rightarrow i}B_{ji}I_{ij}h\nu_{ij}n_j \mbox{,}
\label{eq:fs3}
\end{equation}
where $A_{ij}$ is the Einstein coefficient for spontaneous emission for the transition from level $i\rightarrow j$, $B_{ij}$ is the  Einstein coefficient for stimulated emission ($i>j$) or for absorption ($i<j$), $h\nu_{ij}$ is the energy difference between levels $i$ and $j$, $I_{ij}$ is the radiation intensity at frequency $\nu_{ij}$, and $n_i$ is the number density, or level population, of ions in state $i$.

We solve for the level populations, $n_i$, by assuming statistical equilibrium,
\begin{equation}
n_i \sum_{j\ne i} R_{ij} = \sum_{j\ne i} n_j R_{ji} \mbox{,}
\label{eq:fs4}
\end{equation}
where the sum runs over all possible transitions and
\begin{equation}
 R_{ij} = 
  \begin{cases}
   A_{ij} + C_{ij} + B_{ij}I_{ij} & \text{if } i > j \mbox{ ,} \\
   C_{ij} + B_{ij} I_{ij}     &   \text{if } i < j \mbox{ .}
  \end{cases}
  \label{eq:fs5}
\end{equation}
Here, $C_{ij}$ represents the total rate of collisional de-excitations ($i>j$) or excitations ($j>i$). The collisional de-excitation rate is given by 
\begin{equation}
C_{ij} = \sum_k \gamma_{ij,k}n_k \mbox{,}
\label{eq:fs6}
\end{equation}
where $\gamma_{ij,k}$ is the collisional de-excitation rate for collisions with species $k$ and $n_k$ is the number density of the collider. We use the de-excitation rates (and other atomic data) given in \citet{Glover07} which include collisions with neutral and molecular hydrogen, protons, and electrons as sources of excitations and de-excitations. Finally, the excitation and de-excitation rates are related by (for $i>j$)
\begin{equation}
C_{ji} = C_{ij} \frac{g_i}{g_j}\mathrm{exp}\left(-\frac{h\nu_{ij}}{\kb T}\right) \mbox{,}
\label{eq:fs7}
\end{equation}
where $g_{i}$ is the statistical weight of level $i$ and $T$ is the gas temperature.

\begin{figure*}
\begin{center}$
\begin{array}{ccc}
\includegraphics[scale=0.4, trim = 0 0 15 0]{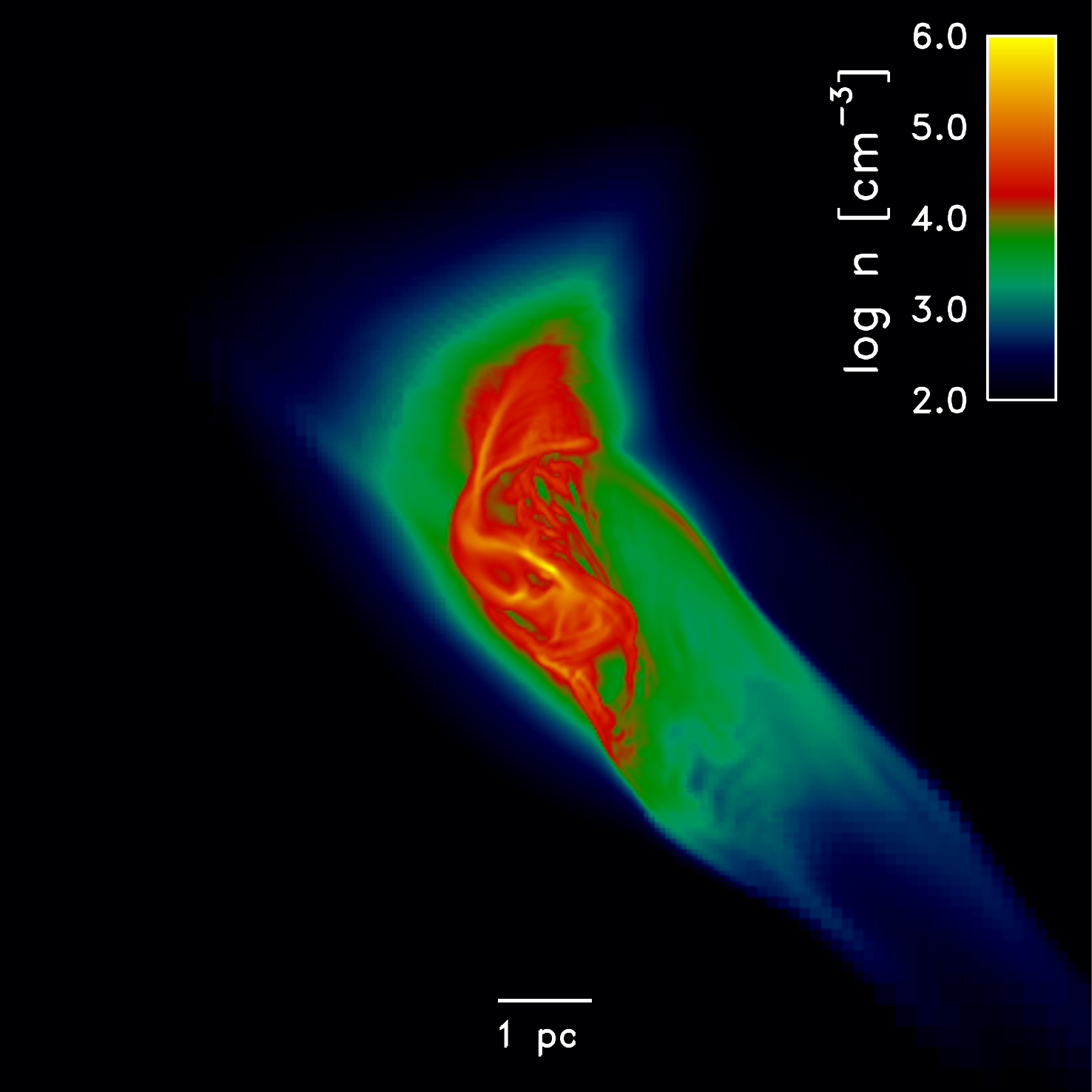} &
\includegraphics[scale=0.4, trim = 0 0 15 0]{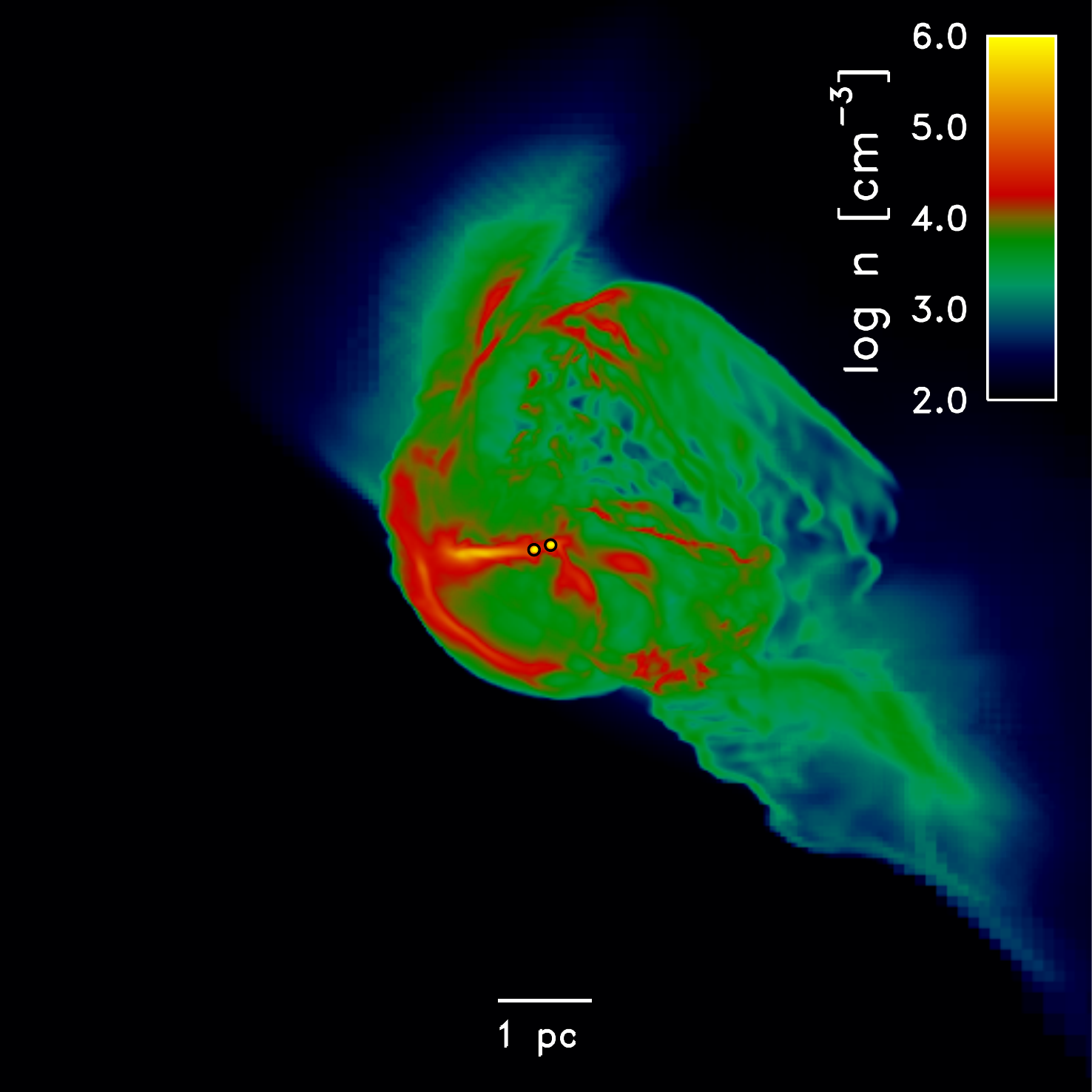} &
\includegraphics[scale=0.4, trim = 0 0 15 0]{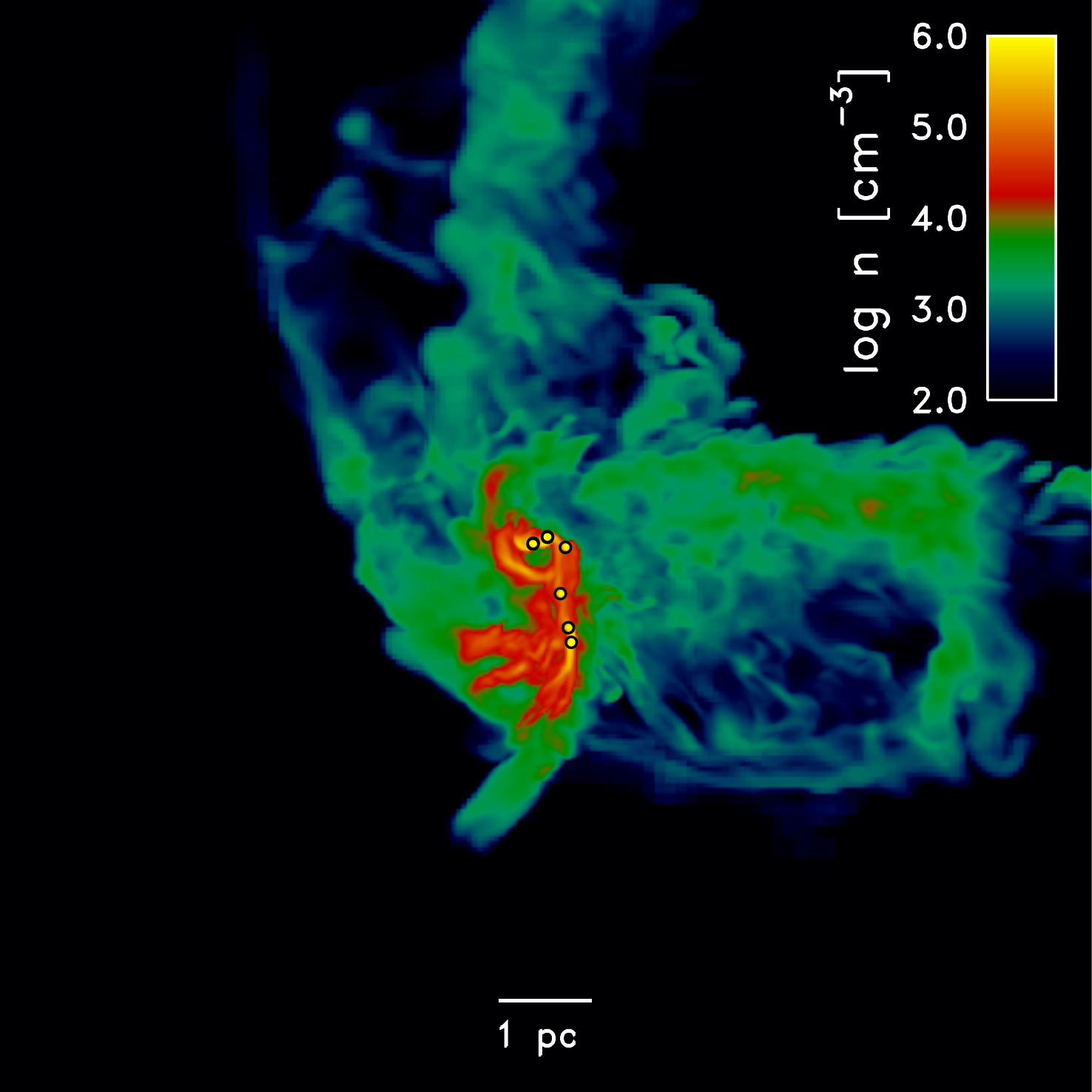} \\
\includegraphics[scale=0.4, trim = 0 0 15 0]{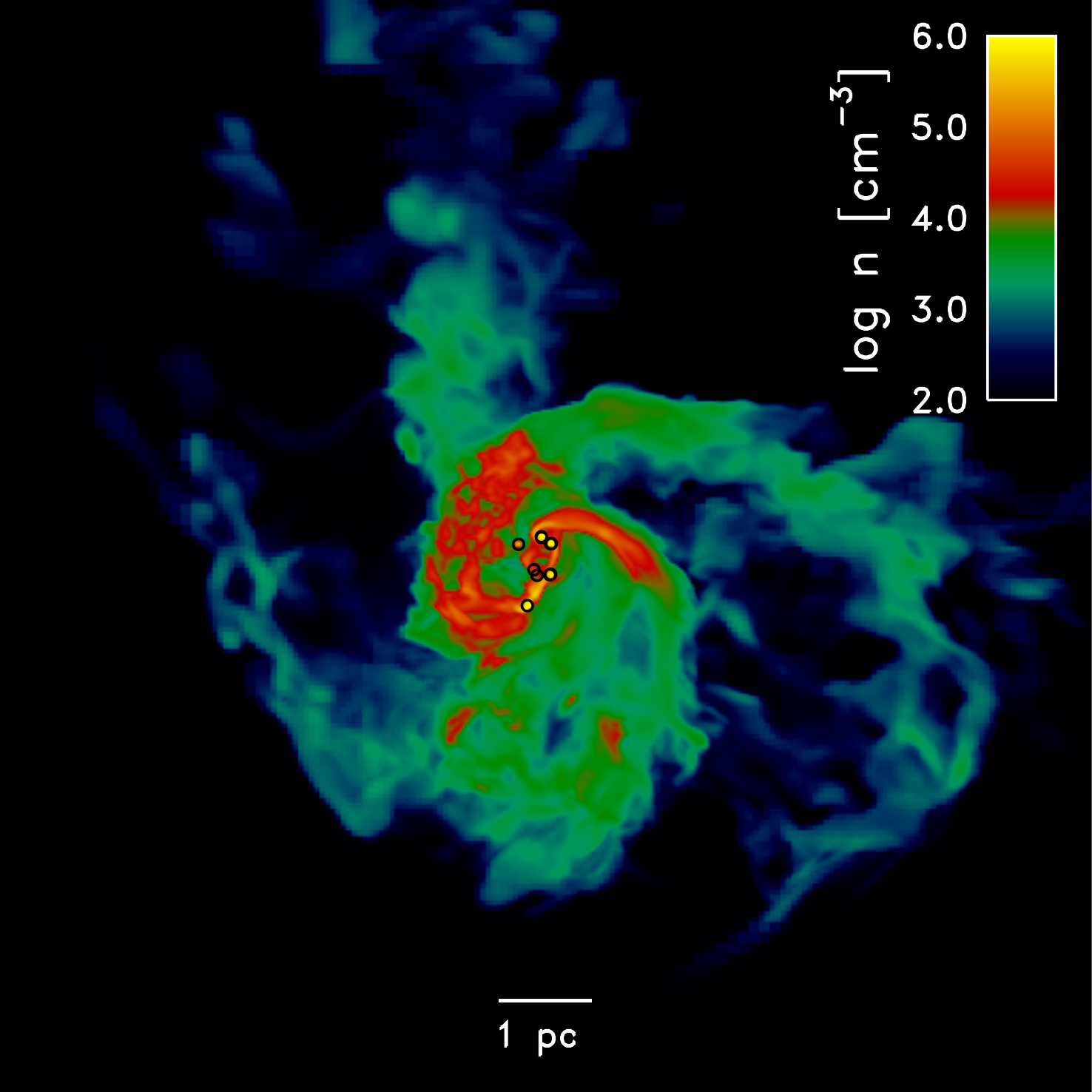} &
\includegraphics[scale=0.4, trim = 0 0 15 0]{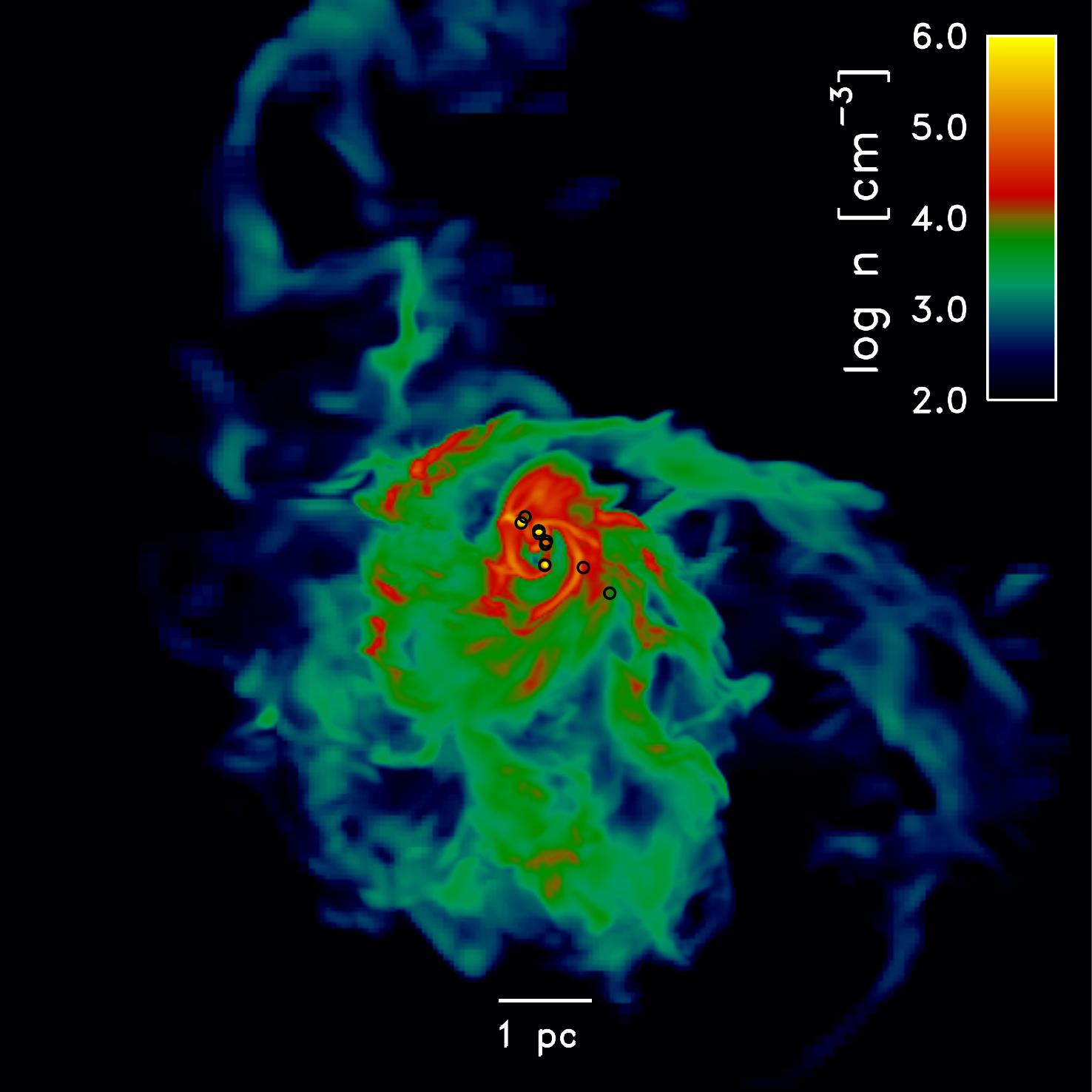} &
\includegraphics[scale=0.4, trim = 0 0 15 0]{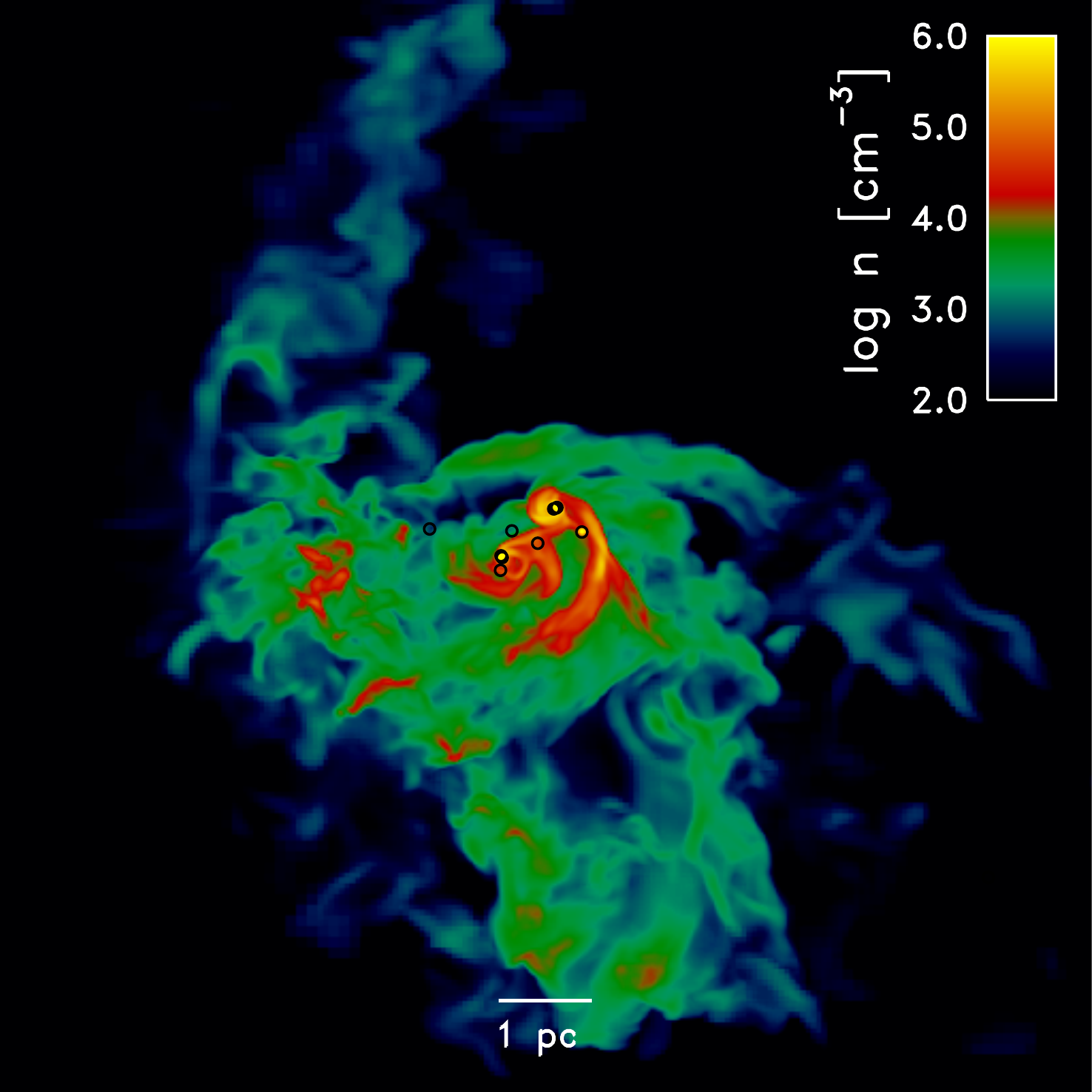} \\
\end{array}$
\end{center}
\caption{Snapshots showing mass-weighted line-of-sight density projections from the $10^{-2}\,Z_\odot$ run. Black circles represent sink particles, with the circle radius equal to the particle accretion radius. The top left panel shows the state of the simulation just before sink particle formation. After the gas has reached the CMB temperature, thermal instability and strong, quasi-isothermal shocks produce a sheet-like morphology (seen face-on in the top-left panel) and filamentary striations. These features are transient and within $\sim2\,\myr$ the star-forming cloud becomes more disordered and has an apparent net rotation. The consecutive snapshots from left-to-right and top-to-bottom are separated by a time interval of $0.8\, \myr$.}
\label{fig:mn2_morph_dens}
\end{figure*}

\begin{figure*}
\begin{center}$
\begin{array}{ccc}
\includegraphics[scale=0.4, trim = 0 0 15 0]{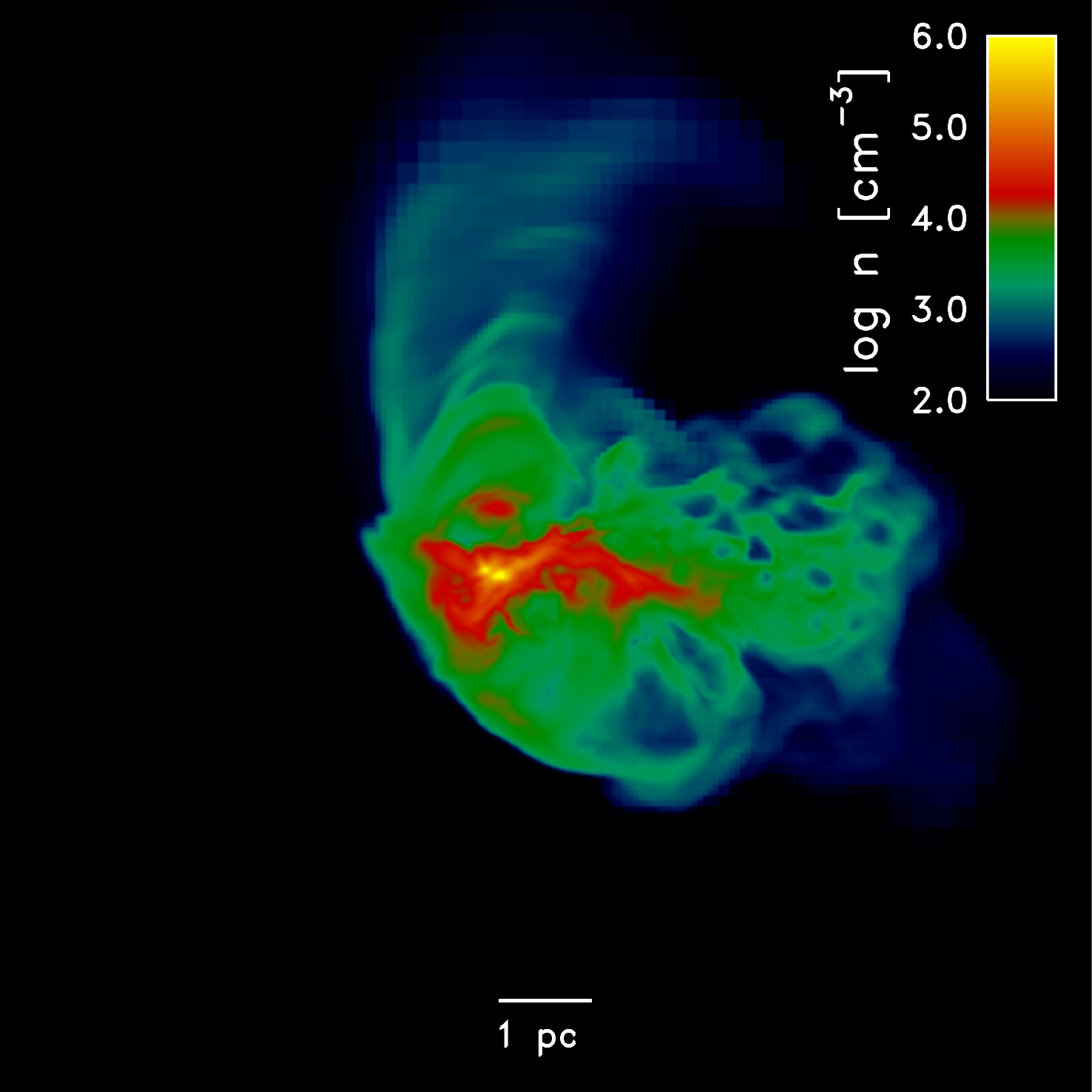} &
\includegraphics[scale=0.4, trim = 0 0 15 0]{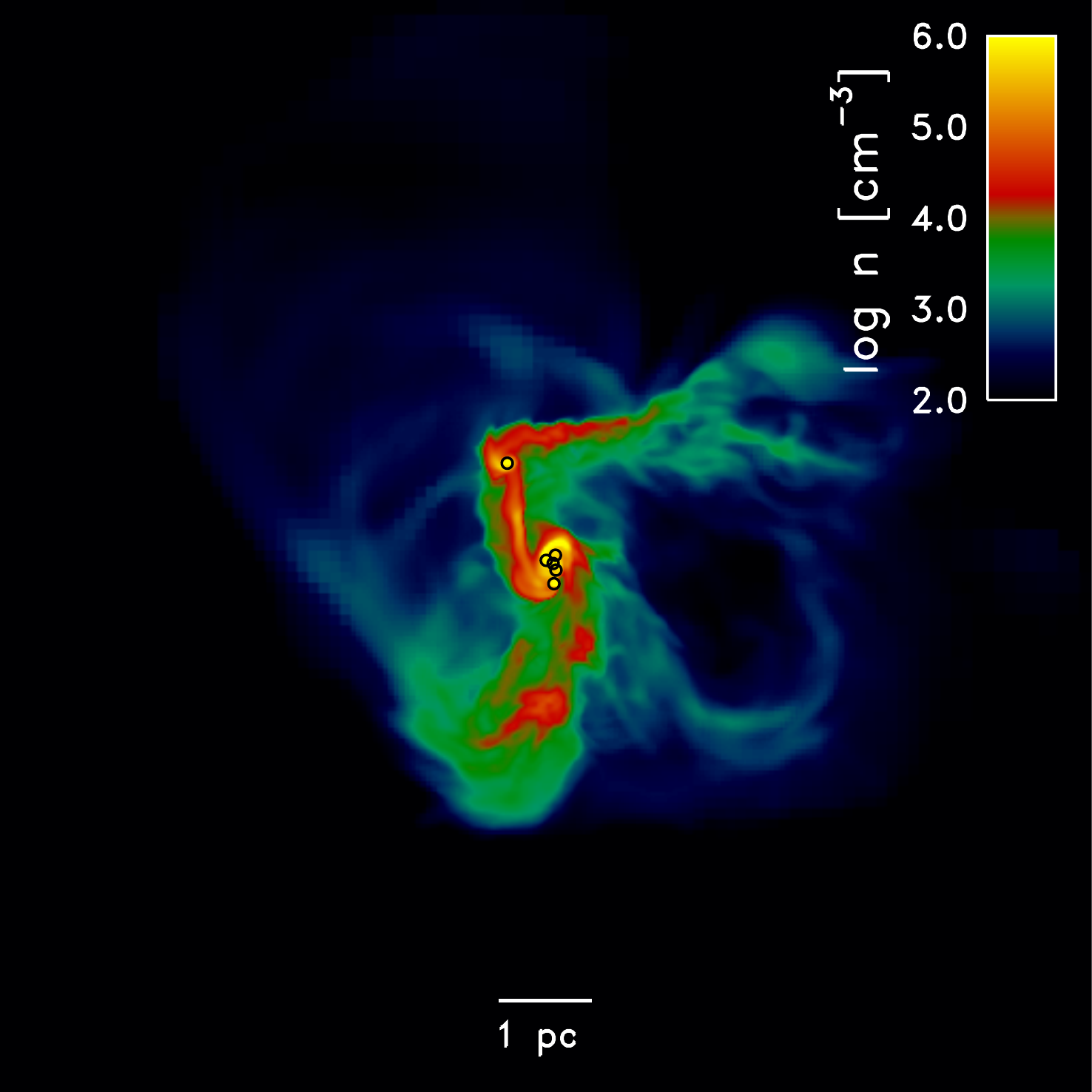} &
\includegraphics[scale=0.4, trim = 0 0 15 0]{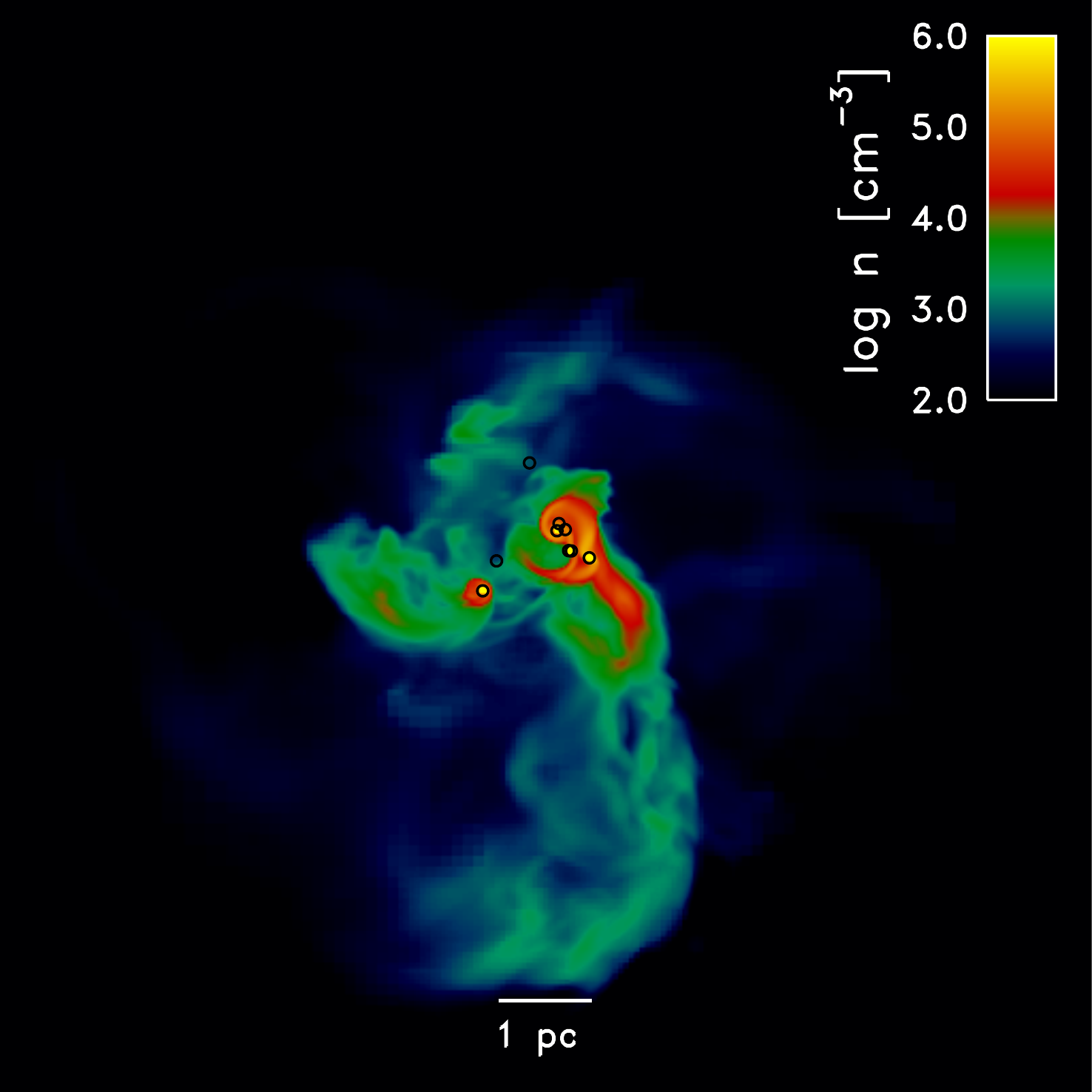} \\
\includegraphics[scale=0.4, trim = 0 0 15 0]{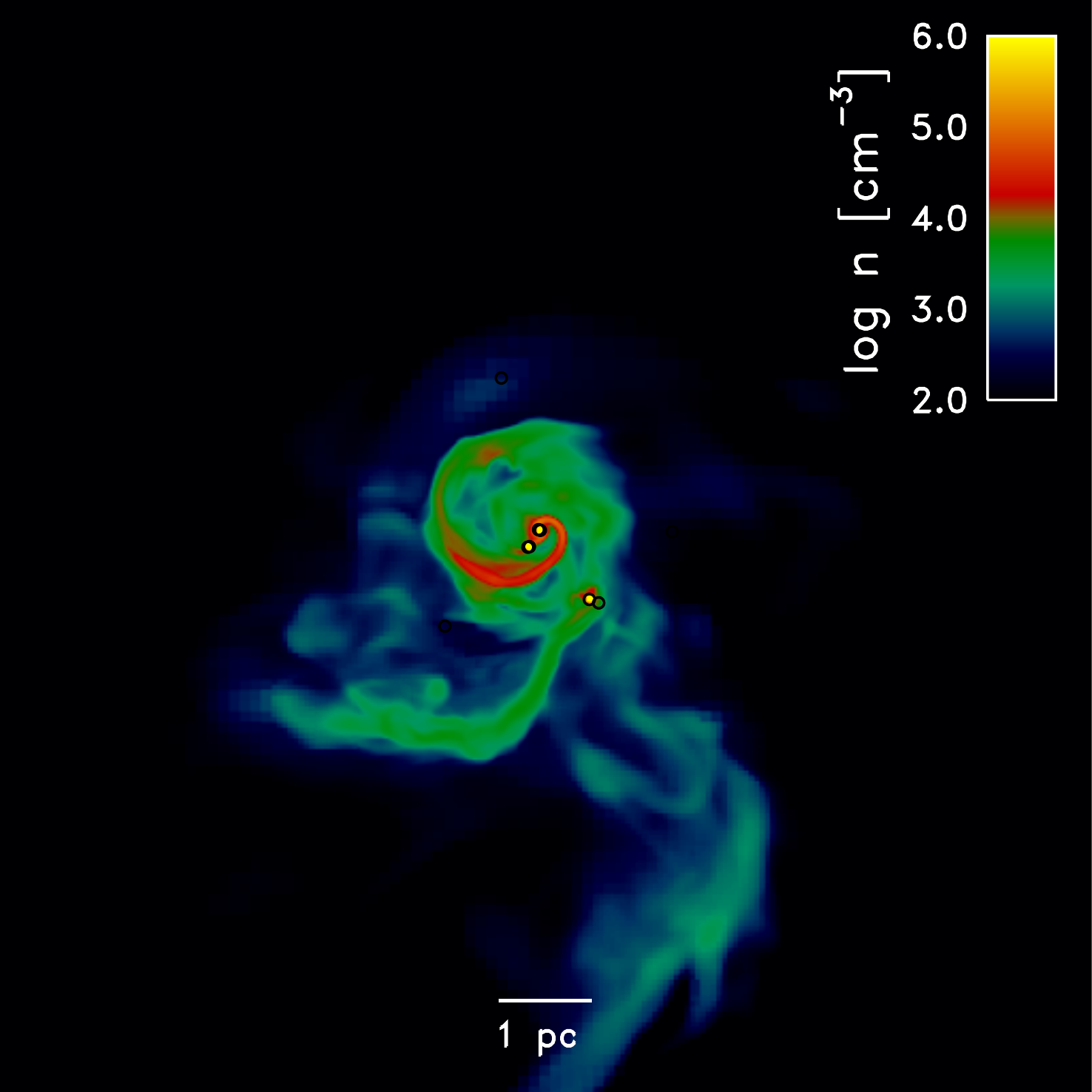} &
\includegraphics[scale=0.4, trim = 0 0 15 0]{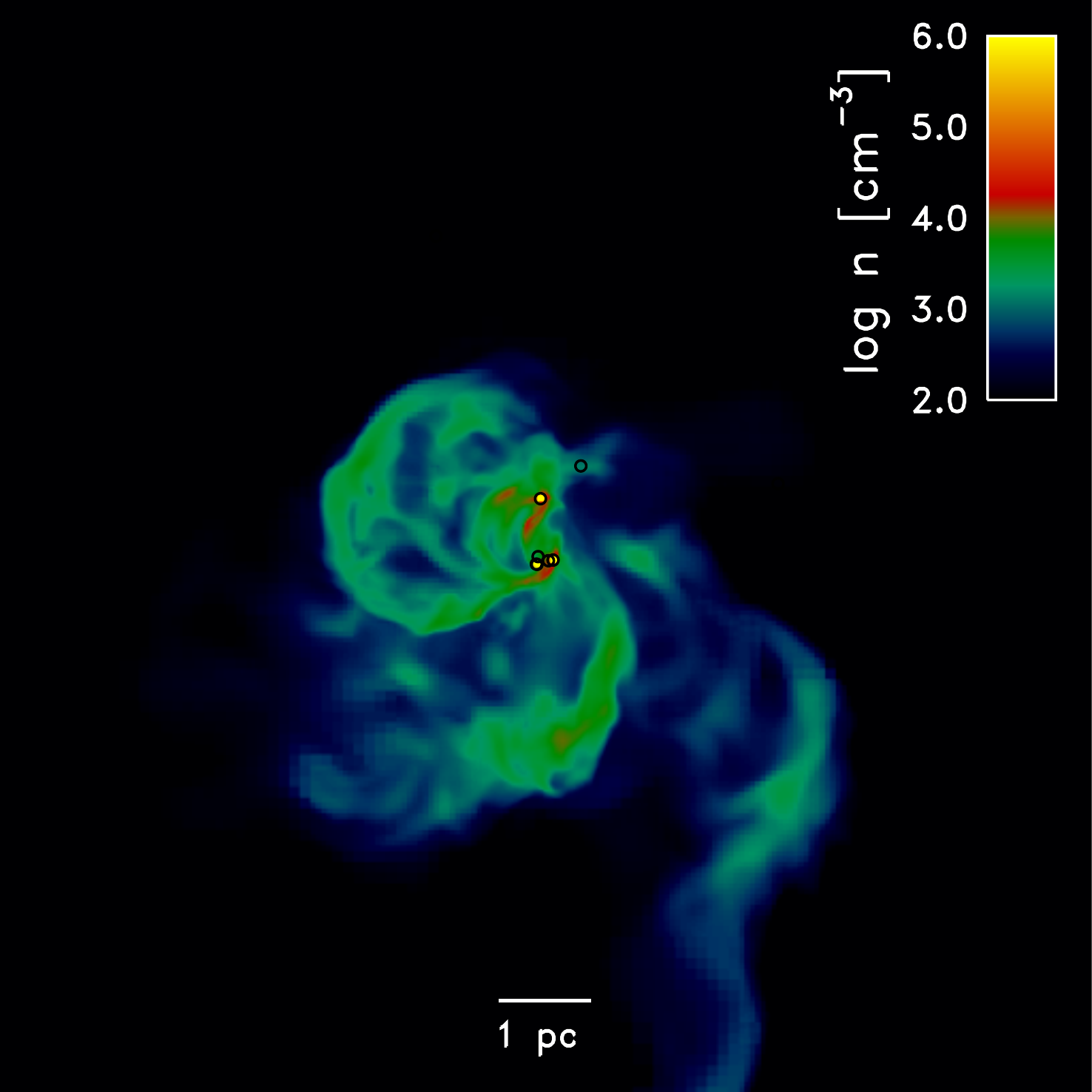} &
\includegraphics[scale=0.4, trim = 0 0 15 0]{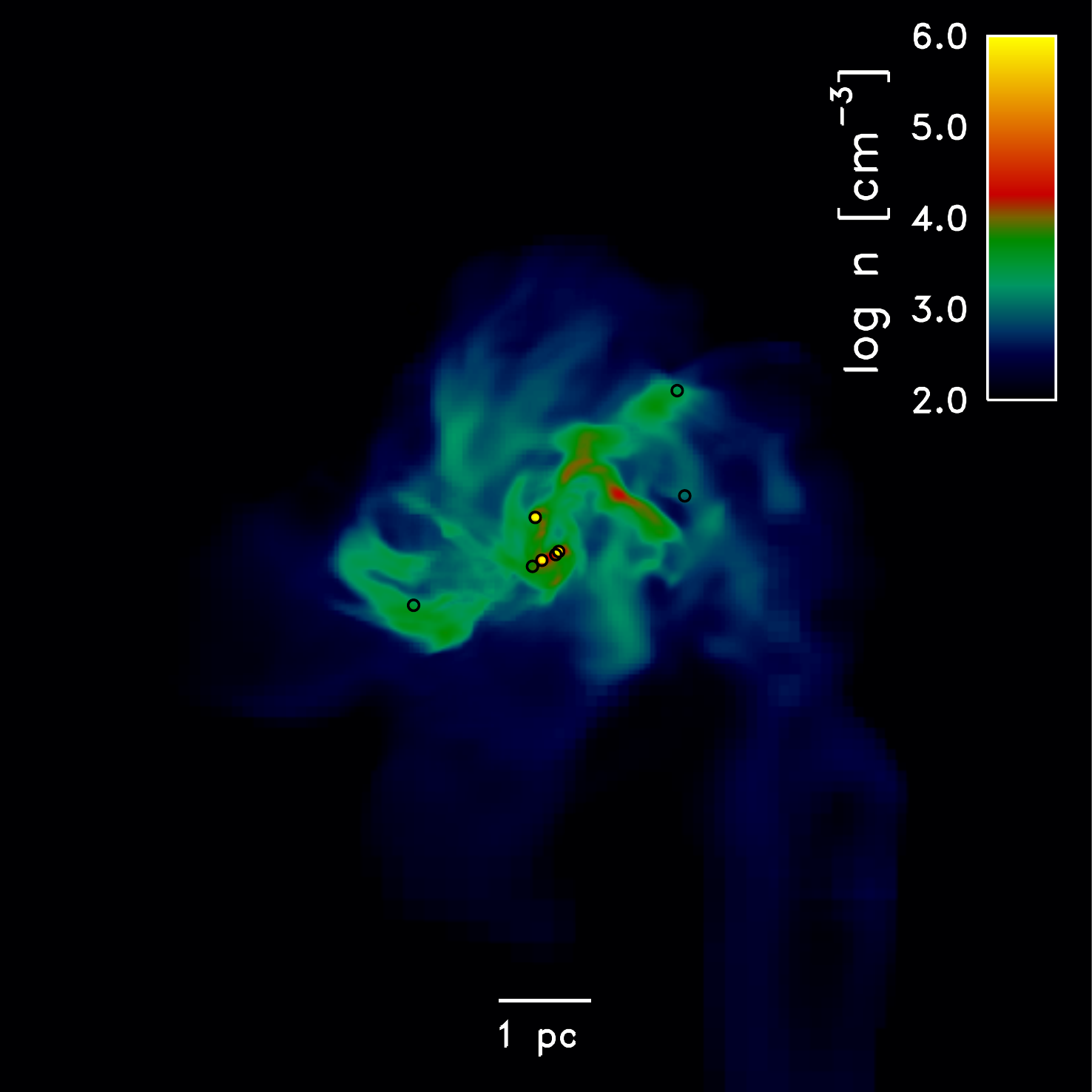} \\
\end{array}$
\end{center}
\caption{Same as Figure \ref{fig:mn2_morph_dens} but for the $10^{-3}\,\zsun$ run.}
\label{fig:mn3_morph_dens}
\end{figure*}

This formulation of metal fine-structure line cooling implicitly includes a CMB imposed temperature floor by including photon absorption terms with a radiation intensity $I_{\nu} = B_{\nu}(\tcmb)$. Below $\tcmb$, fine-structure cooling becomes heating via CMB photon absorption. For other coolants, notably $\hd$, we use an effective cooling rate of the form
\begin{equation}
\Lambda_{\mathrm{eff}}(T) = \Lambda(T) - \Lambda(\tcmb) \mbox{ ,}
\label{eq:lambda_eff}
\end{equation}
which mimics the effect of the CMB temperature floor.

\subsection{Sink Particles}
\label{sec:sink_particles}

We utilize sink particles to follow the evolution of gas undergoing gravitational collapse over multiple free-fall times. Sink particles were originally introduced by \citet{Bate95} in smoothed-particle hydrodynamics (SPH) simulations and were first adapted into a grid-based Eulerian setting by \citet{Krumholz04}. By accreting gas directly from the grid, sink particles effectively impose an upper-limit on gas density and a lower-limit on the simulation timestep. Without sink particles, the timestep would become prohibitively short as gas reached increasingly high densities. 

The limitations of the sink particle approach have been discussed extensively in the literature \citep[e.g.,][]{Bate03,Bate09,Bate10,Federrath10}. According to the original implementation of \citet{Bate95}, when a sink particle forms all SPH particles within its accretion radius are removed and their mass is added to the new sink particle. This results in a sharp pressure discontinuity around the sink particle and can lead to grossly overestimated accretion rates for subsonic gas flow. This problem can be overcome by imposing a boundary pressure at the sink particle accretion radius or by gradually, rather than instantaneously, accreting SPH particle mass \citep{Hubber13}. This problem is not as severe in Eulerian implementations, like ours,  since sink particle formation and the subsequent gas accretion does not result in a sharp pressure discontinuity. Other limitations include the effects of treating sink particles as softened extended objects and issues concerning angular momentum conservation and viscous torques during gas accretion. Finally, it should be stressed that inside the accretion radius of a sink particle all knowledge of the flow is lost and unless sinks form at a density much greater than the opacity limit for fragmentation, which is not the case in the present simulation, there is the possibility of unresolved sub-fragmentation.

We use the sink particle implementation from \citet{SafranekShrader12}, which is originally from \citet{Federrath10}, with slight modifications. Specifically, we no longer use a constant density cutoff for sink particle formation. Instead, we utilize a criterion similar to that in \citet{Krumholz04} where the sink particle formation is triggered by the failure to properly resolve the Jeans length. By mandating that the Jeans length (Equation \ref{eq:lj}) be resolved by at least $N$ grid cells, we can write the density threshold for sink particle creation as 
\begin{equation}
\rhoj = \frac{\pi \cs^2}{GN^2\Delta x^2}  \mbox{ ,}
\label{eq:rho_j}
\end{equation}
where $\cs$ is the sound speed and $\Delta x$ is the physical grid size at the highest level of refinement  ($l_{\mathrm{max}} = 20$). In practice, we always resolve the Jeans length by 24 grid cells until the highest level of refinement is reached where we set $N=4$ for sink particle creation. If a cell has density $\rho > \rhoj$ we flag it and perform a series of additional checks for sink particle creation. These include checking for convergence of the gas flow, $\bmath{\nabla}\cdot\bmath{v}<0$, that the cell is a local gravitational potential minimum, and that a small control volume around the cell (typically 2 -- 3 cells) is gravitationally bound. These checks have been shown to be very important for avoiding the spurious creation of sinks by structures not undergoing runaway free-fall collapse \citep{Federrath10}. If all these conditions are fulfilled, a sink particle is created in the cell centre. The sink's initial mass $(\rho-\rhoj)\Delta x^3$ is deducted from the cell mass, effectively capping the cell density at $\rhoj$. Gas accretion onto existing sink particles is handled in a similar way; cells with $\rho>\rhoj$ inside a sink particle's accretion radius, $\racc$, have a portion of their mass transferred to the sink particle if the gas is gravitationally bound to the sink particle and the radial component of the cell's velocity is directed towards the sink.


We evolve the systems for $\sim 4 \,\myr$ past the formation of the first sink particle. It is important to note that sink particles do not represent individual stars. Instead, sink particles are utilized as a computational tool making it possible evolve a self-gravitating, collapsing system for many free-fall times. In physical terms, they are approximately related to a pre-stellar core or clump that is destined to become a small stellar association \citep[e.g.,][]{Bergin07}. The properties of the sink particles, such as spatial distribution, mass, and accretion rates, thus do have physical significance. Higher-resolution studies are in progress that will shed light on the high-density fragmentation behavior of the sinks.

In the simulations here we take the sink accretion radius to be the Jeans length (Equation \ref{eq:lj}) at the highest level of refinement, $\racc = 4\Delta x = 0.06 \,\pc \approx 12,000\,\au$, where $\Delta x$ is the cell spacing at the highest level of refinement. The softening length for the sink-gas gravitational interaction is set to $\rsoft=\racc/2$ to ensure that the gravitational softening does not interfere with accretion onto the sink. Sink-sink interactions are softened when the distance between the two particles is less than $\Delta x/2 = \racc / 8$. While the density threshold $\rhoj$ is temperature dependent, we note that at a typical temperature of $50\,\kelvin\approx\tcmb(z=16)$, the threshold is $\rhoj = 6.3\times10^{-18} \mathrm{g}\,\cc$. Finally, we do not allow merging between sinks, even if the sink particles are gravitationally bound to each other. We comment on the validity of this choice in Section \ref{sec:caveats}.

\section{Results}
\label{sec:results}

\begin{figure*}
\begin{center}$
\begin{array}{ccc}
\includegraphics[scale=0.4, trim = 0 0 15 0]{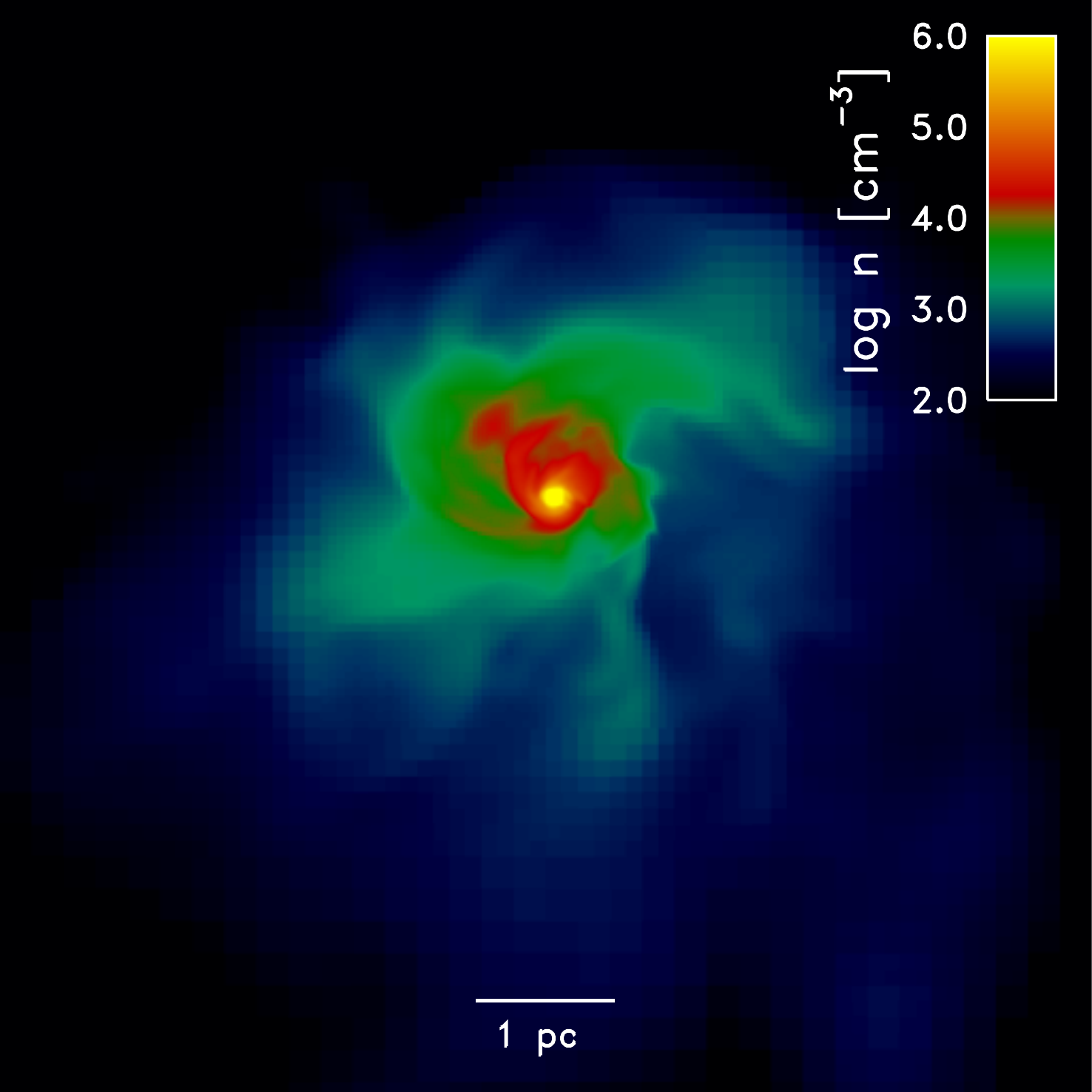} &
\includegraphics[scale=0.4, trim = 0 0 15 0]{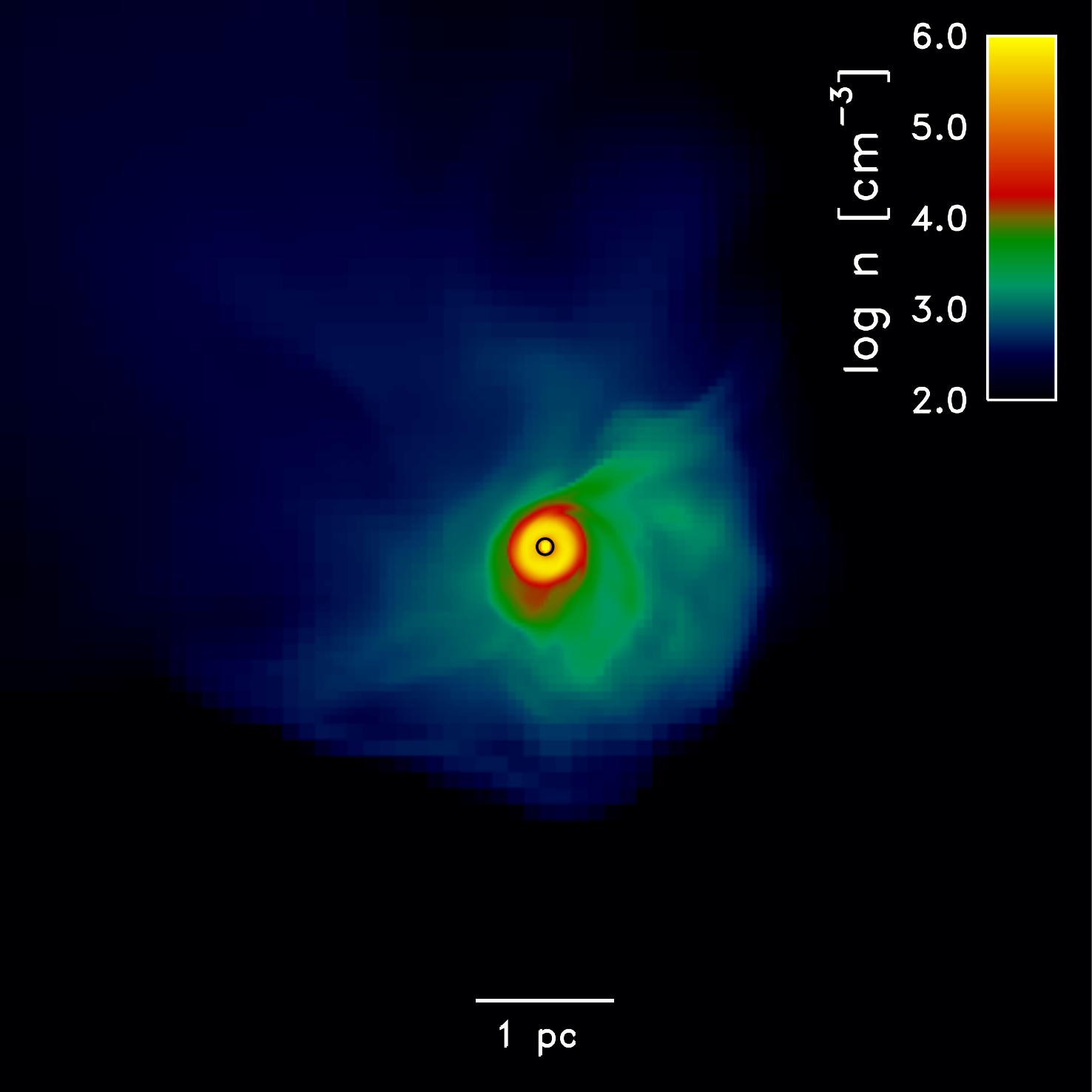} &
\includegraphics[scale=0.4, trim = 0 0 15 0]{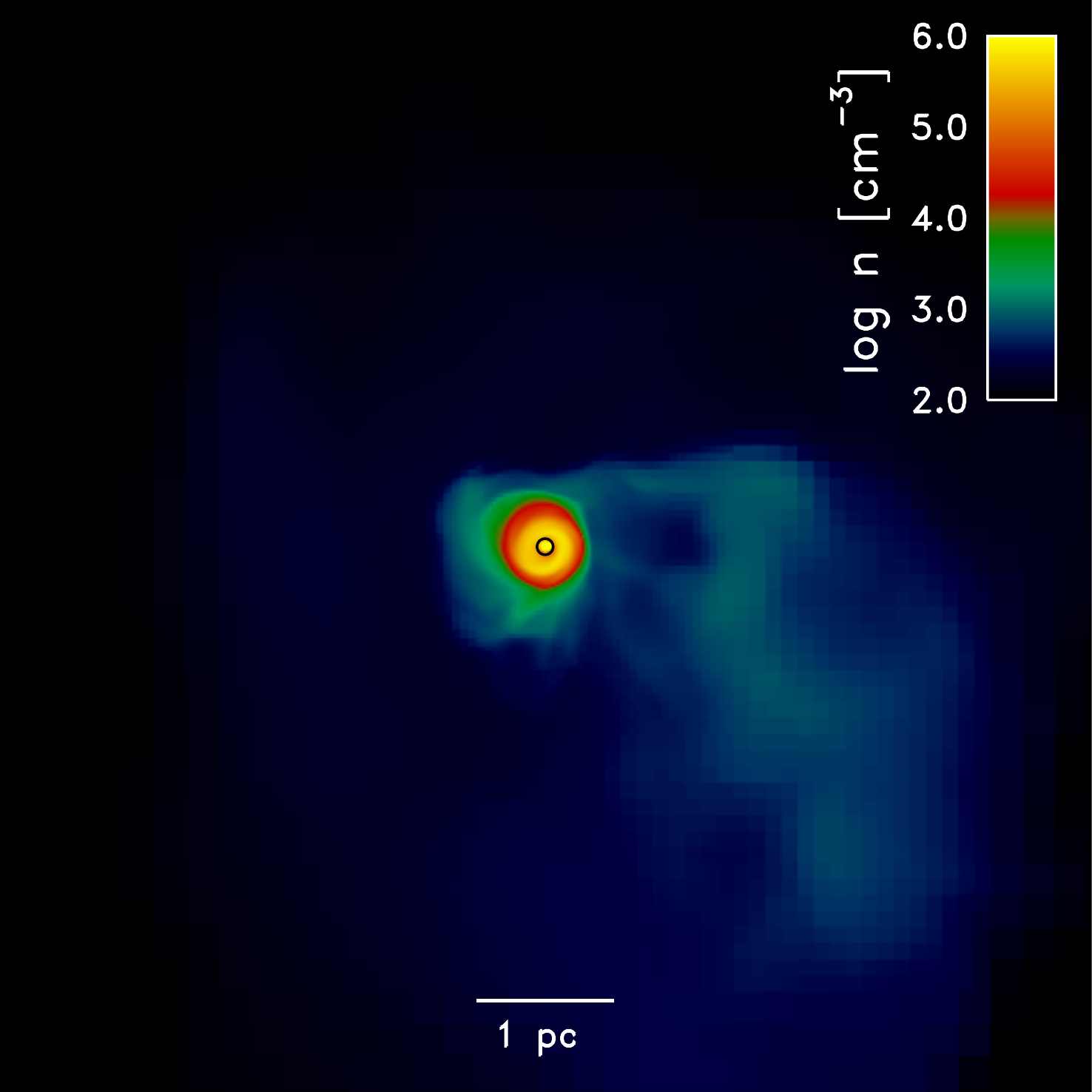} \\
\end{array}$
\end{center}
\caption{Same as Figure \ref{fig:mn2_morph_dens} but for the $10^{-4}\,\zsun$ run. The snapshots are approximately $2\,\myr$ apart. Differences between this and the higher metallicity runs (Figures \ref{fig:mn2_morph_dens} and \ref{fig:mn3_morph_dens}) are evident. Here, only a single sink forms and is surrounded by a non-fragmenting gas disk of radius $\sim0.5\,\pc$.}
\label{fig:mn4_morph_dens}
\end{figure*}

As in \citet{SafranekShrader12}, a spatially constant LW background with intensity $\jlw=100$ strongly suppresses the formation of $\htwo$ and thus prevents the collapse of gas in haloes unable to cool by atomic $\lya$ emission. Once a halo grows massive enough so that its virial temperature is $\tvir\approx10^4\,\kelvin$, gas begins to isothermally collapse at $T\sim8000\,\kelvin$. When gas reaches a density of $\sim100\,\cc$, we increase the gas metallicity within $\sim500\,\pc$ of the halo's point of maximum density to a constant, non-zero value. This occurs at a redshift of $z=15.8$ in a halo with virial mass $M_{\mathrm{vir}}=1.4\times10^7\,\msun$, maximum circular velocity $v_{\mathrm{circ}}=12\,\kms$, and virial radius $r_{\mathrm{vir}}\approx350\,\pc$. The halo at this point is shown in the top panels of Figure \ref{fig:halo_slice}. 

We choose three different values for this metallicity, $Z= 10^{-2}$, $10^{-3}$, and $10^{-4}\,Z_\odot$. The limiting values bracket the metallicity needed for metal fine-structure lines to significantly alter the thermodynamic evolution of gas and promote fragmentation \citep[e.g.,][]{Bromm01}. They are also physically reasonable given the expected generation and dispersal of metals in the first star forming halos  \citep[e.g.,][]{Wise08,Greif10,Ritter12}, and the metallicities of local, metal-poor ultra-faint dwarf spheroidal galaxies \citep[e.g.,][]{Kirby11,Frebel12}. The three simulations discussed here are identical except for the value of the metallicity.

 \begin{figure}
\includegraphics[scale=0.62, clip, trim = 10 5 0 5 ]{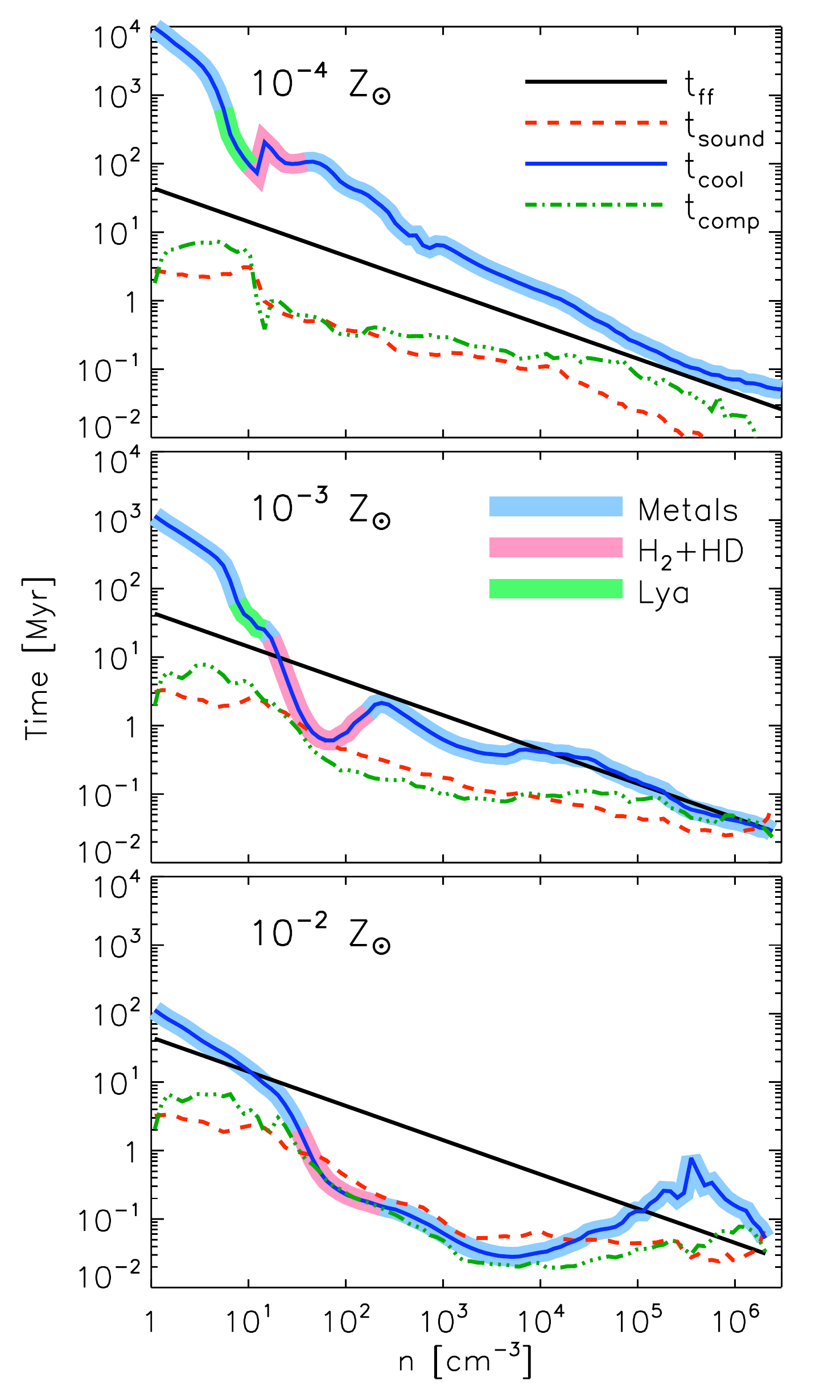}
\caption{Various characteristic time scales as a function of gas density in the $10^{-4}$ (top), $10^{-3}$ (middle), and $10^{-2}\,\zsun$ run (bottom). Shown is the free-fall time (solid black line), the sound crossing time of the pressure scale height (red dashed line --- see text), the compression time (green dot-dashed line --- see text) and the cooling time $\tcool = 3 n \kb T / 2\Lambda$ (blue solid line enclosed by a thicker line). The color of the thicker line enclosing the cooling time indicates the identity of the most effective coolant. Light blue refers to metal line cooling, light red is the cooling by $\htwo$ and HD, and green is $\lya$ cooling. }
\label{fig:metaltimescales}
\end{figure}

\subsection{Overall Evolution}
\label{sec:overall_evol}

In all three simulations, the non-zero metallicity greatly enhances the cooling rate.  The dominant cooling processes are fine-structure line emission by O and C$^+$. The onset of enhanced cooling allows the gas in the halo's center to cease its quasi-hydrostatic isothermal contraction and begin a collapse where temperature decreases with increasing density. In the $10^{-2}$ and $10^{-3}\,\zsun$ runs, the metal cooling induced collapse proceeds isobarically, remaining in pressure equilibrium with the surrounding warm halo gas. In the $10^{-4}\,\zsun$ run radiative cooling is not as efficient and the collapsing gas is overpressurized with respect to the surrounding gas. Therefore the collapse takes longer, by roughly a factor of two, for sink particles to form in this simulation compared with the higher metallicity runs. Gas in both the $10^{-2}$ and $10^{-3}\,Z_\odot$ runs reaches the CMB temperature, $\tcmb\sim50\,\kelvin$, in approximately one local free-fall time, although the larger cooling rate in the $10^{-2}\,\zsun$ run leads to much stronger CMB coupling and nearly isothermal evolution for $10^4\,\cc<n<10^6\,\cc$. In the $10^{-4}\,\zsun$ run, the gas temperature remains  above $\tcmb$, instead reaching a minimum temperature of $T\sim100\,\kelvin$. Eventually, gas in all three simulations reaches the density for sink particle formation given in Equation \ref{eq:rho_j}. We show density-temperature phase plots at the time of the first sink particle formation in Figure\ref{fig:dens_temp_multiplot}. Past the formation of the first sink particle we evolve each run for $\approx4\,\myr$. Given the characteristic density of the cold, dense regions in each run, $n\sim10^4\,\cc$, this corresponds to roughly eight free-fall times. In this time period, fragmentation was widespread in the $10^{-2}$ and $10^{-3}\,\zsun$ runs which formed 11 and 9 sink particles, respectively. Only one sink particle formed in the $10^{-4}\,\zsun$ run.

 \subsubsection{Morphology and Density Evolution}
 \label{sec:morph_and_dens_evol}

The subsequent collapse in the $10^{-2}\,\zsun$ run becomes nearly isothermal at a density of $n\sim10^3-10^4\,\cc$ and $T=45\,\kelvin = \tcmb$. The resulting structure is a cold, dense pocket of gas embedded within the turbulent, hot ($T\sim10^4\,\kelvin$) gas of the halo. This cold, dense region contains approximately $400\,\msun$ in gas and is $\sim1-2\,\pc$ in physical size. Compression of the cold by the warm medium produces a strong, quasi-isothermal shock with Mach number $\mach\equiv v/\cs\sim3$. The morphology of the cooled region evolves from roughly spherical to sheet-like and is seeded with filamentary density perturbations. These are the well known `thermal pancakes,' an outcome of nonlinear effects developing in the aftermath of thermal instability \citep[e.g.,][]{Kritsuk02}.  We see this in projection in the top-left panel of Figure \ref{fig:mn2_morph_dens} --- in this view the sheet-like compressed region is seen face-on. The filamentary morphology is created entirely by the strong shock and resulting instabilities after the gas hit $\tcmb$. We note this is very similar to the model in which cold atomic and molecular clouds form as a result of supersonic flow convergence and thermal instability in a warm neutral medium \citep[e.g.,][]{Hennebelle99,Koyama00,Hartmann01,Heitsch05,VazquezSemadeni06,Hennebelle07,Heitsch08}.

The density enhancement from the isothermal shock elevates the peak gas density to $\sim10^{5}\,\cc$. At this point, self-gravity becomes dominant in the highest density gas and two nearby sites of fragmentation emerge nearly concurrently, separated by $\sim0.1\,\pc$. This point in the simulation is shown in the top-middle panel of Figure \ref{fig:mn2_morph_dens}. These sites of fragmentation form two sink particles that rapidly create a tight binary pair. One of these sinks remains the most massive for the remainder of the simulation. We consider the onset of star formation to correspond to the time of the first sink particle formation. This occurs $4.8\,\myr$ after the gas metallicity was assigned a non-zero value and corresponds to one free-fall time when evaluated at the density triggering metal introduction, $n=100\,\cc$.

 \begin{figure}
\includegraphics[scale=0.55, clip, trim = 15 5 0 20 ]{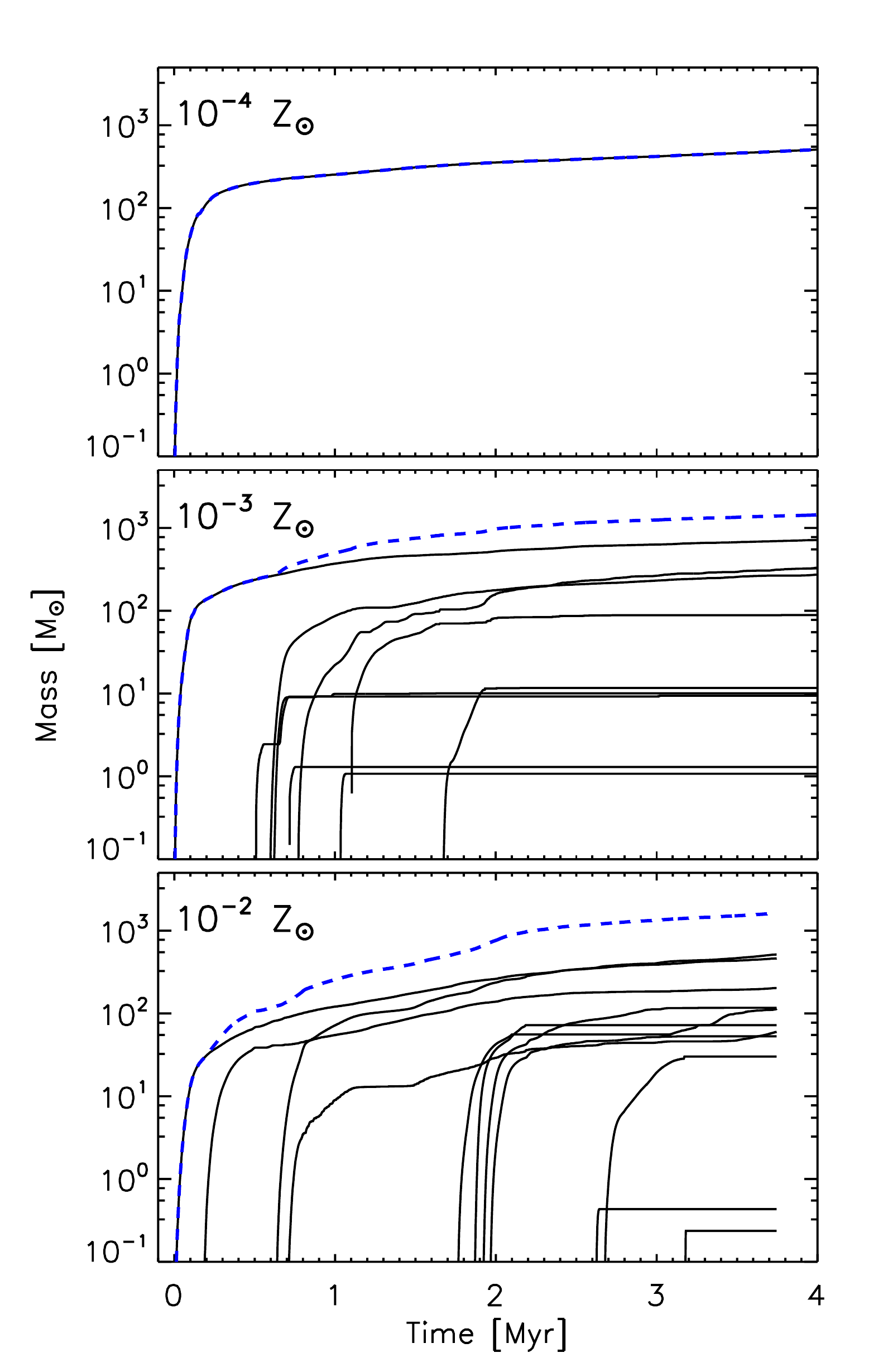}
\caption{Masses of individual sink particles (black lines) and total mass in sink particles (blue dashed line) in the $10^{-4}$ (top), $10^{-3}$ (middle), and the $10^{-2}\,Z_\odot$ (bottom) runs. Even though the $10^{-2}$ and $10^{-3}\,\zsun$ runs formed similar numbers and total masses of sink particles, new sink particle formation in the $10^{-3}\,\zsun$ run effectively ceased after $\sim2\,\myr$.}
\label{fig:sinkmass}
\end{figure}

The morphology of the cold, dense gas changes substantially over the course of the ensuing $\sim4\,\myr$ as can be seen in the remaining panels of Figure \ref{fig:mn2_morph_dens}. Only the densest regions, $n>10^4\,\cc$ survive this subsequent hydrodynamic bounce. Over the first $3.8\,\myr$ after the first sink particle formation, approximately $1700\,\msun$ of gas becomes incorporated into a total of 11 sink particles, a value that can also be considered a firm upper limit to the mass in stars that would have formed. Most of this mass, $\sim1000\,\msun$, exists in the two most massive sink particles (the first and third to form), with masses of $460$ and $510\,\msun$, respectively. The average sink particle mass is $\approx145\,\msun$, though this does include two sink particles whose masses are $0.2$ and $0.4\,\msun$. These sink masses are near the grid mass resolution of the simulation ($\rho\Delta x^3\sim0.3\msun$), though we argue in Section \ref{sec:sink_particle_accretion} that these are genuine gravitationally collapsed structures and not numerical artifacts. The average accretion rate onto sink particles was $\approx10^{-4}\,\msunperyr$, consistent with a few times the characteristic accretion rate in gas clumps collapsing after Jeans instability $\cs^3/G$ \citep[e.g.,][]{Shu77,Larson03} evaluated at $T=50\,\kelvin$. The sink particles are distributed in a region of approximately $1\,\pc$ in size. We stop the simulation $\approx4\,\myr$ after the formation of the first sink particle because this is roughly the time scale at which we would expect stellar radiative feedback to significantly affect the subsequent evolution. 


The $10^{-3}\,\zsun$ run is qualitatively similar to the $10^{-2}\,\zsun$ run. We show the density evolution for that run in Figure \ref{fig:mn3_morph_dens} and a representative density-temperature diagram in Figure \ref{fig:subfig_mn3}.  At $4\,\myr$ after the formation of the first sink particle there is a similar number of and amount of mass incorporated in sink particles as in the $10^{-2}\,\zsun$ run. The gas, as in the $10^{-2}\,\zsun$ run, hits the CMB temperature floor at a density of $n\sim10^4\,\cc$, but, unlike in the higher metallicity run, because of the smaller cooling rate, begins heating up again at still higher densities. It is difficult to assess the impact of the CMB temperature floor --- one-zone models would suggest that even in the absence of the CMB, gas at this metallicity would reach a minimum temperature of $T\sim50\,\kelvin$ \citep{Omukai05}. Nine sink particles form within the course of $4\,\myr$ with a total mass of $1450\,\msun$. Unlike in the $10^{-2}\,\zsun$ run, most of the fragmentation occurred in a disk that assembled around the first sink particle. The disk became locally gravitationally unstable and underwent pervasive fragmentation (seen in the top-middle panel of Figure \ref{fig:mn3_morph_dens}). This can be contrasted with fragmentation resulting from the collapse and break up of filamentary features in the $10^{-2}\,\zsun$ run.

Another major difference between the $10^{-2}$ and $10^{-3}\,Z_\odot$ runs is the amount of cold gas available for sink particle accretion. As can be seen in the progression of the panels in Figure \ref{fig:mn3_morph_dens}, the dense gas supply is continually diminishing and gas above a density of $n\sim10^{4}\,\cc$ is virtually non existent in the last panel when sink particles have been accreting for $4\,\myr$. Quantitatively, after $\approx4\,\myr$, there is $5700\,\msun$ of cold gas (defined such that $n>100\,\cc$ and $T<1000\,\kelvin$) in the $10^{-2}\,\zsun$ run and only $1500\,\msun$ in the $10^{-3}\,\zsun$ run. Either the sink particles in this run are much more effective at accreting gas, or more likely the supply of cold, dense gas from the halo is unable to meet the demand  of sink particle accretion. We return to this point in Section \ref{sec:sink_particle_accretion}.

We show the density evolution in the $10^{-4}\,\zsun$ run in Figure \ref{fig:mn4_morph_dens}.  We also show a representative density-temperature diagram in Figure \ref{fig:subfig_mn4}. Gas reaches a minimum temperature of $T\sim100\,\kelvin$ at $n\sim10^{4}\,\cc$ where the Jeans mass is $\sim1000\,\msun$. While the initial cooling that prompted the collapse was due to metal fine-structure emission, $\htwo$ did play a significant role due to self-shielding allowing the molecular abundance to increase (see Equation \ref{eq:fsh}). This is identical in many respects to the thermodynamical evolution of collapsing metal-free gas and likely leads to a similar outcome \citep[see, e.g.,][]{SafranekShrader12}. Given the warmer temperature of the dense gas, the strong shock present in the higher metallicity simulations was absent following the collapse, the subsequent collapse remained relatively spherical, and it did not exhibit sheet-like or filamentary density perturbations. Only a single sink particle formed surrounded by a gravitationally stable, non-fragmenting gaseous disk formed in the simulation. In $4\,\myr$, the lone sink particle reached a mass of $520\,\msun$, had an average accretion rate of $\approx10^{-4}\,\msunperyr$, and a peak accretion rate of $\approx10^{-3}\,\msunperyr$. This is a much smaller accretion rate than the peak rate found in collapsing primordial gas \citep[e.g.,][]{Abel02,Yoshida06,OShea07}, a discrepancy predominantly due to the much higher time and spatial resolution in those studies and their ability to capture the onset of collapse at higher densities. If stellar mass scale fragmentation occurred at higher densities than resolved in the simulation, this would result in an extremely compact stellar association, or cluster, with size less than the sink particle accretion radius, $0.01\,\pc$ --- this estimate, however, neglects the internal dynamical evolution of the cluster which could eject stars onto more extended orbits \citep[e.g.,][]{Omukai08}.

The importance of following the evolution for multiple free-fall times, enabled by our use of sink particles, past the first point of collapse becomes evident from an inspection of the morphology evolution plots (Figures \ref{fig:mn2_morph_dens}, \ref{fig:mn3_morph_dens}, and \ref{fig:mn4_morph_dens}). This is most apparent in the $10^{-2}$ and $10^{-3}\,\zsun$ runs where the gas morphology at the point of the first sink particle formation is not at all representative of its longer-term, $\sim10^6\,\yr$, evolution. The initial isobaric cooling phase and isothermal collapse towards sink particle formation produces a dense, compact region with significant filamentary density perturbations. Within $1\,\myr$ the region containing cold, dense gas becomes more diffuse. The further evolution, driven by the accretion of gas and sink particle dynamics produces a disordered, intermittent, filamentary flow morphology, typical of supersonic turbulence \citep[e.g.,][]{Kritsuk07,Federrath10a}. The majority of the sink particles, and thus stellar objects, form from density fluctuations not present at the point of first collapse. The $10^{-4}\,\zsun$ run, though, does not experience any significant morphological evolution after sink particle creation.

 \subsubsection{Characteristic Time Scales}
 \label{sec:timescales}

To better understand the differences between the three runs, in Figure \ref{fig:metaltimescales} we show characteristic time scales related to the dynamical and thermal evolution of the gas. We plot the free-fall time, $\tff = (3\pi/32G\rho)^{1/2}$, the compression time, $t_{\mathrm{comp}} = \rho / |\vect{\nabla}\cdot(\rho\vect{v})|$, the sound crossing time,  $t_{\mathrm{sound}} = \lambda_{P} / \cs$, where $\lambda_{P}$ is the pressure scale length given by $\lambda_{P} = P/|\vect{\nabla}P|$, and the cooling time, $t_{\mathrm{cool}} = 3n\kb T / 2\Lambda$. We choose to treat the sound crossing time as a local quantity since sound waves need not propagate across an extended region to maintain isobaricity, only across one pressure scale length. The time scales are computed via mass-weighting in logarithmically-spaced density bins. For the cooling time, we additionally denote the most dominant coolant in the particular density bin (see figure).

Gas is expected to behave quasi-isobarically when $t_{\mathrm{sound}}\lesssim\tff,t_{\mathrm{cool}}$. This is indeed the case in the $10^{-2}$ and $10^{-3}\,\zsun$ runs at densities between $\sim10\,\cc$ and $\sim10^4\,\cc$. As evident in Figures \ref{fig:subfig_mn3} and \ref{fig:subfig_mn2} the gas cooling from $T\approx10^4\,\kelvin$ to $\tcmb$ does indeed proceed quasi-isobarically. In the $10^{-4}\,\zsun$ run, however, the ordering of time scales is $t_{\mathrm{sound}} \ll \tff,t_{\mathrm{cool}}$, suggesting that the gas is overpressured with the respect to its surroundings, with gravity, rather than ambient warm gas pressure, driving the compression. This is supported by the density-temperature slopes in Figure \ref{fig:subfig_mn4}, and by the longer time gravitational collapse took after the gas metallicity was increased in the $10^{-4}\,\zsun$ run as compared to the higher metallicity simulations. The time scales therefore help explain the different outcomes in the $10^{-4}\,\zsun$ run that exhibited no fragmentation and the higher metallicity runs with pervasive fragmentation.

Figure \ref{fig:metaltimescales} also highlights the relative importance of molecular and metal cooling \citep[see, e.g.,][]{Glover13}. When the gas is first endowed with a non-zero metallicity, the increased cooling rate pushes the gas past a threshold where it ceases its isothermal collapse and begins an evolution in which temperature decreases with increasing density and the adiabatic index is sub-isothermal, $d\ln P/d\ln \rho<1$. This increases the $\htwo$ self-shielding factor (Equation \ref{eq:fsh}), the $\htwo$ abundance rapidly increases, and $\htwo$ becomes the dominant coolant in each simulation, albeit only over a small range of densities. Once the gas temperature drops below $\sim200\,\kelvin$, however, $\htwo$ loses its cooling efficacy and metal line-cooling once again dominates. We note that when $\htwo$ is the dominant coolant, its cooling rate is never more than a factor of $\sim3$ greater than the metal line cooling rate.

The metal abundance, perhaps surprisingly, plays a significant role in regulating the $\htwo$ abundance. Our treatment of carbon and silicon as singly ionized significantly elevates the free electron abundance at densities $n>100\,\cc$, an effect more pronounced in the higher metallicity simulations. In the $10^{-2}\,\zsun$ run these additional free electrons actually set a lower limit of $\abunde\sim2\times10^{-6}$ on the free electron abundance. Given that the formation of $\htwo$ through the gas phase $\mathrm{H}^{-}$ channel depends sensitively on $\abunde$, it is not surprising that the primordial coolant would depend on the gas metallicity.

\subsection{Sink Particle Formation, Growth, and Mass Function}
\label{sec:sink_particle_accretion}

 \begin{figure*}
\includegraphics[scale=0.7, clip, trim = 0 0 0 0 ]{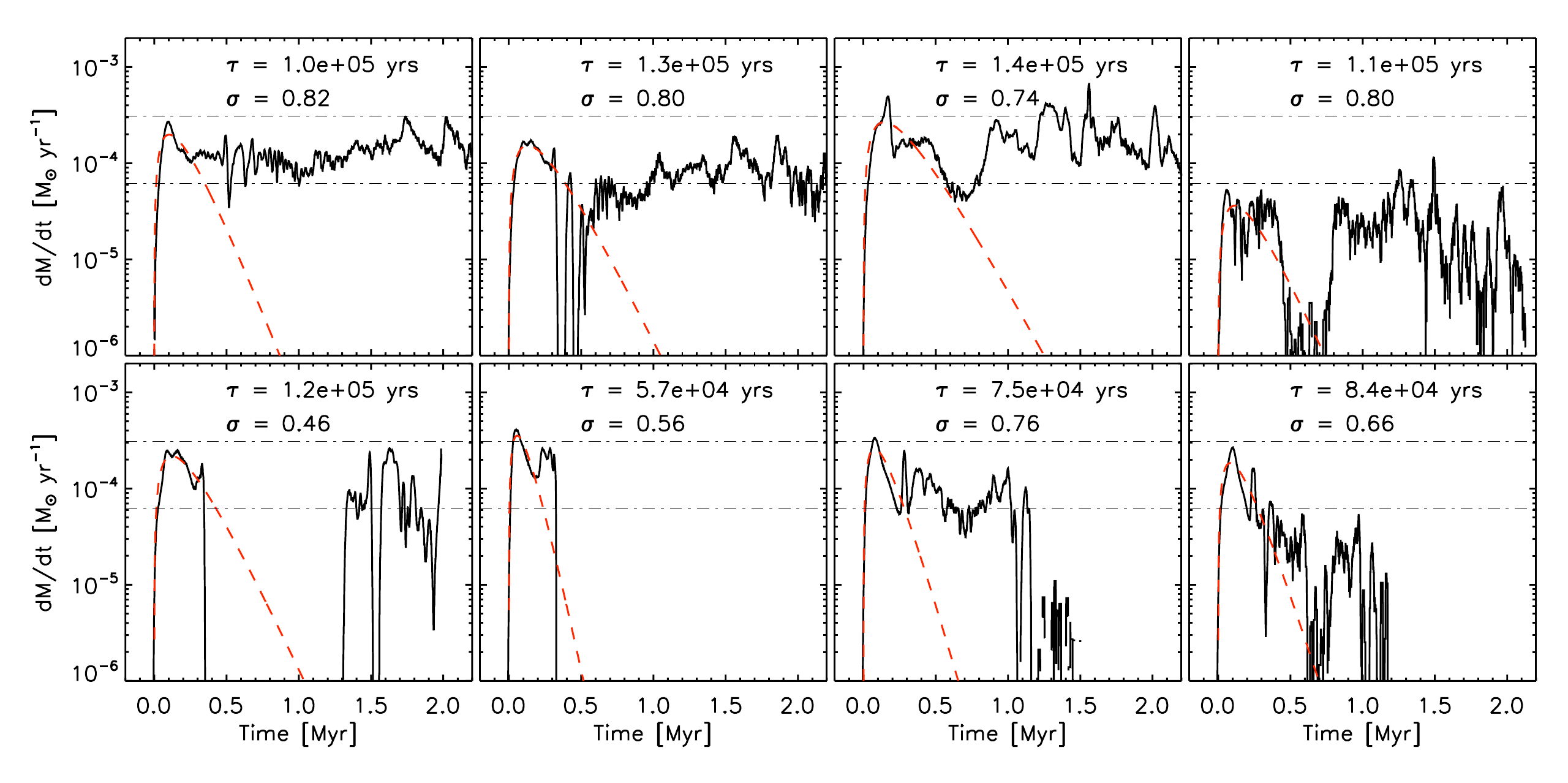}
\caption{Accretion rates of the first eight sink particles to form in the $10^{-2}\,Z_\odot$ run in the first $\sim2\,\myr$. The top and bottom horizontal dot-dashed lines show the fiducial accretion rate $\cs^3/G$ evaluated at $T=100\,\kelvin$ and $T=50\,\kelvin$, respectively. The dashed red line is a fit to the first $0.5\,\textrm{Myr}$ of accretion (Equation \ref{eq:mdot_fit}) with the value of fitting parameter $\tau$ show at the top of each panel.}
\label{fig:sink_accrete2}
\end{figure*}

 We emphasize that the sink particles here do not represent single stars, but instead are small stellar associations. We take the sink particle mass as a proxy for the stellar mass of the association. This mass is a firm upper limit as we do not include any form of stellar feedback that would likely act to reduce the gas-to-star conversion efficiency. The total mass accretion rate onto protostellar objects in one association, which is approximated by the sink particle gas accretion rate, has implications for protostellar evolution \citep[e.g.,][]{Omukai03}. We calculate the sink particle accretion rates by dividing the mass accreted during a timestep by the length of the timestep. To derive any insight on the nature of individual protostars we would need to probe higher densities up to the opacity limit for fragmentation \citep{Low76,Rees76}.

In Figure \ref{fig:sinkmass} we show the total mass in sink particles over time in the three simulations. The two higher metallicity runs reach a total sink mass of $\approx1500\,\msun$ in $4\,\myr$. The $10^{-4}\,\zsun$ run converts roughly a factor of 3 lower gas mass, $500\,\msun$, into one sink particle $4\,\myr$ after its formation. In all runs, most of the sink mass is in a few large mass ($>500\,\msun$) objects. The average sink accretion rate is $\approx4\times10^{-4}\,\msunperyr$ in the two higher metallicity simulations and $\approx10^{-4}\,\msunperyr$ in the lowest metallicity simulation. The general trend is that the first sink particles to form remain the most massive, though there are exceptions.

In the $10^{-2}$ and $10^{-3}\,\zsun$ runs, many sink particles appear to permanently stop accreting after less than $\sim10^5$ years of growth --- this is most apparent in the middle panel of Figure \ref{fig:sinkmass}. It is possible these ephemerally accreting sink particles do not represent physical gravitationally bound structures but are numerical artifacts; many of their masses lie far below the typical Jeans mass where fragmentation is likely to set in ($\mj\sim50-100\,\msun$). These sinks, though, did pass a stringent suite of tests for creation (see Section \ref{sec:sink_particles}) that includes only forming sink particles from a gravitationally bound converging gas flow. Thus it seems that these objects are collapsed physical structures and that their mass growth was terminated by a physical mechanism, likely strong tidal interactions with more massive sink particles, though higher resolution simulations are needed to establish this conclusively.

\begin{figure}
\includegraphics[scale=0.55, clip, trim = 25 5 0 20 ]{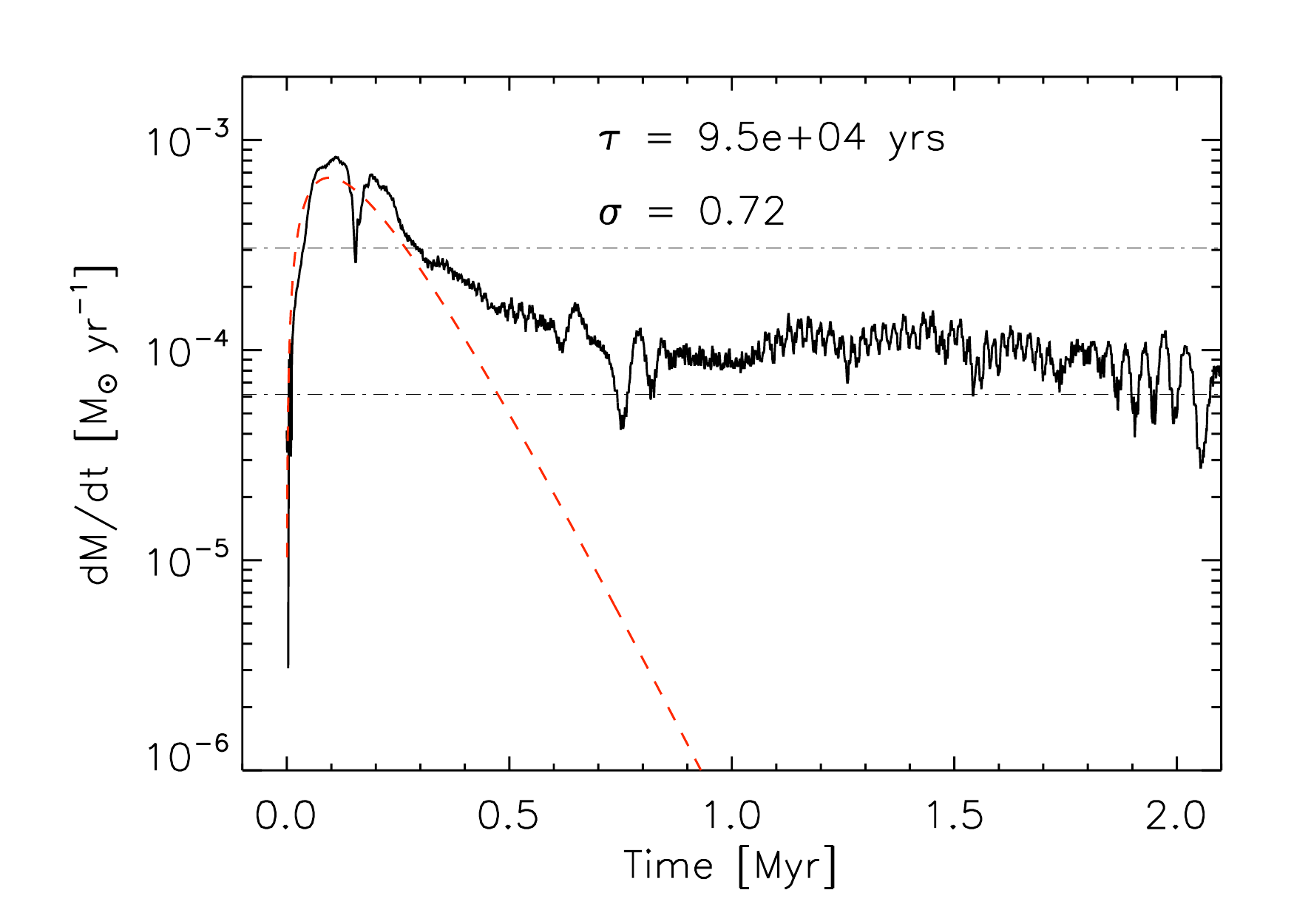}
\caption{Same as Figure \ref{fig:sink_accrete2}, but for the $10^{-4}\,Z_\odot$ run and smoothed on a $10^4,\textrm{yr}$ time scale. Only one sink particle formed in the simulation. }
\label{fig:sink_accrete4}
\end{figure}

We can understand the formation and growth of sink particles further by inspecting their individual accretion rates. In Figures \ref{fig:sink_accrete2} and \ref{fig:sink_accrete4} we do so for the first eight sink particles formed in each simulation (except for the $10^{-4}\,\zsun$ run where only one sink formed). The individual sink particle accretion rates in the $10^{-3}\,\zsun$ simulation are qualitatively similar to the $10^{-2}\,\zsun$ run and are not shown. We also show a simple fit meant to capture the accretion rate history of protostellar objects forming via gravo-turbulent fragmentation suggested by \citet{Schmeja04},
\begin{equation}
\dot{M}(t) = \dot{M}_0\frac{e}{\tau}\,t\,e^{-t/\tau} \mbox{,}
\label{eq:mdot_fit}
\end{equation}
where $\dot{M}_0$ and $\tau$ are fitting parameters and $t$ is time measured from sink particle formation. We estimate the quality of the fit similarly to \citet{Schmeja04} by computing the normalized standard deviation
\begin{equation}
\sigma = \sqrt{\frac{1}{n-1}  \sum\limits_{i=1}^n \left[1-\frac{\dot{M}(t_i)}{\dot{M}_{\mathrm{fit}}(t_i)} \right]^2 }  \mbox{,}
\label{eq:mdot_fit}
\end{equation}
where $\dot{M}_{\mathrm{fit}}(t)$ is the fit from Equation \ref{eq:mdot_fit}, $\dot{M}(t)$ is the accretion rate extracted from our simulations, $n$ is the total number of data points, and we compute $\sigma$ for a time period of $5\tau$ for each sink particle. In this $5\tau$ time period, we find $\sigma$ to be typically between $0.5-0.8$. Most sinks possess an initial accretion rate peak which lasts for $\sim5\times10^4 - 10^5$ years. This peak accretion rate is typically on the order of $\dot{M}_{\rm J}=\mj/t_{\mathrm{ff}}\sim\cs^3/G$. In this time period, the sink particle accretes the initial Jeans unstable gas clump that triggered its formation. In the $10^{-2}$ and $10^{-3}\zsun$ runs this mass is between $20-100\,\msun$ depending on the exact temperature and density when fragmentation set in. This evolution of the accretion rate can be understood by considering the initial reservoir of the gas the sink particles accrete as a Bonnor-Ebert sphere or a Plummer-like density profile consisting of a flat plateau at small radii and a power law envelope at larger radii \citep[e.g.,][]{Whitworth01}. This first peak in $\dot{M}(t)$ is the accretion of the flat, inner plateau. For the majority of the sink particles, this early phase lasting $\sim0.5\,\myr$ is fit reasonably well by Equation (\ref{eq:mdot_fit}) as shown by the small standard deviation, but the fit becomes significantly worse as time passes and the accretion rate either stays relatively constant or instantaneously drops to zero.

Almost all the sink particle accretion rates show a similar early time ($\lesssim0.5$ Myr) behavior --- a rapid rise to a peak value followed by a drop to a lower (or zero) accretion rate. It is reasonable to ask whether this general behavior is physical or is instead a numerical artifact from instantaneous gas pressure loss following sink particle formation \citep[e.g.,][]{Bate95}. As discussed in Section \ref{sec:sink_particles}, Eulerian sink particle implementations do not suffer from this issue to the same degree that SPH implementations do. Additionally, numerous numerical studies of protostellar collapse have found rapidly peaking accretion rates followed by a decrease \citep[e.g.,][]{Hunter77,Basu97,Whitworth01,Motoyama03,Schmeja04}. While it is possible the accretion rates we measure are slightly overestimated by the reduction of gas density around sink particles, this general behavior we see is physically expected.

Beyond the first accretion rate peak, many sink particles in the $10^{-2}$ and $10^{-3}\,\zsun$ runs, especially the earliest ones formed, show sustained gas accretion over many Myr. Others do not display sustained gas accretion but instead have their accretion terminated within a few $10^5$ years after their formation. This sharp cutoff in the accretion rates is primarily the result of these sinks experiencing a close encounter with a more massive sink particle, also seen in idealized simulations of present-day \citep[e.g.,][]{Klessen00,Klessen01} and Pop III \citep{Clark11} star formation. These strong dynamical encounters tend to eject less massive sinks to the outskirts of the cold gas environment where gas densities are smaller and accretion is thus lowered or completely shut off. Given the density and velocity dependence of the Bondi-Hoyle accretion rate, $\dot{M}_{\mathrm{BH}}\propto \rho v^{-3}$, the ejected sinks effectively stop accreting.


Overall, it seems that the mode of gaseous accretion onto sink particles is a hybrid between monolithic collapse \citep[e.g.,][]{McKee03}, where the stellar mass (and mass function) is ultimately determined by the mass of gravitationally unstable cores, and competitive accretion \citep[e.g.,][]{Bonnell01}, where stellar masses are the result of many cores competing for the same reservoir of gas resulting in more massive (though rarer) objects near the cluster center accreting more such that the `rich get richer.' The less massive sink particles, typically the latter ones to form, can only accrete a portion of their initial Jeans-unstable gas clumps before dynamical interactions with other sink particles terminate gas accretion. Higher mass sink particles from more massive Jeans-unstable clumps and then continue accreting from the dense gas in the central region of the cluster over many Myr.

\begin{figure}
\includegraphics[scale=0.62, clip, trim = 20 10 10 0 ]{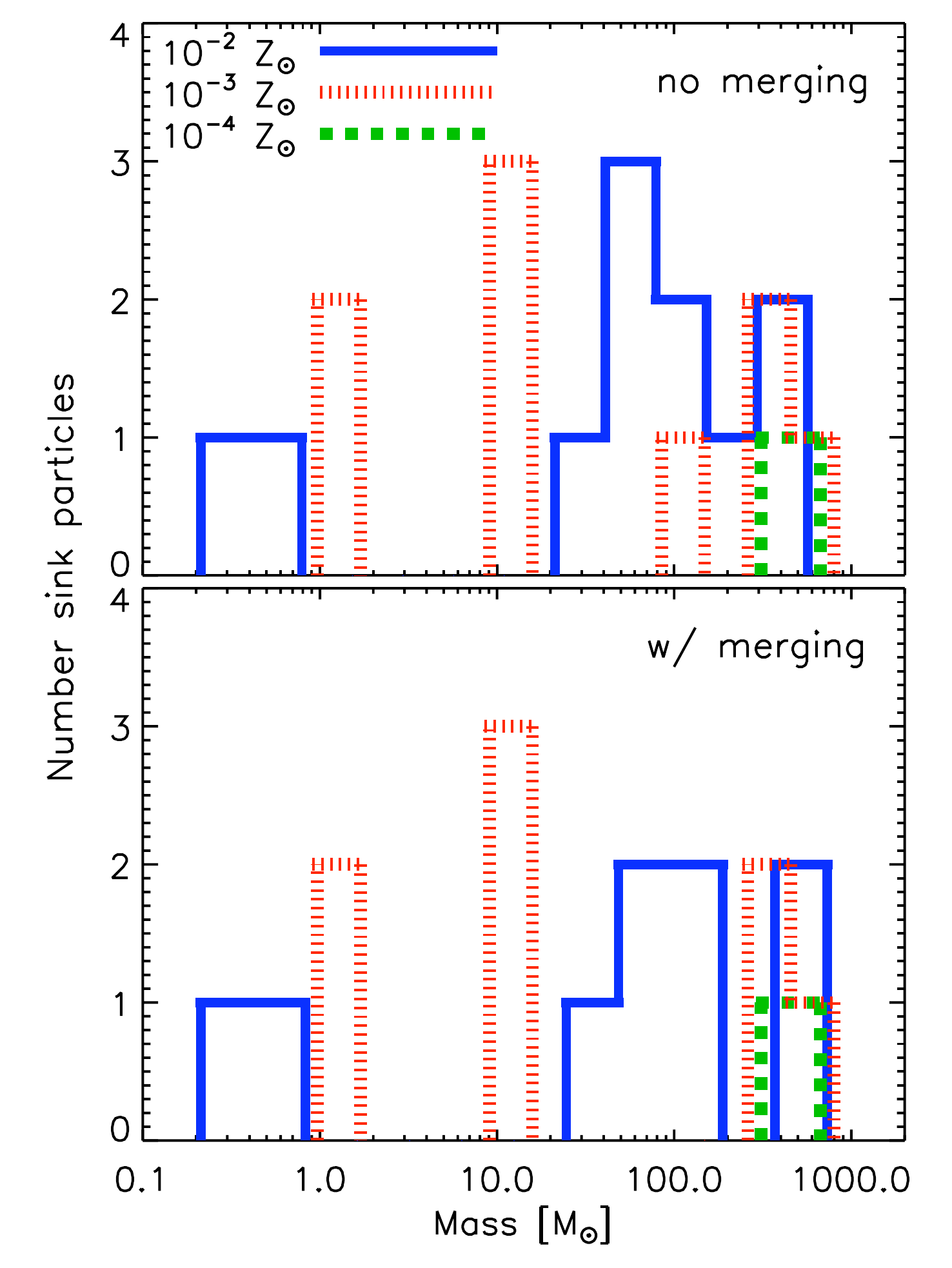}
\caption{Distribution of sink particle masses in the $10^{-2}$ (blue solid line), $10^{-3}$ (red short-dashed line), and $10^{-4}\,\zsun$ (green dotted line) runs $4\,\myr$ after the first sink particle formation. The top panel is the mass function extracted directly from the simulations.  In the bottom panel, in post-processing, we merged any sink particles within each other's accretion radius. Sink merging performed in post-processing does not drastically affect the mass spectrum.}
\label{fig:sink_mass_spectrum}
\end{figure}

The top panel of Figure \ref{fig:sink_mass_spectrum} shows the sink particle mass function at the end of each run after star formation has progressed for $\approx 4\,\myr$. The higher metallicity simulations show a preferred mass scale around $\sim50\,\msun$, roughly consistent with the Jeans mass at the point at which the gas density hits the CMB temperature and fragmentation is expected to be thermodynamically suppressed. While this mass function does not represent the final stellar IMF, observations have shown there to be a connection between the pre-stellar clump mass function and the ultimate stellar IMF \citep[e.g.,][]{Motte98,Beuther04,Andre10}, suggesting that the power-law slope toward higher masses, only marginally evident in the $10^{-2}\,\zsun$ run, may translate directly to the slope of the ultimate stellar IMF. We caution that the inclusion of radiative feedback from accreting protostars has the potential to suppress fragmentation and thus alter the resultant mass function, particularly at the low mass end \citep[e.g.,][]{Krumholz07,Urban10,Bate12}. The bottom panel of Figure \ref{fig:sink_mass_spectrum} shows the mass function in which sink particles located within each others accretion radius are merged in post-processing and are only counted as a single particle. Doing this, four sink particles in the $10^{-2}\,\zsun$ run merge into two sinks, while in the $10^{-3}\,\zsun$ run only one pair of particles merges. As is clear, this merging procedure has little effect on the resultant mass function. Finally, we emphasize that the mass function is sensitive to the resolution (which determines the sink particle formation density) and the particular choice of sink particle parameters (such as the accretion radius and the sink-gas softening length) as discovered in our initial exploratory simulations.


\subsection{Density Probability Distribution Function}
\label{sec:pdf}
 
\begin{figure}
\includegraphics[scale=0.55, clip, trim = 0 10 10 0 ]{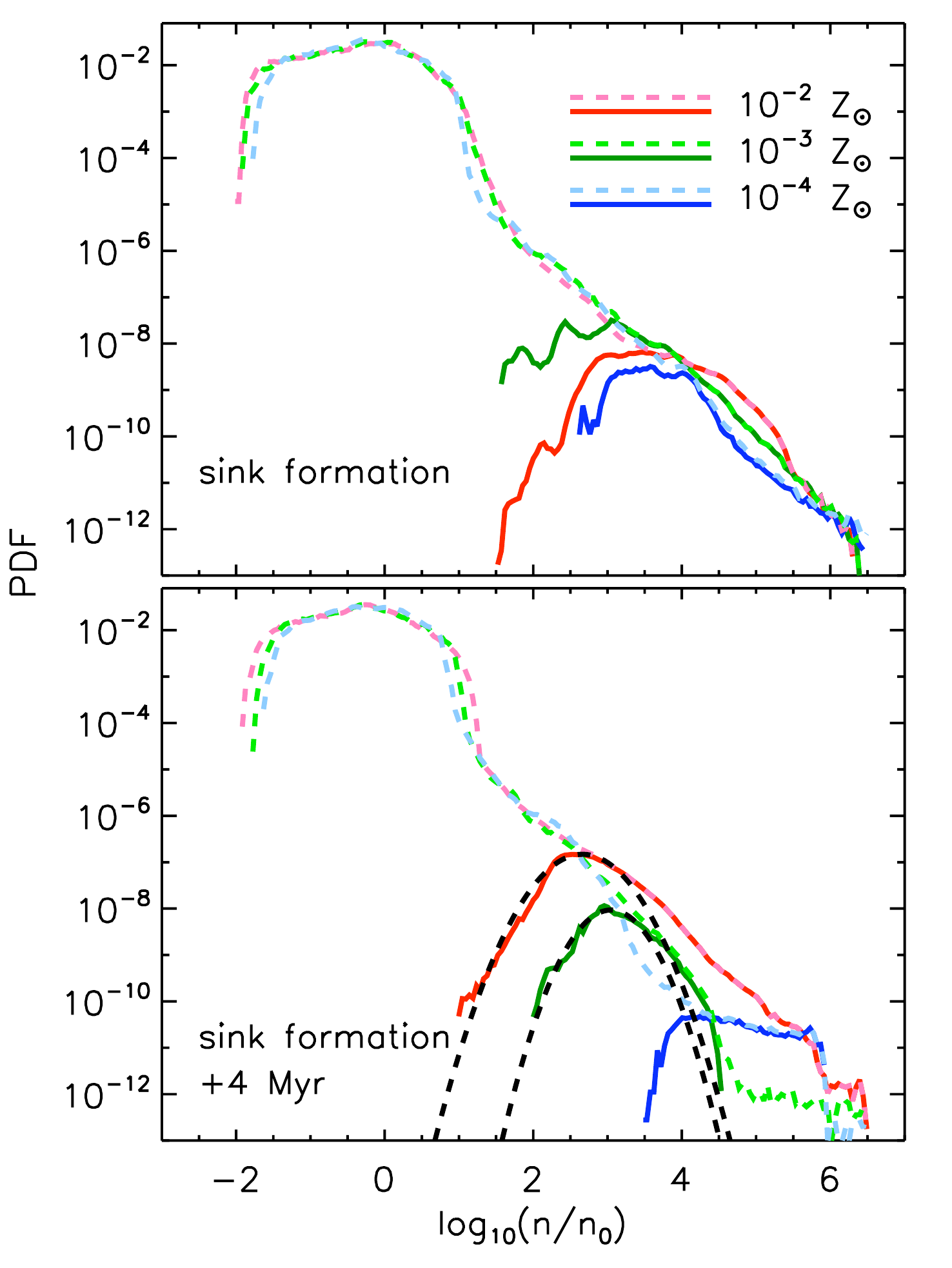}
\caption{Gas density PDFs at the onset of sink particle formation (top panel) and $4$ Myr later (bottom panel). We show PDFs from the $10^{-2}$ (red lines), $10^{-3}$ (green lines), and $10^{-4}\,\zsun$ (blue lines) runs. We separately plot PDFs computed using all gas within the virial radius of the halo (dashed lines) and only gas with $T<200\,\kelvin$ and $n>100\,\cc$ (solid lines). For the all-gas PDFs, the average density is $n_{0}\approx1\,\cc$ at both times. In the bottom panel, we show lognormal fits to the $10^{-2}$ and $10^{-3}\,\zsun$ run cold gas PDFs which have widths of $\sigma=0.86$ and $0.71$, respectively. Power law tails, evident in the top panel and in the $10^{-4}\,\zsun$ run in the bottom panel, are an anticipated outcome of self-gravity. In the $10^{-3}\,\zsun$ run, however, this power-law tail disappears within $\sim2\,\myr$ after the first sink particle formation as the sink particles deplete the cold, dense gas. }
\label{fig:density_pdf}
\end{figure}

The gas density probability distribution function (PDF), $p(\rho)$, is defined such that $p(\rho)d\rho$ represents the probability that a given parcel of gas has density within the range $[\rho,\rho+d\rho]$. It is particularly sensitive to the character and strength of turbulence that imprints a specific turbulent signature on the gas density PDF. We show volume-weighted gas density PDFs for the three simulations and two different times in Figure \ref{fig:density_pdf}. We show the PDF for all gas within the virial radius of the halo and only for gas with temperature $T<200\,\kelvin$ and density $n>100\,\cc$, the latter representing the gas that has cooled and is participating in star formation.

The density PDFs in Figure \ref{fig:density_pdf} exhibit a number of interesting features. The high density, cold gas PDFs peak at densities roughly 3 orders of magnitude higher than that of the full density PDFs. The cold gas PDFs also represent a significant `bump' in the full density PDF, mainly seen in the bottom panel. This is clearly reminiscent of the density structure in thermally bistable turbulent flows \citep[e.g.,][]{Gazol10,Siefried11,Saury13}. The sharp increase in the peak density of the cold gas PDF in the $10^{-4}\,\zsun$ run is due to the $T<200\,\kelvin$ temperature cutoff which was selected with the higher metallicity simulations in mind, meant to cleanly separate the cold and hot phases that are separated by an isobarically unstable cooling phase. In the $10^{-4}\,\zsun$ run, is it not possible to cleanly distinguish between the ambient halo gas and the cooling gas.

It is well established that supersonic, isothermal turbulence tends to produce a lognormal gas density PDF \citep[e.g.,][]{VazquezSemadeni94,Scalo98,Passot98,Ostriker01,Federrath08}. As in \citet{SafranekShrader12}, we write the lognormal distribution as 
 \begin{equation}
p(s)ds = \frac{1}{(2\pi\sigma^2)^{1/2}} \mathrm{exp}\left[-\frac{1}{2}\left(\frac{s-\bar{s}}{\sigma}\right)^2\right]ds \mbox{,}
\label{eq:lognormal}
\end{equation}
 where $s = \mathrm{ln}(n/n_{0})$ is the logarithmic density contrast and $\sigma$ is the lognormal width. The average of the logarithmic density contrast is related to $\sigma$ via $\bar{s} = -\sigma^2 / 2$, effectively reducing Equation \ref{eq:lognormal} to a one parameter model. Only the cold gas PDFs for the $10^{-2}$ and $10^{-3}\,\zsun$ runs $4\,\myr$ after sink particle formation are reasonably well fit by the lognormal distribution. We plot a lognormal fit to these two PDFs in the bottom panel of Figure \ref{fig:density_pdf}. The main deviation from the lognormal shape is the power-law tail at high density contrasts, understood to be a result of self-gravity \citep[e.g.,][]{Kritsuk11} or the peculiar dynamics of bistable turbulent flows \citep{Gazol10,Siefried11,Saury13,Federrath13}.

\subsection{Star Formation Efficiency}
\label{sec:efficiency}

 \begin{figure}
\includegraphics[scale=0.52, clip, trim = 30 5 0 20 ]{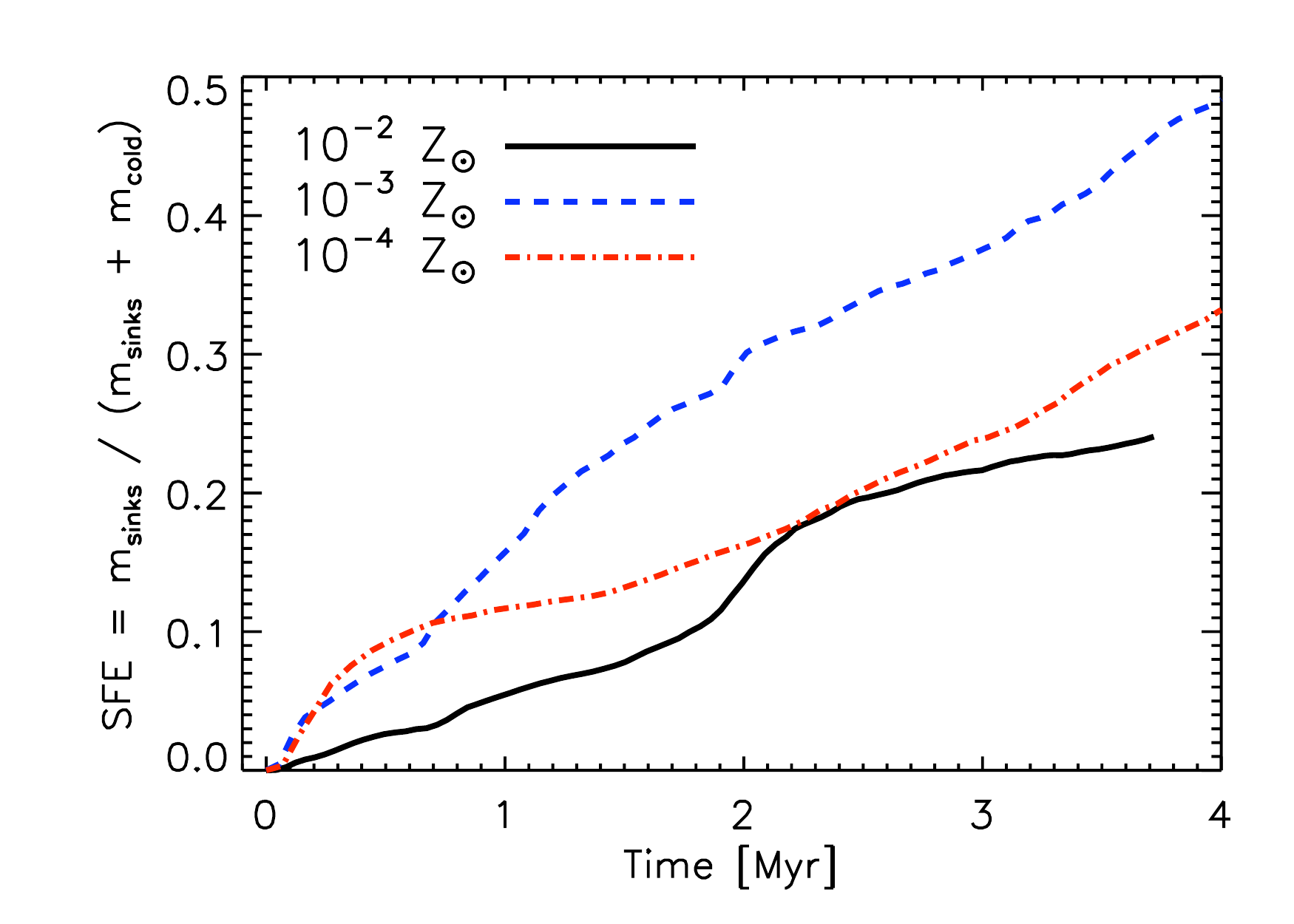}
\caption{Star formation efficiency as a function of time in the three different metallicity runs. We define the star formation efficiency as $\mathrm{SFE}(t)\equiv M_{\mathrm{sinks}}(t) / (M_{\mathrm{gas,cold}}(t) + M_{\mathrm{sinks}}(t))$ where $M_{\mathrm{gas,cold}}$ is the total mass of gas with temperature $T<1000\,\kelvin$ and density $n>10\,\cc$. This efficiency measures the efficiency with which the cold and dense gas is being incorporated in stars. Since all the curves increase with time, the sink particles are depleting the cold gas supply faster than gas is able to cool, implying that eventually, the star-forming cluster becomes gas-starved. }
\label{fig:sink_sfe}
\end{figure}

As discussed in Section \ref{sec:sink_particle_accretion}, the lowest metallicity simulation converts less mass, by roughly a factor of 3, into sink particles than the higher metallicity simulations. Is this smaller mass the result of inefficient accretion of gas that has already cooled and collapsed, or is it instead due to inefficient cooling and collapse starving the sink particle? To answer this, we compute the star formation efficiency (SFE), which we define as 
\begin{equation}
\mathrm{SFE} = \frac{M_{\mathrm{sinks}}}{M_{\mathrm{sinks}} + M_{\mathrm{gas,cold}}} \mbox{,}
\label{eq:sfe}
\end{equation}
where $M_{\mathrm{gas,cold}}$ is the total mass in gas that has cooled and collapsed as a result of metal fine-structure emission, defined to have temperature $T<1000\,\kelvin$ and density $n>10\,\cc$, and again we are taking sink particle mass as a proxy for the total stellar mass in the stellar association that the sink represents. Given that in our simulations gas both flows from the halo to the star forming region, and from the star forming region into sink particles, Equation \ref{eq:sfe} is better construed as a measure of how effective sink particles accrete gas that has already cooled and collapsed. We show the time evolution of the SFE in Figure \ref{fig:sink_sfe}. The similarity in the SFE between the three runs is most striking, differing by no more than a factor of $\sim3$ at any time. This suggests that sink particles, regardless of gas metallicity, are equally efficient at accreting cold, dense gas. The primary difference stems from how gas is supplied to the star forming region --- in the lowest metallicity simulation, gas cooling is inefficient and accretion onto sinks is modulated by gas cooling. In high metallicity gas ($Z\sim10^{-2}\,\zsun$), the total stellar mass is modulated by the hydrodynamics of gas accretion onto prestellar cores, not by the ability of the gas to cool.

\section{Caveats and Limitations}
\label{sec:caveats}

The most significant approximation used in the simulations is our metal enrichment strategy. Recall that when the target halo has entered the $\lya$ cooling-enabled isothermal collapse, we endowed gas in the vicinity of the halo with a non-zero metallicity. The metallicity was made spatially uniform, implying that metal mixing acted with perfect efficiency, which is certainly not the case. Realistically, the mixing is not instantaneous and could potentially result in concurrent Pop III and Pop II star formation \citep[e.g.,][]{deAvillez02,Pan11,Pan13}. The process of metal mixing may begin in hydrodynamic instabilities of supernovae \citep[e.g.,][]{Joggerst09} and supernova remnants \citep[e.g.,][]{Madau01,Ritter12} and following turbulent transport is ultimately completed on microscopic scales via molecular diffusion. Since molecular diffusion is effective only on microscopic spatial scales \citep[e.g.,][]{Oey03,Pan07}, turbulence is paramount for efficient mixing of primordial and metal-enriched gas. There is evidence that the process of virialization and the inflow of dark matter directed `cold accretion' streams in atomic cooling haloes can be effective in driving supersonic turbulent motions \citep{Wise07,Greif08,Prieto12,Latif13} and thus accelerating the mixing of primordial and metal enriched gas. \citet{Frebel12} suggested that a signature of incomplete mixing can be found in the chemical inhomogeneity of metal-poor stellar populations, particularly those in ultra-faint dwarf galaxies.

The fact that metals were not present during the initial virialization process of the atomic cooling halo may also have consequences. If the high density, cold accretion streams that account for the majority of baryonic accretion into the halo had been metal enriched we would have expected them to have significantly lower temperatures than if the gas was primordial. Given the lower temperatures, the stream termination shocks would be more compressive and this could affect the virialization and star formation processes. Our choice of the time to raise the gas metallicity to a non-zero value may also directly set the characteristic fragmentation mass. In all simulations the metals were introduced when the maximum gas density in the halo reached $100\,\cc$. Increasing the metallicity of the gas at a lower (higher) peak density threshold would have let the gas hit the CMB floor at a smaller (larger) density, thus increasing (decreasing) the characteristic fragmentation mass scale. Our main result, though, concerning efficient fragmentation above $Z=10^{-4}\,\zsun$ is robust in this respect, as is the typical gas accretion rate and total cluster mass. Additionally, actual protostellar fragments must ultimately emerge at higher densities in the presence of thermal gas-dust coupling and would be unaffected by this caveat.

Our simulations did not include any prescriptions for stellar feedback, processes well known to play significant roles in star formation, allowing us to isolate the process of gas collapse and fragmentation in the presence of metal cooling. We run the simulations for $4$ Myr past the formation of the first sink particle, a timescale meant to capture only the initial starburst. In this time we do not expect supernovae, stellar winds, or stellar evolutionary effects to play significant roles given these processes operate on longer timescales ($\sim3-100$ Myr). Radiative feedback from accreting protostars should influence the outcome of turbulent fragmentation in the first metal-enriched star forming regions. Detailed radiation hydrodynamic simulations \citep[e.g.,][]{Krumholz07,Bate12} have shown that the primary impact of radiative heating is the suppression of fragmentation. This results in fewer low-mass brown dwarf-like objects \citep[e.g.,][]{Offner09} and may enable the formation of high mass stars in dense star forming regions \citep[e.g.,][]{Krumholz12}. This suppression may be less effective in the present simulations given the much smaller metallicity and dust abundance \citep[though see][]{Myers11}. Simulations of metal-free star formation that include prescriptions for radiation from accreting protostars \citep[e.g.,][]{Smith11,Greif11} find fragmentation to be delayed by the feedback, but not suppressed. In addition to radiative feedback, accreting protostars are known to posses energetic jets that are capable of driving turbulence and reducing the rate of star formation in actively star forming regions \citep{Li06,Matzner07,Wang10,Cunningham11,Krumholz12}. These jets may be especially significant in the regime considered here since our results suggest these earliest stellar clusters formed in relatively dense environments ($\sim0.01-1\,\pc$) where the effect from these outflows would be more pronounced.

We do not include dust grains in our chemical model. This eliminates dust catalyzed $\htwo$ formation and dust-gas collisional coupling that can act to heat or cool the gas. The neglect of dust-gas coupling is well justified since this does not occur at densities $< \sim10^{8}\,\cc$ that we resolve \citep[e.g.,][]{Omukai05}. The dust-grain-catalyzed formation of $\htwo$ may have more significant thermal consequences. However, given that much of the star forming gas in our simulations has temperatures below $100\,\kelvin$ where $\htwo$ cooling is ineffective, this would completely eliminate $\htwo$ as an effective coolant even if it did form in much higher abundance. The formation of $\htwo$, though, is an exothermic reaction and thus elevated $\htwo$ formation from dust grains, while not augmenting the cooling rate, may act as a heating source.

Finally, it is worth stressing that the results of the simulations here are resolution dependent. At higher or lower resolutions, resolving higher and lower maximum densities,, different physical processes will dominate the gas thermodynamics. Gas-dust coupling would have a significant thermodynamic effect on the gas at much higher densities ($n\gtrsim10^{10}\,\cc$). At lower densities ($n\lesssim10^3\,\cc$) there would be almost no difference between the different metallicities considered here (see Figure \ref{fig:dens_temp_multiplot}). Our resolution was chosen to most effectively study the effect that metal fine-structure cooling has on the fragmentation of collapsing gas concentrations. The results of the simulations are also sensitive to the sink particle parameters we used (such as the accretion radius, softening length, density for sink creation, etc.). In preparation for the production simulations presented here, exploratory simulations were performed that varied the resolution and the sink particle parameters. Modification of the gas-sink softening length and accretion radius had the largest effect on sink particle creation, growth, and dynamics. Overall, the main difference between these initial simulations and the ones presented here was the number of sink particles created, though never by more than a factor of $\sim1.5$.The $Z=10^{-4}\,\zsun$ simulations always produced only one sink particle independent of the resolution or sink particle parameters.

\section{Discussion and Summary}
\label{sec:summary}

In this work, we have focused on the effect that metallic fine-structure line cooling has on the process and outcome of star formation in a high-redshift atomic cooling halo. We have run three high-resolution, zoomed cosmological simulations until the assembly of a $M_{\mathrm{vir}}\approx2\times10^7\,\msun$ halo at redshift $\sim16$ capable of atomic cooling. At this point, we endowed gas in the halo with a non-zero metallicity and observed the subsequent collapse and progression of star formation. We utilized the sink particle technique, allowing us to follow the process of star formation for many free-fall times past the first point of gravitational runaway collapse.

Theory predicts that below a metallicity of $Z\sim10^{-3.5}\,\zsun$, fine-structure metal line cooling is not strong enough to significantly alter the thermodynamic and fragmentation behavior of collapsing gas \citep[e.g.,][]{Bromm01}. We indeed confirm this by finding a marked difference in gas morphology and fragmentation between the $10^{-3}$ and $10^{-4}\,\zsun$ runs (as shown in Figures \ref{fig:mn3_morph_dens} and \ref{fig:mn4_morph_dens}). In the higher metallicity runs, $10^{-2}$ and $10^{-3}\,\zsun$, the gas acquires a disordered, turbulent flow morphology. In the $10^{-4}\,\zsun$ run, cooling was not effective enough to allow the gas to reach $\tcmb$, the thermal Jeans mass stayed relatively high, and no fragmentation occurred. We attribute this to two key reasons.  First, the lower cooling rate in the low metallicity simulation produces gas with higher temperatures and a correspondingly higher thermal Jeans mass, decreasing the propensity for fragmentation. Additionally, the higher Mach number in the higher metallicity runs where gas was able to reach $\tcmb$ led to larger compressibility and a greater propensity for dynamical instabilities to induce small scale density fluctuations. This was not the case in the $10^{-4}\,\zsun$ run where the collapse took longer (by roughly a factor of 2), maintained a large degree of spherical symmetry, and underwent no fragmentation.

The accretion rates of individual sink particles are well approximated by $\cs^3/G$, potentially over many millions of years. Quite generically, sink particle accretion rates begin low and increase to some maximum value within a few $\times10^4$ years. This first phase ends when the initial Jeans unstable gas clump is accreted by the sink. Sink particles destined to become the most massive have their accretion rate fall off by only a factor of $2-10$ after this phase, but stay on order of $\cs^3/G$ for many Myr. Lower mass sinks have their accretion terminated shortly after consuming all or a portion of the initially Jeans unstable gas clump, mainly through dynamical encounters with larger mass sinks. This suggests protostellar accretion is a combination of  relatively quick collapse ($t\sim10^5\,\yrs$) and sustained accretion as the stellar association moves throughout the parent gas cloud. Sink particle masses are controlled by both the physical properties of the parent gas cloud, such as the Jeans mass when the gas hits the $\tcmb$, and dynamical encounters that regulate the long-term accretion rate. The creation of stellar mass fragments must predominantly occur at densities above our resolution limit, possibly as a result of dust-gas coupling \citep{Schneider06,Dopcke13} or through the fragmentation of gravitationally unstable disks \citep{Stacy10,Greif11}. Overall, it seems most appropriate to describe the process of gaseous fragmentation and sink particle growth in the higher metallicity simulations as a hybrid between competitive accretion and monolithic collapse.

We do not detect suppression of fragmentation at high metallicities due to the CMB temperature floor suggested by \citet{Smith09}. Gas in the $10^{-2}\,\zsun$ run does become more strongly coupled to the CMB temperature than the $10^{-3}\,\zsun$ run, but continues to fragment. In fact, both runs with metallicities above the critical value for efficient fine-structure cooling, $Z_{\mathrm{crit}}\approx10^{-3.5}\,\zsun$, exhibited roughly the same fragmentation behavior as discussed in Section \ref{sec:results}, although the primary cause of the fragmentation was different between the two runs as discussed in Section \ref{sec:morph_and_dens_evol}. Given the clump-finding approach of \citet{Smith09} to characterize the amount of fragmentation, our differing results are unsurprising. Shown in Figure \ref{fig:sinkmass}, fragmentation and sink particle formation is continuous over many Myr, well after the initial runaway gravitational collapse. It is possible that another run with metallicity between $10^{-4}$ and $10^{-3}\,\zsun$ would have demonstrated an elevated degree of fragmentation, but even so this would suggest that the `metallicity regulated' star formation mode of \citet{Smith09} would exist only under a very narrow range of metallicities and would not be a significant star formation pathway.

We also find disagreement with the simulations of \citet{Jappsen07,Jappsen09,Jappsen09a} who find no metallicity threshold that distinguishes Pop III and Pop II star formation, and instead argue it is the initial conditions (e.g., level of rotation, initial temperature, degree of turbulence) that moderate the process of gas fragmentation, whereas we unambiguously identify a metallicity threshold in controlling fragmentation. \citet{Jappsen07} argues that $\htwo$ cooling is partly responsible for eliminating a Pop III-II metallicity threshold given that it may be a dominant coolant even in the presence of metals. In all three of the simulations presented here, $\htwo$ was a significant coolant even given the strong Lyman-Werner background we included, though there was still a clear metallicity threshold for widespread fragmentation between $10^{-4}$ and $10^{-3}\,\zsun$. The disagreement additionally stems from the differing redshift at which the collapse occurs. In \citet{Jappsen09}, this redshift was $25$ where the CMB temperature is $\sim70\,\kelvin$, as opposed to a CMB temperature of $\tcmb(z\approx16)\sim50\,\kelvin$ in the present work. Given Jeans mass scale fragmentation sets in when gas reaches $\tcmb$ and the temperature dependance of the Jeans mass, $M_{\mathrm{J}}\propto T^{3/2}$, this difference in redshift can potentially be significant in suppressing fragmentation \citep[see also][]{Clarke03}.

In previous work, \citet{SafranekShrader10}, we presented a semi-analytical, one-zone model for fragmentation in metal-enriched atomic cooling halos with results directly applicable to the present study. The model assumed cold-flow accretion into a $M_{\mathrm{vir}}=10^{8}\,\msun$ halo at $z=10$. These cold flows, when shocked in the halo's interior, were heated to a temperature of $T=10^{4}\,\kelvin$, compressed to a density $n=4\times10^{3}\,\cc$, and began to isobarically cool. Continued gas accretion resulted in the compressed layer increasing in size until the sound crossing time across the fragment exceeded its free-fall time, our criteria for the onset of fragmentation. While highly simplified, it supported the conclusion, here and elsewhere, that above $10^{-4}\,\zsun$ gas will experience fragmentation due to metal line-cooling. \citet{SafranekShrader10} did predict a much smaller fragmentation mass scale than found here. We attribute this discrepancy to an overestimate of the post-shock density and the lower redshift and cooler CMB temperature. We also argued that gas with metallicity $10^{-2}\,\zsun$ should not evolve isobarically given its short cooling time compared to the sound crossing time across the compressed post-shock layer. We find this is not the case if considering only the sound crossing time of one local pressure scale height. Overall, it is clear that the process of gas fragmentation in high-redshift atomic cooling halos is complex and difficult to analyze using one-zone models.

Extrapolating beyond the results of the simulation here, it seems plausible that two different fragmentation episodes occur in low metallicity gas at high redshifts. The first occurs at relatively low densities, $n\sim10^{3}-10^{4}\,\cc$, when gas reaches the CMB floor as a result of metal line cooling. As shown in the simulations here, this mode is capable of forming a compact cluster of size $\sim1\,\pc$ and fragment masses on the order of $50-100\,\msun$. Solar mass fragments, however, are not expected to form as a result of metal-line-induced fragmentation, especially at high redshifts, $z>10$, since the CMB sets a thermodynamic lower limit on the temperature that gas can reach. The second fragmentation episode occurs at much higher densities, $n\sim10^{10}-10^{14}\,\cc$, when dust grains and gas thermally couple. This dust-induced fragmentation is capable of producing solar mass scale fragments \citep[e.g.,][]{Clark08,Dopcke13} that have the potential to survive until the present day. Whether these two modes of fragmentation acted in tandem, or were individually responsible for separate stellar populations \citep[e.g.,][]{Norris13} is yet to be determined. An intriguing possibility, supported by the simulations presented here, is that dust-induced fragmentation is responsible for isolated solar mass Pop II stars, while the formation of a bona-fide stellar cluster requires metal fine-structure line cooling operating at lower densities and larger spatial scales \citep[e.g.,][]{Karlsson13}.

Higher resolution studies will probe densities $n\gtrsim10^{14}\,\cc$ approaching the final opacity limit for fragmentation. These future studies are crucial to fully assess the impact of metallicity and dust on the Pop III to II transition and constrain the resulting stellar IMF. Simulations that focus on the mechanical, radiative, and chemical feedback from this first metal-enriched stellar generation will provide a clearer picture of stellar assembly in the first galaxies and hints to what next generation telescopes, such as the JWST, will observe.

\section*{Acknowledgments}

CSS is grateful to Christoph Federrath, Jeremy Ritter, and Meghann Agarwal for valuable help in technical and theoretical matters. The author also thanks Harriet Dinerstein, Steve Finkelstein, Jacob Hummel, Paul Ricker, John Scalo, and Raffaella Schneider for useful comments and insights. We thank the anonymous referee for an extremely thorough and helpful critique of the original text. We acknowledge Robi Banerjee and Christopher Lindner for providing software used to produce some of the visualizations here. The FLASH code was in part developed by the DOE-supported Alliance Center for Astrophysical Thermonuclear Flashes (ACS) at the University of Chicago. The authors acknowledge the Texas Advanced Computing Center (TACC) at The University of Texas at Austin for providing HPC resources under XSEDE allocation TG-AST120024. CSS is especially grateful to generous support provided by the NASA Earth and Space Science Fellowship (NESSF) program. This study was supported in part by NSF grant AST-1009928 and by the NASA grant NNX09AJ33G. This research has made use of NASA's Astrophysics Data System.

\bibliography{metallicity_influence2}

\hyphenation{Post-Script Sprin-ger}
\begin{thebibliography}{176}
\expandafter\ifx\csname natexlab\endcsname\relax\def\natexlab#1{#1}\fi

\bibitem[{{Abel} {et~al}\mbox{.}(1997){Abel}, {Anninos}, {Zhang}, \&
  {Norman}}]{Abel97}
{Abel} T., {Anninos} P., {Zhang} Y., {Norman} M.~L., 1997, \na, 2, 181

\bibitem[{{Abel} {et~al}\mbox{.}(2000){Abel}, {Bryan}, \& {Norman}}]{Abel00}
{Abel} T., {Bryan} G.~L., {Norman} M.~L., 2000, \apj, 540, 39

\bibitem[{{Abel} {et~al}\mbox{.}(2002){Abel}, {Bryan}, \& {Norman}}]{Abel02}
{Abel} T., {Bryan} G.~L., {Norman} M.~L., 2002, Science, 295, 93

\bibitem[{{Andr{\'e}} {et~al}\mbox{.}(2010){Andr{\'e}}, {Men'shchikov},
  {Bontemps}, {K{\"o}nyves}, {Motte}, {Schneider}, {Didelon}, {Minier},
  {Saraceno}, {Ward-Thompson}, {di Francesco}, {White}, {Molinari}, {Testi},
  {Abergel}, {Griffin}, {Henning}, {Royer}, {Mer{\'{\i}}n}, {Vavrek}, {Attard},
  {Arzoumanian}, {Wilson}, {Ade}, {Aussel}, {Baluteau}, {Benedettini},
  {Bernard}, {Blommaert}, {Cambr{\'e}sy}, {Cox}, {di Giorgio}, {Hargrave},
  {Hennemann}, {Huang}, {Kirk}, {Krause}, {Launhardt}, {Leeks}, {Le Pennec},
  {Li}, {Martin}, {Maury}, {Olofsson}, {Omont}, {Peretto}, {Pezzuto}, {Prusti},
  {Roussel}, {Russeil}, {Sauvage}, {Sibthorpe}, {Sicilia-Aguilar}, {Spinoglio},
  {Waelkens}, {Woodcraft}, \& {Zavagno}}]{Andre10}
{Andr{\'e}} P. {et~al.}, 2010, \aap, 518, L102

\bibitem[{{Aykutalp} \& {Spaans}(2011)}]{Aykutalp11}
{Aykutalp} A., {Spaans} M., 2011, \apj, 737, 63

\bibitem[{{Bailin} {et~al}\mbox{.}(2010){Bailin}, {Stinson}, {Couchman},
  {Harris}, {Wadsley}, \& {Shen}}]{Ballin10}
{Bailin} J., {Stinson} G., {Couchman} H., {Harris} W.~E., {Wadsley} J., {Shen}
  S., 2010, \apj, 715, 194

\bibitem[{{Basu}(1997)}]{Basu97}
{Basu} S., 1997, \apj, 485, 240

\bibitem[{{Bate}(2009)}]{Bate09}
{Bate} M.~R., 2009, \mnras, 392, 590

\bibitem[{{Bate}(2012)}]{Bate12}
{Bate} M.~R., 2012, \mnras, 419, 3115

\bibitem[{{Bate} {et~al}\mbox{.}(2003){Bate}, {Bonnell}, \& {Bromm}}]{Bate03}
{Bate} M.~R., {Bonnell} I.~A., {Bromm} V., 2003, \mnras, 339, 577

\bibitem[{{Bate} {et~al}\mbox{.}(1995){Bate}, {Bonnell}, \& {Price}}]{Bate95}
{Bate} M.~R., {Bonnell} I.~A., {Price} N.~M., 1995, \mnras, 277, 362

\bibitem[{{Bate} {et~al}\mbox{.}(2010){Bate}, {Lodato}, \& {Pringle}}]{Bate10}
{Bate} M.~R., {Lodato} G., {Pringle} J.~E., 2010, \mnras, 401, 1505

\bibitem[{{Belokurov} {et~al}\mbox{.}(2007){Belokurov}, {Zucker}, {Evans},
  {Kleyna}, {Koposov}, {Hodgkin}, {Irwin}, {Gilmore}, {Wilkinson}, {Fellhauer},
  {Bramich}, {Hewett}, {Vidrih}, {De Jong}, {Smith}, {Rix}, {Bell}, {Wyse},
  {Newberg}, {Mayeur}, {Yanny}, {Rockosi}, {Gnedin}, {Schneider}, {Beers},
  {Barentine}, {Brewington}, {Brinkmann}, {Harvanek}, {Kleinman}, {Krzesinski},
  {Long}, {Nitta}, \& {Snedden}}]{Belokurov07}
{Belokurov} V. {et~al.}, 2007, \apj, 654, 897

\bibitem[{{Bergin} \& {Tafalla}(2007)}]{Bergin07}
{Bergin} E.~A., {Tafalla} M., 2007, \araa, 45, 339

\bibitem[{{Beuther} \& {Schilke}(2004)}]{Beuther04}
{Beuther} H., {Schilke} P., 2004, Sci, 303, 1167

\bibitem[{{Bland-Hawthorn} {et~al}\mbox{.}(2010){Bland-Hawthorn}, {Karlsson},
  {Sharma}, {Krumholz}, \& {Silk}}]{BlandHawthorn10}
{Bland-Hawthorn} J., {Karlsson} T., {Sharma} S., {Krumholz} M., {Silk} J.,
  2010, \apj, 721, 582

\bibitem[{{Bonnell} \& {Bastien}(1993)}]{Bonnell93}
{Bonnell} I., {Bastien} P., 1993, \apj, 406, 614

\bibitem[{{Bonnell} {et~al}\mbox{.}(2001){Bonnell}, {Bate}, {Clarke}, \&
  {Pringle}}]{Bonnell01}
{Bonnell} I.~A., {Bate} M.~R., {Clarke} C.~J., {Pringle} J.~E., 2001, \mnras,
  323, 785

\bibitem[{{Bouwens} {et~al}\mbox{.}(2011){Bouwens}, {Illingworth}, {Oesch},
  {Labb{\'e}}, {Trenti}, {van Dokkum}, {Franx}, {Stiavelli}, {Carollo},
  {Magee}, \& {Gonzalez}}]{Bouwens11}
{Bouwens} R.~J. {et~al.}, 2011, \apj, 737, 90

\bibitem[{{Bromm}(2013)}]{Bromm13}
{Bromm} V., 2013, ArXiv e-prints, 1305.5178

\bibitem[{{Bromm} {et~al}\mbox{.}(2002){Bromm}, {Coppi}, \& {Larson}}]{Bromm02}
{Bromm} V., {Coppi} P.~S., {Larson} R.~B., 2002, \apj, 564, 23

\bibitem[{{Bromm} {et~al}\mbox{.}(2001){Bromm}, {Ferrara}, {Coppi}, \&
  {Larson}}]{Bromm01}
{Bromm} V., {Ferrara} A., {Coppi} P.~S., {Larson} R.~B., 2001, \mnras, 328, 969

\bibitem[{{Bromm} \& {Loeb}(2003)}]{Bromm03}
{Bromm} V., {Loeb} A., 2003, \nat, 425, 812

\bibitem[{{Caffau} {et~al}\mbox{.}(2011){Caffau}, {Bonifacio}, {Fran{\c c}ois},
  {Sbordone}, {Monaco}, {Spite}, {Spite}, {Ludwig}, {Cayrel}, {Zaggia},
  {Hammer}, {Randich}, {Molaro}, \& {Hill}}]{Caffau11}
{Caffau} E. {et~al.}, 2011, \nat, 477, 67

\bibitem[{{Chabrier}(2003)}]{Chabrier03}
{Chabrier} G., 2003, \pasp, 115, 763

\bibitem[{{Chatzopoulos} \& {Wheeler}(2012)}]{Chatzopoulos12}
{Chatzopoulos} E., {Wheeler} J.~C., 2012, \apj, 748, 42

\bibitem[{{Clark} {et~al}\mbox{.}(2008){Clark}, {Glover}, \&
  {Klessen}}]{Clark08}
{Clark} P.~C., {Glover} S.~C.~O., {Klessen} R.~S., 2008, \apj, 672, 757

\bibitem[{{Clark} {et~al}\mbox{.}(2011{\natexlab{a}}){Clark}, {Glover},
  {Klessen}, \& {Bromm}}]{Clark11}
{Clark} P.~C., {Glover} S.~C.~O., {Klessen} R.~S., {Bromm} V.,
  2011{\natexlab{a}}, \apj, 727, 110

\bibitem[{{Clark} {et~al}\mbox{.}(2011{\natexlab{b}}){Clark}, {Glover},
  {Smith}, {Greif}, {Klessen}, \& {Bromm}}]{Clark11a}
{Clark} P.~C., {Glover} S.~C.~O., {Smith} R.~J., {Greif} T.~H., {Klessen}
  R.~S., {Bromm} V., 2011{\natexlab{b}}, Sci, 331, 1040

\bibitem[{{Clarke} \& {Bromm}(2003)}]{Clarke03}
{Clarke} C.~J., {Bromm} V., 2003, \mnras, 343, 1224

\bibitem[{{Couchman} \& {Rees}(1986)}]{Couchman86}
{Couchman} H.~M.~P., {Rees} M.~J., 1986, \mnras, 221, 53

\bibitem[{{Cunningham} {et~al}\mbox{.}(2011){Cunningham}, {Klein}, {Krumholz},
  \& {McKee}}]{Cunningham11}
{Cunningham} A.~J., {Klein} R.~I., {Krumholz} M.~R., {McKee} C.~F., 2011, \apj,
  740, 107

\bibitem[{{de Avillez} \& {Mac Low}(2002)}]{deAvillez02}
{de Avillez} M.~A., {Mac Low} M.-M., 2002, \apj, 581, 1047

\bibitem[{{Dopcke} {et~al}\mbox{.}(2011){Dopcke}, {Glover}, {Clark}, \&
  {Klessen}}]{Dopcke11}
{Dopcke} G., {Glover} S.~C.~O., {Clark} P.~C., {Klessen} R.~S., 2011, \apjl,
  729, L3

\bibitem[{{Dopcke} {et~al}\mbox{.}(2013){Dopcke}, {Glover}, {Clark}, \&
  {Klessen}}]{Dopcke13}
{Dopcke} G., {Glover} S.~C.~O., {Clark} P.~C., {Klessen} R.~S., 2013, \apj,
  766, 103

\bibitem[{{Draine} \& {Bertoldi}(1996)}]{Draine96}
{Draine} B.~T., {Bertoldi} F., 1996, \apj, 468, 269

\bibitem[{{Dunlop}(2013)}]{Dunlop13}
{Dunlop} J.~S., 2013, in Astrophysics and Space Science Library, Vol. 396,
  Astrophysics and Space Science Library, {Wiklind} T., {Mobasher} B., {Bromm}
  V., eds., p. 223

\bibitem[{{Dwek} \& {Cherchneff}(2011)}]{Dwek11}
{Dwek} E., {Cherchneff} I., 2011, \apj, 727, 63

\bibitem[{{Elmegreen}(1993)}]{Elmegreen93}
{Elmegreen} B.~G., 1993, \apjl, 419, L29

\bibitem[{{Evans} {et~al}\mbox{.}(2009){Evans}, {Dunham}, {J{\o}rgensen},
  {Enoch}, {Mer{\'{\i}}n}, {van Dishoeck}, {Alcal{\'a}}, {Myers},
  {Stapelfeldt}, {Huard}, {Allen}, {Harvey}, {van Kempen}, {Blake}, {Koerner},
  {Mundy}, {Padgett}, \& {Sargent}}]{Evans09}
{Evans}, II N.~J. {et~al.}, 2009, \apjs, 181, 321

\bibitem[{{Federrath} {et~al}\mbox{.}(2010{\natexlab{a}}){Federrath},
  {Banerjee}, {Clark}, \& {Klessen}}]{Federrath10}
{Federrath} C., {Banerjee} R., {Clark} P.~C., {Klessen} R.~S.,
  2010{\natexlab{a}}, \apj, 713, 269

\bibitem[{{Federrath} \& {Klessen}(2012)}]{Federrath12}
{Federrath} C., {Klessen} R.~S., 2012, \apj, 761, 156

\bibitem[{{Federrath} \& {Klessen}(2013)}]{Federrath13}
{Federrath} C., {Klessen} R.~S., 2013, \apj, 763, 51

\bibitem[{{Federrath} {et~al}\mbox{.}(2008){Federrath}, {Klessen}, \&
  {Schmidt}}]{Federrath08}
{Federrath} C., {Klessen} R.~S., {Schmidt} W., 2008, \apjl, 688, L79

\bibitem[{{Federrath} {et~al}\mbox{.}(2010{\natexlab{b}}){Federrath},
  {Roman-Duval}, {Klessen}, {Schmidt}, \& {Mac Low}}]{Federrath10a}
{Federrath} C., {Roman-Duval} J., {Klessen} R.~S., {Schmidt} W., {Mac Low}
  M.-M., 2010{\natexlab{b}}, \aap, 512, A81

\bibitem[{{Federrath} {et~al}\mbox{.}(2011){Federrath}, {Sur}, {Schleicher},
  {Banerjee}, \& {Klessen}}]{Federrath11}
{Federrath} C., {Sur} S., {Schleicher} D.~R.~G., {Banerjee} R., {Klessen}
  R.~S., 2011, \apj, 731, 62

\bibitem[{{Finkelstein} {et~al}\mbox{.}(2012){Finkelstein}, {Papovich}, {Ryan},
  {Pawlik}, {Dickinson}, {Ferguson}, {Finlator}, {Koekemoer}, {Giavalisco},
  {Cooray}, {Dunlop}, {Faber}, {Grogin}, {Kocevski}, \&
  {Newman}}]{Finkelstein12}
{Finkelstein} S.~L. {et~al.}, 2012, \apj, 758, 93

\bibitem[{{Frebel} \& {Bromm}(2012)}]{Frebel12}
{Frebel} A., {Bromm} V., 2012, \apj, 759, 115

\bibitem[{{Frebel} {et~al}\mbox{.}(2007){Frebel}, {Johnson}, \&
  {Bromm}}]{Frebel07}
{Frebel} A., {Johnson} J.~L., {Bromm} V., 2007, \mnras, 380, L40

\bibitem[{{Fryxell} {et~al}\mbox{.}(2000){Fryxell}, {Olson}, {Ricker},
  {Timmes}, {Zingale}, {Lamb}, {MacNeice}, {Rosner}, {Truran}, \&
  {Tufo}}]{Fryxell00}
{Fryxell} B. {et~al.}, 2000, \apjs, 131, 273

\bibitem[{{Gazol} \& {Kim}(2010)}]{Gazol10}
{Gazol} A., {Kim} J., 2010, \apj, 723, 482

\bibitem[{{Glover} \& {Clark}(2013)}]{Glover13}
{Glover} S.~C.~O., {Clark} P.~C., 2013, ArXiv e-prints, 1305.7365

\bibitem[{{Glover} \& {Jappsen}(2007)}]{Glover07}
{Glover} S.~C.~O., {Jappsen} A.-K., 2007, \apj, 666, 1

\bibitem[{{Greif} {et~al}\mbox{.}(2012){Greif}, {Bromm}, {Clark}, {Glover},
  {Smith}, {Klessen}, {Yoshida}, \& {Springel}}]{Greif12}
{Greif} T.~H., {Bromm} V., {Clark} P.~C., {Glover} S.~C.~O., {Smith} R.~J.,
  {Klessen} R.~S., {Yoshida} N., {Springel} V., 2012, \mnras, 424, 399

\bibitem[{{Greif} {et~al}\mbox{.}(2010){Greif}, {Glover}, {Bromm}, \&
  {Klessen}}]{Greif10}
{Greif} T.~H., {Glover} S.~C.~O., {Bromm} V., {Klessen} R.~S., 2010, \apj, 716,
  510

\bibitem[{{Greif} {et~al}\mbox{.}(2008){Greif}, {Johnson}, {Klessen}, \&
  {Bromm}}]{Greif08}
{Greif} T.~H., {Johnson} J.~L., {Klessen} R.~S., {Bromm} V., 2008, \mnras, 387,
  1021

\bibitem[{{Greif} {et~al}\mbox{.}(2011){Greif}, {Springel}, {White}, {Glover},
  {Clark}, {Smith}, {Klessen}, \& {Bromm}}]{Greif11}
{Greif} T.~H., {Springel} V., {White} S.~D.~M., {Glover} S.~C.~O., {Clark}
  P.~C., {Smith} R.~J., {Klessen} R.~S., {Bromm} V., 2011, \apj, 737, 75

\bibitem[{{Haiman} {et~al}\mbox{.}(1996){Haiman}, {Thoul}, \&
  {Loeb}}]{Haiman96}
{Haiman} Z., {Thoul} A.~A., {Loeb} A., 1996, \apj, 464, 523

\bibitem[{{Hartmann} {et~al}\mbox{.}(2001){Hartmann}, {Ballesteros-Paredes}, \&
  {Bergin}}]{Hartmann01}
{Hartmann} L., {Ballesteros-Paredes} J., {Bergin} E.~A., 2001, \apj, 562, 852

\bibitem[{{Heger} {et~al}\mbox{.}(2003){Heger}, {Fryer}, {Woosley}, {Langer},
  \& {Hartmann}}]{Heger03}
{Heger} A., {Fryer} C.~L., {Woosley} S.~E., {Langer} N., {Hartmann} D.~H.,
  2003, \apj, 591, 288

\bibitem[{{Heitsch} {et~al}\mbox{.}(2005){Heitsch}, {Burkert}, {Hartmann},
  {Slyz}, \& {Devriendt}}]{Heitsch05}
{Heitsch} F., {Burkert} A., {Hartmann} L.~W., {Slyz} A.~D., {Devriendt}
  J.~E.~G., 2005, \apjl, 633, L113

\bibitem[{{Heitsch} {et~al}\mbox{.}(2008){Heitsch}, {Hartmann}, {Slyz},
  {Devriendt}, \& {Burkert}}]{Heitsch08}
{Heitsch} F., {Hartmann} L.~W., {Slyz} A.~D., {Devriendt} J.~E.~G., {Burkert}
  A., 2008, \apj, 674, 316

\bibitem[{{Heitsch} {et~al}\mbox{.}(2001){Heitsch}, {Mac Low}, \&
  {Klessen}}]{Heitsch01}
{Heitsch} F., {Mac Low} M.-M., {Klessen} R.~S., 2001, \apj, 547, 280

\bibitem[{{Hennebelle} {et~al}\mbox{.}(2007){Hennebelle}, {Audit}, \&
  {Miville-Desch{\^e}nes}}]{Hennebelle07}
{Hennebelle} P., {Audit} E., {Miville-Desch{\^e}nes} M.-A., 2007, \aap, 465,
  445

\bibitem[{{Hennebelle} \& {P{\'e}rault}(1999)}]{Hennebelle99}
{Hennebelle} P., {P{\'e}rault} M., 1999, \aap, 351, 309

\bibitem[{{Hosokawa} {et~al}\mbox{.}(2011){Hosokawa}, {Omukai}, {Yoshida}, \&
  {Yorke}}]{Hosokawa11}
{Hosokawa} T., {Omukai} K., {Yoshida} N., {Yorke} H.~W., 2011, Sci, 334, 1250

\bibitem[{{Hubber} {et~al}\mbox{.}(2013){Hubber}, {Walch}, \&
  {Whitworth}}]{Hubber13}
{Hubber} D.~A., {Walch} S., {Whitworth} A.~P., 2013, \mnras, 430, 3261

\bibitem[{{Hunter}(1977)}]{Hunter77}
{Hunter} C., 1977, \apj, 218, 834

\bibitem[{{Jappsen} {et~al}\mbox{.}(2007){Jappsen}, {Glover}, {Klessen}, \&
  {Mac Low}}]{Jappsen07}
{Jappsen} A.-K., {Glover} S.~C.~O., {Klessen} R.~S., {Mac Low} M.-M., 2007,
  \apj, 660, 1332

\bibitem[{{Jappsen} {et~al}\mbox{.}(2009{\natexlab{a}}){Jappsen}, {Klessen},
  {Glover}, \& {Mac Low}}]{Jappsen09a}
{Jappsen} A.-K., {Klessen} R.~S., {Glover} S.~C.~O., {Mac Low} M.-M.,
  2009{\natexlab{a}}, \apj, 696, 1065

\bibitem[{{Jappsen} {et~al}\mbox{.}(2005){Jappsen}, {Klessen}, {Larson}, {Li},
  \& {Mac Low}}]{Jappsen05}
{Jappsen} A.-K., {Klessen} R.~S., {Larson} R.~B., {Li} Y., {Mac Low} M.-M.,
  2005, \aap, 435, 611

\bibitem[{{Jappsen} {et~al}\mbox{.}(2009{\natexlab{b}}){Jappsen}, {Mac Low},
  {Glover}, {Klessen}, \& {Kitsionas}}]{Jappsen09}
{Jappsen} A.-K., {Mac Low} M.-M., {Glover} S.~C.~O., {Klessen} R.~S.,
  {Kitsionas} S., 2009{\natexlab{b}}, \apj, 694, 1161

\bibitem[{{Joggerst} {et~al}\mbox{.}(2009){Joggerst}, {Woosley}, \&
  {Heger}}]{Joggerst09}
{Joggerst} C.~C., {Woosley} S.~E., {Heger} A., 2009, \apj, 693, 1780

\bibitem[{{Karlsson} {et~al}\mbox{.}(2013){Karlsson}, {Bromm}, \&
  {Bland-Hawthorn}}]{Karlsson13}
{Karlsson} T., {Bromm} V., {Bland-Hawthorn} J., 2013, Rev. Mod. Phys., 85, 809

\bibitem[{{Kirby} {et~al}\mbox{.}(2011){Kirby}, {Lanfranchi}, {Simon}, {Cohen},
  \& {Guhathakurta}}]{Kirby11}
{Kirby} E.~N., {Lanfranchi} G.~A., {Simon} J.~D., {Cohen} J.~G., {Guhathakurta}
  P., 2011, \apj, 727, 78

\bibitem[{{Klessen}(2001{\natexlab{a}})}]{Klessen01a}
{Klessen} R.~S., 2001{\natexlab{a}}, \apj, 556, 837

\bibitem[{{Klessen}(2001{\natexlab{b}})}]{Klessen01}
{Klessen} R.~S., 2001{\natexlab{b}}, \apjl, 550, L77

\bibitem[{{Klessen} \& {Burkert}(2000)}]{Klessen00}
{Klessen} R.~S., {Burkert} A., 2000, \apjs, 128, 287

\bibitem[{{Klessen} {et~al}\mbox{.}(2012){Klessen}, {Glover}, \&
  {Clark}}]{Klessen12}
{Klessen} R.~S., {Glover} S.~C.~O., {Clark} P.~C., 2012, \mnras, 421, 3217

\bibitem[{{Klessen} {et~al}\mbox{.}(2000){Klessen}, {Heitsch}, \& {Mac
  Low}}]{Klessen00a}
{Klessen} R.~S., {Heitsch} F., {Mac Low} M.-M., 2000, \apj, 535, 887

\bibitem[{{Komatsu} {et~al}\mbox{.}(2011){Komatsu}, {Smith}, {Dunkley},
  {Bennett}, {Gold}, {Hinshaw}, {Jarosik}, {Larson}, {Nolta}, {Page},
  {Spergel}, {Halpern}, {Hill}, {Kogut}, {Limon}, {Meyer}, {Odegard}, {Tucker},
  {Weiland}, {Wollack}, \& {Wright}}]{Komatsu11}
{Komatsu} E. {et~al.}, 2011, \apjs, 192, 18

\bibitem[{{Koyama} \& {Inutsuka}(2000)}]{Koyama00}
{Koyama} H., {Inutsuka} S.-I., 2000, \apj, 532, 980

\bibitem[{{Kritsuk} \& {Norman}(2002)}]{Kritsuk02}
{Kritsuk} A.~G., {Norman} M.~L., 2002, \apjl, 569, L127

\bibitem[{{Kritsuk} {et~al}\mbox{.}(2007){Kritsuk}, {Norman}, {Padoan}, \&
  {Wagner}}]{Kritsuk07}
{Kritsuk} A.~G., {Norman} M.~L., {Padoan} P., {Wagner} R., 2007, \apj, 665, 416

\bibitem[{{Kritsuk} {et~al}\mbox{.}(2011){Kritsuk}, {Norman}, \&
  {Wagner}}]{Kritsuk11}
{Kritsuk} A.~G., {Norman} M.~L., {Wagner} R., 2011, \apjl, 727, L20

\bibitem[{{Kroupa}(2002)}]{Kroupa02}
{Kroupa} P., 2002, Sci, 295, 82

\bibitem[{{Krumholz} {et~al}\mbox{.}(2007){Krumholz}, {Klein}, \&
  {McKee}}]{Krumholz07}
{Krumholz} M.~R., {Klein} R.~I., {McKee} C.~F., 2007, \apj, 656, 959

\bibitem[{{Krumholz} {et~al}\mbox{.}(2012){Krumholz}, {Klein}, \&
  {McKee}}]{Krumholz12}
{Krumholz} M.~R., {Klein} R.~I., {McKee} C.~F., 2012, \apj, 754, 71

\bibitem[{{Krumholz} {et~al}\mbox{.}(2004){Krumholz}, {McKee}, \&
  {Klein}}]{Krumholz04}
{Krumholz} M.~R., {McKee} C.~F., {Klein} R.~I., 2004, \apj, 611, 399

\bibitem[{{Kuhlen} \& {Faucher-Gigu{\`e}re}(2012)}]{Kuhlen12}
{Kuhlen} M., {Faucher-Gigu{\`e}re} C.-A., 2012, \mnras, 423, 862

\bibitem[{{Lada} \& {Lada}(2003)}]{Lada03}
{Lada} C.~J., {Lada} E.~A., 2003, \araa, 41, 57

\bibitem[{{Larson}(1985)}]{Larson85}
{Larson} R.~B., 1985, \mnras, 214, 379

\bibitem[{{Larson}(2003)}]{Larson03}
{Larson} R.~B., 2003, Rep. Prog. Phys., 66, 1651

\bibitem[{{Larson}(2005)}]{Larson05}
{Larson} R.~B., 2005, \mnras, 359, 211

\bibitem[{{Latif} {et~al}\mbox{.}(2013){Latif}, {Schleicher}, {Schmidt}, \&
  {Niemeyer}}]{Latif13}
{Latif} M.~A., {Schleicher} D.~R.~G., {Schmidt} W., {Niemeyer} J., 2013,
  \mnras, 432, 668

\bibitem[{{Latif} {et~al}\mbox{.}(2012){Latif}, {Schleicher}, \&
  {Spaans}}]{Latif12}
{Latif} M.~A., {Schleicher} D.~R.~G., {Spaans} M., 2012, \aap, 540, A101

\bibitem[{{Li} {et~al}\mbox{.}(2003){Li}, {Klessen}, \& {Mac Low}}]{Li03}
{Li} Y., {Klessen} R.~S., {Mac Low} M.-M., 2003, \apj, 592, 975

\bibitem[{{Li} \& {Nakamura}(2006)}]{Li06}
{Li} Z.-Y., {Nakamura} F., 2006, \apjl, 640, L187

\bibitem[{{Low} \& {Lynden-Bell}(1976)}]{Low76}
{Low} C., {Lynden-Bell} D., 1976, \mnras, 176, 367

\bibitem[{{Mac Low} \& {Klessen}(2004)}]{MacLow04}
{Mac Low} M.-M., {Klessen} R.~S., 2004, Rev. Mod. Phys., 76, 125

\bibitem[{{MacDonald} {et~al}\mbox{.}(2013){MacDonald}, {Lawlor}, {Anilmis}, \&
  {Rufo}}]{MacDonald13}
{MacDonald} J., {Lawlor} T.~M., {Anilmis} N., {Rufo} N.~F., 2013, \mnras, 431,
  1425

\bibitem[{{Machida} {et~al}\mbox{.}(2008){Machida}, {Omukai}, {Matsumoto}, \&
  {Inutsuka}}]{Machida08}
{Machida} M.~N., {Omukai} K., {Matsumoto} T., {Inutsuka} S.-i., 2008, \apj,
  677, 813

\bibitem[{{Madau} {et~al}\mbox{.}(2001){Madau}, {Ferrara}, \& {Rees}}]{Madau01}
{Madau} P., {Ferrara} A., {Rees} M.~J., 2001, \apj, 555, 92

\bibitem[{{Maio} {et~al}\mbox{.}(2010){Maio}, {Ciardi}, {Dolag}, {Tornatore},
  \& {Khochfar}}]{Maio10}
{Maio} U., {Ciardi} B., {Dolag} K., {Tornatore} L., {Khochfar} S., 2010,
  \mnras, 407, 1003

\bibitem[{{Matzner}(2007)}]{Matzner07}
{Matzner} C.~D., 2007, \apj, 659, 1394

\bibitem[{{McKee} \& {Ostriker}(2007)}]{McKee07}
{McKee} C.~F., {Ostriker} E.~C., 2007, \araa, 45, 565

\bibitem[{{McKee} \& {Tan}(2003)}]{McKee03}
{McKee} C.~F., {Tan} J.~C., 2003, \apj, 585, 850

\bibitem[{{Motoyama} \& {Yoshida}(2003)}]{Motoyama03}
{Motoyama} K., {Yoshida} T., 2003, \mnras, 344, 461

\bibitem[{{Motte} {et~al}\mbox{.}(1998){Motte}, {Andre}, \& {Neri}}]{Motte98}
{Motte} F., {Andre} P., {Neri} R., 1998, \aap, 336, 150

\bibitem[{{Myers} {et~al}\mbox{.}(2011){Myers}, {Krumholz}, {Klein}, \&
  {McKee}}]{Myers11}
{Myers} A.~T., {Krumholz} M.~R., {Klein} R.~I., {McKee} C.~F., 2011, \apj, 735,
  49

\bibitem[{{Norris} {et~al}\mbox{.}(2013){Norris}, {Yong}, {Bessell},
  {Christlieb}, {Asplund}, {Gilmore}, {Wyse}, {Beers}, {Barklem}, {Frebel}, \&
  {Ryan}}]{Norris13}
{Norris} J.~E. {et~al.}, 2013, \apj, 762, 28

\bibitem[{{Oey}(2003)}]{Oey03}
{Oey} M.~S., 2003, \mnras, 339, 849

\bibitem[{{Offner} {et~al}\mbox{.}(2009){Offner}, {Klein}, {McKee}, \&
  {Krumholz}}]{Offner09}
{Offner} S.~S.~R., {Klein} R.~I., {McKee} C.~F., {Krumholz} M.~R., 2009, \apj,
  703, 131

\bibitem[{{Omukai}(2000)}]{Omukai00}
{Omukai} K., 2000, \apj, 534, 809

\bibitem[{{Omukai} \& {Palla}(2003)}]{Omukai03}
{Omukai} K., {Palla} F., 2003, \apj, 589, 677

\bibitem[{{Omukai} {et~al}\mbox{.}(2008){Omukai}, {Schneider}, \&
  {Haiman}}]{Omukai08}
{Omukai} K., {Schneider} R., {Haiman} Z., 2008, \apj, 686, 801

\bibitem[{{Omukai} {et~al}\mbox{.}(2005){Omukai}, {Tsuribe}, {Schneider}, \&
  {Ferrara}}]{Omukai05}
{Omukai} K., {Tsuribe} T., {Schneider} R., {Ferrara} A., 2005, \apj, 626, 627

\bibitem[{{O'Shea} \& {Norman}(2007)}]{OShea07}
{O'Shea} B.~W., {Norman} M.~L., 2007, \apj, 654, 66

\bibitem[{{Ostriker} {et~al}\mbox{.}(2001){Ostriker}, {Stone}, \&
  {Gammie}}]{Ostriker01}
{Ostriker} E.~C., {Stone} J.~M., {Gammie} C.~F., 2001, \apj, 546, 980

\bibitem[{{Padoan} \& {Nordlund}(1999)}]{Padoan99}
{Padoan} P., {Nordlund} {\AA}., 1999, \apj, 526, 279

\bibitem[{{Padoan} \& {Nordlund}(2002)}]{Padoan02}
{Padoan} P., {Nordlund} {\AA}., 2002, \apj, 576, 870

\bibitem[{{Padoan} \& {Nordlund}(2011)}]{Padoan11}
{Padoan} P., {Nordlund} {\AA}., 2011, \apj, 730, 40

\bibitem[{{Pan} \& {Scalo}(2007)}]{Pan07}
{Pan} L., {Scalo} J., 2007, \apjl, 654, L29

\bibitem[{{Pan} {et~al}\mbox{.}(2011){Pan}, {Scannapieco}, \& {Scalo}}]{Pan11}
{Pan} L., {Scannapieco} E., {Scalo} J., 2011, ArXiv e-prints, 1110.0571

\bibitem[{{Pan} {et~al}\mbox{.}(2013){Pan}, {Scannapieco}, \& {Scalo}}]{Pan13}
{Pan} L., {Scannapieco} E., {Scalo} J., 2013, ArXiv e-prints, 1306.4663

\bibitem[{{Passot} \& {V{\'a}zquez-Semadeni}(1998)}]{Passot98}
{Passot} T., {V{\'a}zquez-Semadeni} E., 1998, \pre, 58, 4501

\bibitem[{{Pawlik} {et~al}\mbox{.}(2011){Pawlik}, {Milosavljevi{\'c}}, \&
  {Bromm}}]{Pawlik11}
{Pawlik} A.~H., {Milosavljevi{\'c}} M., {Bromm} V., 2011, \apj, 731, 54

\bibitem[{{Peters} {et~al}\mbox{.}(2012){Peters}, {Schleicher}, {Klessen},
  {Banerjee}, {Federrath}, {Smith}, \& {Sur}}]{Peters12}
{Peters} T., {Schleicher} D.~R.~G., {Klessen} R.~S., {Banerjee} R., {Federrath}
  C., {Smith} R.~J., {Sur} S., 2012, \apjl, 760, L28

\bibitem[{{Plewa} \& {M{\"u}ller}(1999)}]{Plewa99}
{Plewa} T., {M{\"u}ller} E., 1999, \aap, 342, 179

\bibitem[{{Prieto} {et~al}\mbox{.}(2012){Prieto}, {Jimenez}, \&
  {Mart{\'{\i}}}}]{Prieto12}
{Prieto} J., {Jimenez} R., {Mart{\'{\i}}} J., 2012, \mnras, 419, 3092

\bibitem[{{Rees}(1976)}]{Rees76}
{Rees} M.~J., 1976, \mnras, 176, 483

\bibitem[{{Richardson} {et~al}\mbox{.}(2013){Richardson}, {Scannapieco}, \&
  {Gray}}]{Richardson13}
{Richardson} M.~L.~A., {Scannapieco} E., {Gray} W.~J., 2013, ArXiv e-prints,
  1308.5684

\bibitem[{{Ritter} {et~al}\mbox{.}(2012){Ritter}, {Safranek-Shrader}, {Gnat},
  {Milosavljevi{\'c}}, \& {Bromm}}]{Ritter12}
{Ritter} J.~S., {Safranek-Shrader} C., {Gnat} O., {Milosavljevi{\'c}} M.,
  {Bromm} V., 2012, \apj, 761, 56

\bibitem[{{Robertson} {et~al}\mbox{.}(2013){Robertson}, {Furlanetto},
  {Schneider}, {Charlot}, {Ellis}, {Stark}, {McLure}, {Dunlop}, {Koekemoer},
  {Schenker}, {Ouchi}, {Ono}, {Curtis-Lake}, {Rogers}, {Bowler}, \&
  {Cirasuolo}}]{Robertson13}
{Robertson} B.~E. {et~al.}, 2013, \apj, 768, 71

\bibitem[{{Safranek-Shrader} {et~al}\mbox{.}(2012){Safranek-Shrader},
  {Agarwal}, {Federrath}, {Dubey}, {Milosavljevi{\'c}}, \&
  {Bromm}}]{SafranekShrader12}
{Safranek-Shrader} C., {Agarwal} M., {Federrath} C., {Dubey} A.,
  {Milosavljevi{\'c}} M., {Bromm} V., 2012, \mnras, 426, 1159

\bibitem[{{Safranek-Shrader} {et~al}\mbox{.}(2010){Safranek-Shrader}, {Bromm},
  \& {Milosavljevi{\'c}}}]{SafranekShrader10}
{Safranek-Shrader} C., {Bromm} V., {Milosavljevi{\'c}} M., 2010, \apj, 723,
  1568

\bibitem[{{Salpeter}(1955)}]{Salpeter55}
{Salpeter} E.~E., 1955, \apj, 121, 161

\bibitem[{{Santoro} \& {Shull}(2006)}]{Santoro06}
{Santoro} F., {Shull} J.~M., 2006, \apj, 643, 26

\bibitem[{{Saury} {et~al}\mbox{.}(2013){Saury}, {Miville-Desch{\^e}nes},
  {Hennebelle}, {Audit}, \& {Schmidt}}]{Saury13}
{Saury} E., {Miville-Desch{\^e}nes} M.-A., {Hennebelle} P., {Audit} E.,
  {Schmidt} W., 2013, ArXiv e-prints, 1301.3446

\bibitem[{{Scalo} {et~al}\mbox{.}(1998){Scalo}, {Vazquez-Semadeni}, {Chappell},
  \& {Passot}}]{Scalo98}
{Scalo} J., {Vazquez-Semadeni} E., {Chappell} D., {Passot} T., 1998, \apj, 504,
  835

\bibitem[{{Schmeja} \& {Klessen}(2004)}]{Schmeja04}
{Schmeja} S., {Klessen} R.~S., 2004, \aap, 419, 405

\bibitem[{{Schneider} \& {Omukai}(2010)}]{Schneider10}
{Schneider} R., {Omukai} K., 2010, \mnras, 402, 429

\bibitem[{{Schneider} {et~al}\mbox{.}(2012{\natexlab{a}}){Schneider}, {Omukai},
  {Bianchi}, \& {Valiante}}]{Schneider12a}
{Schneider} R., {Omukai} K., {Bianchi} S., {Valiante} R., 2012{\natexlab{a}},
  \mnras, 419, 1566

\bibitem[{{Schneider} {et~al}\mbox{.}(2006){Schneider}, {Omukai}, {Inoue}, \&
  {Ferrara}}]{Schneider06}
{Schneider} R., {Omukai} K., {Inoue} A.~K., {Ferrara} A., 2006, \mnras, 369,
  1437

\bibitem[{{Schneider} {et~al}\mbox{.}(2012{\natexlab{b}}){Schneider}, {Omukai},
  {Limongi}, {Ferrara}, {Salvaterra}, {Chieffi}, \& {Bianchi}}]{Schneider12}
{Schneider} R., {Omukai} K., {Limongi} M., {Ferrara} A., {Salvaterra} R.,
  {Chieffi} A., {Bianchi} S., 2012{\natexlab{b}}, \mnras, 423, L60

\bibitem[{{Seifried} {et~al}\mbox{.}(2011){Seifried}, {Schmidt}, \&
  {Niemeyer}}]{Siefried11}
{Seifried} D., {Schmidt} W., {Niemeyer} J.~C., 2011, \aap, 526, A14

\bibitem[{{Shu}(1977)}]{Shu77}
{Shu} F.~H., 1977, \apj, 214, 488

\bibitem[{{Smith} \& {Sigurdsson}(2007)}]{Smith07}
{Smith} B.~D., {Sigurdsson} S., 2007, \apjl, 661, L5

\bibitem[{{Smith} {et~al}\mbox{.}(2009){Smith}, {Turk}, {Sigurdsson}, {O'Shea},
  \& {Norman}}]{Smith09}
{Smith} B.~D., {Turk} M.~J., {Sigurdsson} S., {O'Shea} B.~W., {Norman} M.~L.,
  2009, \apj, 691, 441

\bibitem[{{Smith} {et~al}\mbox{.}(2011){Smith}, {Glover}, {Clark}, {Greif}, \&
  {Klessen}}]{Smith11}
{Smith} R.~J., {Glover} S.~C.~O., {Clark} P.~C., {Greif} T., {Klessen} R.~S.,
  2011, \mnras, 414, 3633

\bibitem[{{Spaans} \& {Silk}(2000)}]{Spaans00}
{Spaans} M., {Silk} J., 2000, \apj, 538, 115

\bibitem[{{Stacy} {et~al}\mbox{.}(2010){Stacy}, {Greif}, \& {Bromm}}]{Stacy10}
{Stacy} A., {Greif} T.~H., {Bromm} V., 2010, \mnras, 403, 45

\bibitem[{{Stacy} {et~al}\mbox{.}(2012){Stacy}, {Greif}, \& {Bromm}}]{Stacy12}
{Stacy} A., {Greif} T.~H., {Bromm} V., 2012, \mnras, 422, 290

\bibitem[{{Stecher} \& {Williams}(1967)}]{Stecher67}
{Stecher} T.~P., {Williams} D.~A., 1967, \apjl, 149, L29

\bibitem[{{Sur} {et~al}\mbox{.}(2010){Sur}, {Schleicher}, {Banerjee},
  {Federrath}, \& {Klessen}}]{Sur10}
{Sur} S., {Schleicher} D.~R.~G., {Banerjee} R., {Federrath} C., {Klessen}
  R.~S., 2010, \apjl, 721, L134

\bibitem[{{Susa}(2013)}]{Susa13}
{Susa} H., 2013, ArXiv e-prints, 1304.6794

\bibitem[{{Tegmark} {et~al}\mbox{.}(1997){Tegmark}, {Silk}, {Rees},
  {Blanchard}, {Abel}, \& {Palla}}]{Tegmark97}
{Tegmark} M., {Silk} J., {Rees} M.~J., {Blanchard} A., {Abel} T., {Palla} F.,
  1997, \apj, 474, 1

\bibitem[{{Tornatore} {et~al}\mbox{.}(2007){Tornatore}, {Ferrara}, \&
  {Schneider}}]{Tornatore07}
{Tornatore} L., {Ferrara} A., {Schneider} R., 2007, \mnras, 382, 945

\bibitem[{{Truelove} {et~al}\mbox{.}(1997){Truelove}, {Klein}, {McKee},
  {Holliman}, {Howell}, \& {Greenough}}]{Truelove97}
{Truelove} J.~K., {Klein} R.~I., {McKee} C.~F., {Holliman}, II J.~H., {Howell}
  L.~H., {Greenough} J.~A., 1997, \apjl, 489, L179

\bibitem[{{Tumlinson}(2007)}]{Tumlinson07}
{Tumlinson} J., 2007, \apjl, 664, L63

\bibitem[{{Turk} {et~al}\mbox{.}(2012){Turk}, {Oishi}, {Abel}, \&
  {Bryan}}]{Turk12}
{Turk} M.~J., {Oishi} J.~S., {Abel} T., {Bryan} G.~L., 2012, \apj, 745, 154

\bibitem[{{Urban} {et~al}\mbox{.}(2010){Urban}, {Martel}, \& {Evans}}]{Urban10}
{Urban} A., {Martel} H., {Evans}, II N.~J., 2010, \apj, 710, 1343

\bibitem[{{Vazquez-Semadeni}(1994)}]{VazquezSemadeni94}
{Vazquez-Semadeni} E., 1994, \apj, 423, 681

\bibitem[{{V{\'a}zquez-Semadeni} {et~al}\mbox{.}(2005){V{\'a}zquez-Semadeni},
  {Kim}, \& {Ballesteros-Paredes}}]{VazquezSemadeni05}
{V{\'a}zquez-Semadeni} E., {Kim} J., {Ballesteros-Paredes} J., 2005, \apjl,
  630, L49

\bibitem[{{V{\'a}zquez-Semadeni} {et~al}\mbox{.}(2006){V{\'a}zquez-Semadeni},
  {Ryu}, {Passot}, {Gonz{\'a}lez}, \& {Gazol}}]{VazquezSemadeni06}
{V{\'a}zquez-Semadeni} E., {Ryu} D., {Passot} T., {Gonz{\'a}lez} R.~F., {Gazol}
  A., 2006, \apj, 643, 245

\bibitem[{{Wang} {et~al}\mbox{.}(2010){Wang}, {Li}, {Abel}, \&
  {Nakamura}}]{Wang10}
{Wang} P., {Li} Z.-Y., {Abel} T., {Nakamura} F., 2010, \apj, 709, 27

\bibitem[{{Whalen} {et~al}\mbox{.}(2013){Whalen}, {Johnson}, {Smidt},
  {Meiksin}, {Heger}, {Even}, \& {Fryer}}]{Whalen13}
{Whalen} D.~J., {Johnson} J.~J., {Smidt} J., {Meiksin} A., {Heger} A., {Even}
  W., {Fryer} C.~L., 2013, ArXiv e-prints, 1305.6966

\bibitem[{{Whitworth} \& {Ward-Thompson}(2001)}]{Whitworth01}
{Whitworth} A.~P., {Ward-Thompson} D., 2001, \apj, 547, 317

\bibitem[{{Wise} \& {Abel}(2007)}]{Wise07}
{Wise} J.~H., {Abel} T., 2007, \apj, 665, 899

\bibitem[{{Wise} \& {Abel}(2008)}]{Wise08}
{Wise} J.~H., {Abel} T., 2008, \apj, 685, 40

\bibitem[{{Wise} {et~al}\mbox{.}(2012){Wise}, {Turk}, {Norman}, \&
  {Abel}}]{Wise12}
{Wise} J.~H., {Turk} M.~J., {Norman} M.~L., {Abel} T., 2012, \apj, 745, 50

\bibitem[{{Wolcott-Green} {et~al}\mbox{.}(2011){Wolcott-Green}, {Haiman}, \&
  {Bryan}}]{WolcottGreen11}
{Wolcott-Green} J., {Haiman} Z., {Bryan} G.~L., 2011, \mnras, 418, 838

\bibitem[{{Yoshida} {et~al}\mbox{.}(2006){Yoshida}, {Omukai}, {Hernquist}, \&
  {Abel}}]{Yoshida06}
{Yoshida} N., {Omukai} K., {Hernquist} L., {Abel} T., 2006, \apj, 652, 6

\bibitem[{{Zackrisson} {et~al}\mbox{.}(2011){Zackrisson}, {Rydberg},
  {Schaerer}, {{\"O}stlin}, \& {Tuli}}]{Zackrisson11}
{Zackrisson} E., {Rydberg} C.-E., {Schaerer} D., {{\"O}stlin} G., {Tuli} M.,
  2011, \apj, 740, 13

\bibitem[{{Zackrisson} {et~al}\mbox{.}(2012){Zackrisson}, {Zitrin}, {Trenti},
  {Rydberg}, {Guaita}, {Schaerer}, {Broadhurst}, {{\"O}stlin}, \&
  {Str{\"o}m}}]{Zackrisson12}
{Zackrisson} E. {et~al.}, 2012, \mnras, 427, 2212

\bibitem[{{Zuckerman} \& {Evans}(1974)}]{Zuckerman74}
{Zuckerman} B., {Evans}, II N.~J., 1974, \apjl, 192, L149

\end{thebibliography}
\bibliographystyle{mn2e_fixed}  

\IfFileExists{\jobname.bbl}{}
 {\typeout{}
  \typeout{******************************************}
  \typeout{** Please run "bibtex \jobname" to optain}
  \typeout{** the bibliography and then re-run LaTeX}
  \typeout{** twice to fix the references!}
  \typeout{******************************************}
  \typeout{}
 }

\label{lastpage}

\end{document}